\documentclass[english]{article}
\PassOptionsToPackage{natbib=true}{biblatex}
\usepackage[T1]{fontenc}
\usepackage[latin9]{inputenc}
\usepackage{units}
\usepackage{amsmath}
\usepackage{amsthm}
\usepackage{amssymb}
\usepackage{graphicx}
\usepackage{esint}
\usepackage[all]{xy}
\usepackage{makeidx}
\usepackage{soul}
\usepackage{color}
\makeatletter
\theoremstyle{plain}
\newtheorem{thm}{\protect\theoremname}[section]
\theoremstyle{definition}
\newtheorem{defn}[thm]{\protect\definitionname}
\theoremstyle{plain}
\newtheorem{lem}[thm]{\protect\lemmaname}
\newenvironment{lyxlist}[1]
	{\begin{list}{}
		{\settowidth{\labelwidth}{#1}
		 \setlength{\leftmargin}{\labelwidth}
		 \addtolength{\leftmargin}{\labelsep}
		 }}
	{\end{list}}
\theoremstyle{plain}
\newtheorem{cor}[thm]{\protect\corollaryname}

\@ifundefined{date}{}{\date{}}
\newcommand{\DD}{\mathrm{d}}

\makeatother

\usepackage{babel}
\usepackage[style=authoryear]{biblatex}
\providecommand{\corollaryname}{Corollary}
\providecommand{\definitionname}{Definition}
\providecommand{\lemmaname}{Lemma}
\providecommand{\theoremname}{Theorem}

\makeindex
\begin{document}
\title{Fechnerian Scaling: Dissimilarity Cumulation Theory}
\author{Ehtibar N. Dzhafarov and Hans Colonius }

\maketitle
\tableofcontents{}

\section{Introduction}

\section{Introduction}

\subsection{What is it about?}

\label{sec:Intro}In 1860 Gustav Theodor Fechner published the two-volume
\emph{Elemente der Psychophysik}. From this event one can date scientific
psychology, firmly grounded in mathematics and experimental evidence.
One of the main ideas introduced in Fechner's book is that of measuring
subjective differences between stimuli $\mathbf{a}$ and $\mathbf{b}$
by means of summing (or integrating) just noticeable (or infinitesimal)
differences in the interval of stimuli separating $\mathbf{a}$ and
$\mathbf{b}$. For Fechner, stimuli of a given kind are always represented
by positive reals, so that the interval between them is well-defined.

We use the term ``Fechnerian Scaling''\index{Fechnerian Scaling}
to designate any method of computing distances in a stimulus space
by means of \emph{cumulating} (summing, integrating) values of a \emph{dissimilarity}
\emph{function}\index{dissimilarity function} for pairs of ``neighboring''
stimuli. The term \emph{``dissimilarity cumulation''} can be used
as a synonym of ``Fechnerian Scaling'' or else as designating an
abstract mathematical theory of which Fechnerian Scaling is the main
application.

A \emph{stimulus space}\index{stimulus space} is a set of stimuli
endowed with a structure imposed on this set by an observer's judgments.
Thus, a set of all visible aperture colors such that for each pair
of colors we have a number indicating how often they appear identical
to an observer if presented side by side is an example of a stimulus
space. Stimuli in a stimulus space are referred to as its \emph{points},
and generally are denoted by boldface lowercase letters: $\mathbf{x}_{k},\mathbf{a},\mathbf{b}^{\left(\omega\right)},$
etc. Dissimilarity function is a generalization of the notion of a
\emph{metric}\index{metric}, mapping pairs of stimuli $\left(\mathbf{x},\text{\textbf{y}}\right)$
into nonnegative numbers $D\left(\mathbf{x},\mathbf{y}\right)$. On
a very general level, with minimal assumptions about the structure
of a stimulus space being considered, Fechnerian Scaling\index{Fechnerian Scaling}
is implemented by summing pairwise dissimilarities $D\left(\mathbf{x}_{1},\mathbf{x}_{2}\right)$,
$D\left(\mathbf{x}_{2},\mathbf{x}_{3}\right)$, etc. along \emph{finite
chains} of points $\mathbf{a}=\mathbf{x}_{1},\mathbf{x}_{2},\ldots,\mathbf{x}_{n-1}=\mathbf{b}$.
The distance from $\mathbf{a}$ to \textbf{$\mathbf{b}$} is then
computed as the infimum of these cumulated values over the set of
all such chains. Thus obtained distances from $\mathbf{a}$ to \textbf{$\mathbf{b}$}
and from $\mathbf{b}$ to $\mathbf{a}$ need not be the same, and
to obtain a conventional, symmetric distance, in Fechnerian Scaling
one adds these distances together.

In more specialized stimulus spaces, finite chains can be replaced
with continuous or even continuously differentiable \emph{paths}.
In the latter case the cumulation is replaced with integration along
a path of a certain quantity\emph{, submetric function}\index{submetric function}
$F\left(\mathbf{x},\mathbf{u}\right)$, that depends on the location
$\mathbf{x}$ of a point and the velocity $\mathbf{u}$ with which
it moves along the path. The submetric function is a measure of local
discriminability of $\mathbf{x}$ from its ``immediate'' neighbors
$\mathbf{x}+\mathbf{u}\DD x$, and it can be empirically estimated
by means of one of Fechner's methods for measuring differential thresholds.
The original methods are based on one's ability to compare stimuli
in terms of ``greater than'' with respect to some property (brightness,
loudness, extent, etc.) In more general situations, stimuli can be
compared by a variety of methods based on one's ability to judge whether
two stimuli are the same or different.

The structure defining a stimulus space\index{stimulus space} on
a set of stimuli is always imposed by an observer's judgements of
the stimuli rather than by the way stimuli are measured as physical
objects. In this sense, the structure of stimulus space is a psychological
rather physical construct. For instance, a drawing of human face has
a complex physical description, but if, for example, the faces are
compared in terms of greater-less with respect to some property, such
as ``beauty,'' then, provided certain assumptions are satisfied,
a set of all possible face drawings may form a unidimensional continuum
mappable on an interval of reals. However, physical descriptions of
the stimuli typically have some properties (e.g., order, closeness)
suggestive of the respective properties of the judgements. For instance,
if $\mathbf{a}$ and \textbf{$\mathbf{b}$} have very similar physical
descriptions, one can usually expect the results of their comparisons
with any stimulus $\mathbf{c}$ to also be very similar --- the consideration
we use, e.g., in constructing a differential-geometric version of
Fechnerian Scaling\index{Fechnerian Scaling}.

\subsection{Unidimensional Fechnerian Scaling}

\index{Fechnerian Scaling} Various aspects of Fechner's original
theory are subject to competing interpretations because they are not
presented in his writings with sufficient clarity. The following therefore
is not a historical account. Rather it is a modern theory that preserves
the spirit of Fechner's idea of cumulation of small differences.

Let us assume that stimuli of a particular kind are represented (labeled,
encoded) by values on an interval of positive real numbers $[t,u[$,
where $t$ is the absolute threshold value, and $u$ is an appropriately
defined upper threshold, or infinity. (Throughout this chapter, half-open
or open intervals of reals will always be presented in the form $[t,u[$,
$]t,u]$, $]t,u[$, using only square brackets.) The space structure
on $[t,u[$ is defined by a \emph{psychometric function}\index{psychometric function}
$\gamma\left(\mathbf{x},\mathbf{y}\right)$ that gives us the probability
with which a stimulus $\mathbf{y}$ (represented by a value $y\in[t,u[$)
is judged to be greater than stimulus $\mathbf{x}$ (represented by
a value $x\in[t,u[$). In this special case, it is convenient to simply
replace stimuli with their representations, and write $x,y$ in place
of \textbf{$\mathbf{x},\mathbf{y}$}: 
\begin{equation}
\gamma\left(x,y\right)=\Pr\left[y\text{ is judged to be greater than }x\right].
\end{equation}
We will make the simplifying assumption that 
\begin{equation}
\gamma\left(y,x\right)=1-\gamma\left(x,y\right),\label{eq:balanced gamma}
\end{equation}
with the consequence 
\begin{equation}
\gamma\left(x,x\right)=\nicefrac{1}{2}.\label{eq:balanced gamma 2}
\end{equation}
This will allow us to proceed in this special case without introducing
the notions of observation areas and canonical transformations that
are fundamental for the general theory.

Next, we will make a relatively innocuous assumption that $\gamma\left(x,y\right)$
is strictly increasing in $y$ in the vicinity of $y=x$, and that
it is continuously differentiable in $y$ at $y=x$. That is, the
derivative 
\begin{equation}
F\left(x\right)=\left.\frac{\mathrm{\partial\gamma}\left(x,y\right)}{\mathrm{\partial}y}\right|_{y=x}\label{eq:F(x)}
\end{equation}
exists, is positive, and continuous in $x$. This is the slope of
the psychometric function at its median, and the intuitive meaning
of the differential $F\left(x\right)\mathrm{d}x$ is that it is proportional
to the dissimilarity between $x$ and its ``immediate'' neighbor,
$x+\mathrm{d}x$. We can write this as 
\[
D\left(x,x+\mathrm{d}x\right)=cF\left(x\right)\mathrm{d}x,
\]
where $c$ is a positive constant specific to a given stimulus space.
The intuition of cumulation of differences in this unidimensional
setting is captured by the summation property 
\[
D\left(a,b\right)=D\left(a,c\right)+D\left(c,b\right),
\]
for any $a\leq c\leq b$ in stimulus set $\mathfrak{S}$. It follows
that 
\begin{equation}
D\left(a,b\right)=c\int_{a}^{b}F\left(x\right)\mathrm{d}x.\label{eq:Unter}
\end{equation}
This quantity can be interpreted as the subjective distance between
$a$ and $b$ for any $a\leq b$ in $\mathfrak{S}$. We take the relations
(\ref{eq:F(x)}) and (\ref{eq:Unter}) for the core of the Fechnerian
Scaling in stimulus continua (presented here with simplifying assumptions).

\subsection{Historical Digression: Fechner's Law}

One can easily check that the \emph{logarithmic law} advocated by
Fechner, 
\begin{equation}
D\left(t,x\right)=K\log\frac{x}{t},x\geq t,\label{eq:Fechner law}
\end{equation}
where $K$ is a positive constant, corresponds to 
\begin{equation}
F\left(x\right)=\frac{K}{x},
\end{equation}
which can be viewed as a differential form of the so-called \emph{Weber's
law}. Recall that $t$ designates absolute threshold. 

This is an example of the so-called \emph{psychophysical law}, the
relationship between a physical description of a stimuli $x$, and
the value of $D\left(x,t\right)$, referred to as the \emph{magnitude
of sensation}. In this chapter we attach little importance to this
or other psychophysical laws. In view of the generalization of Fechnerian
Scaling to stimulus spaces with more complex descriptions than real
numbers, such laws have limited scope of applicability. 

Nevertheless, it is appropriate to take a historical detour and look
at how Fechner's law was justified by Fechner himself, in the second
volume of his landmark work\emph{, Elemente der Psychophysik}. The
relationship (\ref{eq:Fechner law}) is referred by Fechner as the
\emph{measurement formula} \emph{(Massformel).} More generally, Fechner's
law can be written as
\begin{equation}
D\left(a,b\right)=D\left(t,b\right)-D\left(t,a\right)=K\log\frac{b}{a},b\geq a\geq t,
\end{equation}
for two stimulus magnitudes $a,b$. Fechner calls this \emph{difference
formula} (\emph{Unterschiedsformel}).

In an addendum to his work \emph{Zen Avesta}, Fechner describes how
the idea of this law occurred to him in the morning of October 22,
1950 (this date is nowadays celebrated as the \emph{Fechner Day}):
he had an insight that an arithmetic progression of sensation magnitude
should correspond to a geometric progression of stimulus magnitudes.
Fechner's insight on that day is all one needs to derive the law,
as logarithm is the only function with non-chaotic behavior that can
transform a geometric progression into an arithmetic one. The derivation
of the law, however, had to wait for 10 more year before it appeared
in vol. 2 of the \emph{Elemente der Psychophysik}, in two different
forms (Chapters 16 and 17).

Unfortunately, the second volume has not been translated into English.
As we learn from a letter written by E. G. Boring to S. Rosenzweig
on February 23, 1968, ``Just now I'm spending long hours working
over translation into English of the second volume of the Fechner's
\emph{Elemente}, because put literally into English it is about as
dull and confusing and sometimes uninterpretable as it always was
in the German. Holt, Rinehart and Winston published the first volume
and someday we will get this second half done, but we do not have
much help after NIH stopped supporting translation. We have to get
it done by little bits.'' It seems that Boring has not completed
this work.

By a historical happenstance, one of Fechner's derivations of his
law was criticized as mathematically incorrect, and the other simply
forgotten. In addition, the law itself was criticized as empirically
incorrect. However, by careful examination of the premises of Fechner's
derivations the mathematical criticisms can be deflected, while empirical
falsifications of the law often involve empirical procedures (e.g.,
direct estimation of sensation magnitudes) that go beyond those Fechner
would consider legitimate. In a paper of rejoinders published in 1877,
Fechner reacts to the criticisms known to him and makes a bold prediction
for the future: ``The tower of Babel was never finished because the
workers could not reach an understanding on how they should build
it; my psychophysical edifice will stand because the workers will
never agree on how to tear it down.''

The difficulty in understanding Fechner's derivations of his logarithmic
law is in that he uses the term ``Weber's law'' in the meaning that
is logically independent of the empirical law established by Ernst
Heinrich Weber (which Fechner, to add to the confusion, also calls
``Weber's law''). According to Weber's law, if $x$ and $x+\Delta x$
are separated by a \emph{just-noticeable difference}, then 
\begin{equation}
\frac{\Delta x}{x}=c^{*},
\end{equation}
where $c^{*}$ is a constant with respect to $x$ (but generally depends
on the stimulus continuum used). In Fechner's mathematical derivations,
however, the term ``Weber's law'' stands for the following statement,
essentially a form of his October 1850 insight :
\begin{quote}
\emph{the subjective dissimilarity $D\left(t,b\right)-D\left(t,a\right)$
between stimuli with physical magnitudes $a$ and $b$ (provided $t\leq a\leq b$)
is determined by the ratio of these magnitudes, $b/a$.}
\end{quote}
We propose calling this statement ``W-principle'' to disentangle
it from Weber's law. The only relationship between the W-principle
and Weber's law can be established through so-called ``\emph{Fechner's
postulate},'' according to which all just-noticeable differences
$\Delta x$ (within a given continuum) are subjectively equal, 
\begin{equation}
D\left(x,x+\Delta x\right)=c.
\end{equation}
Any two of the three statements, Fechner's postulate, Weber's law
(in its usual meaning), and the W-principle implies the third.

In Chapter 17 of the \emph{Elemente}, Fechner derives his law by using
a novel for his time method of \emph{functional equations}. He presents
the W-principle as
\[
\psi\left(b\right)-\psi\left(a\right)=F\left(\frac{b}{a}\right)
\]
where $\psi\left(x\right)$ denotes $D\left(t,x\right)$, and observes
that this implies
\[
F\left(\frac{c}{b}\right)+F\left(\frac{b}{a}\right)=F\left(\frac{c}{a}\right),
\]
for any \emph{$t\leq a\leq b\leq c$}. This in turn means that 
\[
F\left(x\right)+F\left(y\right)=F\left(xy\right),
\]
for any $x,y\geq1$. Fechner recognizes in this the functional equation
introduced only 40 years earlier by Augustin-Louis Cauchy, who showed
that its only continuous solution is
\[
F\left(x\right)=K\log x,x\geq1.
\]
It is known now (Acz\'el, 1987) that continuity can be replaced with
many other regularity assumptions, including monotonicity and nonnegativity,
and that it is sufficient to assume that the equation holds only in
an arbitrarily small vicinity of 1 (i.e., for very similar stimuli
only). It follows that 
\[
\psi\left(b\right)-\psi\left(a\right)=K\log\frac{b}{a},b\geq a\geq t,
\]
which is Fechner's \emph{Unterschiedsformel}.

In Chapter 16 of the \emph{Elemente}, Fechner derives the same relationship
in a different way. He presents the functional equation as
\[
\psi\left(b\right)-\psi\left(a\right)=G\left(\frac{b-a}{a}\right),
\]
and by assuming that $G$ is differentiable at zero gets the differential
equation
\[
\psi'\left(x\right)dx=G'\left(0\right)\frac{dx}{x},
\]
whose integration once again leads to Fechner's logarithmic formula.

The novelty of the method of functional equations in the mid-XIX's
century is probably responsible for the fact that the Chapter 17 derivation
was universally overlooked by Fechner's contemporaries (and then,
as it seems, forgotten altogether). The derivation in Chapter 16,
through differential equations, was, by contrast, common in Fechner's
time, which may be the reason Fechner placed it first. This derivation
has been criticized as mathematically or logically flawed by Fechner's
contemporaries and modern authors alike. The common interpretation
has been that it is based on Fechner's postulate 
\[
\psi\left(x+\Delta x\right)-\psi\left(x\right)=c.
\]
He is thought to have combined this with Weber's law
\[
\frac{\Delta x}{x}=c^{*},
\]
to arrive at
\[
\psi\left(x+\Delta x\right)-\psi\left(x\right)=\frac{c}{c^{*}}\frac{\Delta x}{x}.
\]
Finally, Fechner is thought to have invoked an ``expediency principle''
(\emph{H\"ulfsprinzip}) to illegitimately replace the finite differences
with differentials,
\[
d\psi=\frac{c}{c^{*}}\frac{dx}{x}.
\]
The integration of this equation with the boundary condition $\psi\left(x_{0}\right)=0$
yields
\[
\psi\left(x\right)=\frac{c}{c^{*}}\log\frac{x}{x_{0}}.
\]
It has been pointed out that this derivation is internally contradictory
because it implies
\[
\psi\left(x+\Delta x\right)-\psi\left(x\right)=\frac{c}{c^{*}}\log\frac{x+\Delta x}{x}=\frac{c}{c^{*}}\log\left(1+c^{*}\right),
\]
which is not the same as the postulated
\[
\psi\left(x+\Delta x\right)-\psi\left(x\right)=c.
\]

Boring's characterization of Fechner's book as ``dull and confusing
and sometimes uninterpretable'' being true, it is not easy to refute
this criticism. However, it is clear that Fechner uses neither the
Fechner postulate nor Weber's law in deriving his law, although he
accepts the truth of both. As explained above, he makes use of the
W-principle (which he calls ``Weber's law''). It follows from his
derivation that if Weber's law holds in addition to the W-principle,
then 
\[
\psi\left(x+\Delta x\right)-\psi\left(x\right)=K\log\left(1+c^{*}\right)=c,
\]
which is indeed a constant (Fechner's postulate proved as a theorem).
As Fechner points out in a book of rejoinders, if the Weber fraction
$c^{*}$ is sufficiently small, the constant $K$ \emph{approximately}
equals $c/c^{*}$, as in the criticized formula. The ``expediency
principle'' which Fechner's critics especially disparage seems to
be nothing more than an inept and verbose explanation of the elementary
fact (used in the Chapter 16 derivation) that if a function $f\left(x\right)$
is differentiable at zero, then $df\left(x\right)$ is proportional
to $dx$.

\subsection{\label{sec:Observation-areas}Observation areas and canonical transformation}

\index{observation area} The elementary but fundamental fact is that
if an observer is asked to compare two stimuli, $\mathbf{x}$ and
$\mathbf{y}$, they must differ in some respect that allows the observer
to identify them as two distinct stimuli. For instance, in the pair
written as $\left(\mathbf{x},\mathbf{y}\right)$, the first argument,
$\mathbf{x}$, may denote the stimulus presented chronologically first,
followed by $\mathbf{y}$. Or $\mathbf{x}$ may always be presented
above or to the left of $\mathbf{y}$. In perceptual pairwise comparisons,
the stimuli must differ in their spatial and/or temporal locations,
but the defining properties of $\mathbf{x}$ and $\mathbf{y}$ in
the pair $\left(\mathbf{x},\mathbf{y}\right)$ may vary. Thus, two
line segments to be compared in length may be presented in varying
pairs of distinct spatial locations, but one of the line segments
may always be vertical (and written first in the pair, $\mathbf{x}$)
and the other horizontal (written second, $\mathbf{y}$).

Formally, this means that a stimulus space\index{stimulus space}
involves two stimulus sets rather than one. Denoting them $\mathfrak{S}_{1}^{\star\star}$
(for $\mathbf{x}$-stimuli) and $\mathfrak{S}_{2}^{\star\star}$ (for
$\mathbf{y}$-stimuli), we call them the first and the second \emph{observation
areas,} respectively. The space structure is imposed on the Cartesian
product of these observation areas by a function 
\begin{equation}
\phi^{\star\star}:\mathfrak{S}_{1}^{\star\star}\times\mathfrak{S}_{2}^{\star\star}\rightarrow R,
\end{equation}
where $R$ may be a set of possible responses, or possible probabilities
of a particular response.

We say that two stimuli $\mathbf{x},\mathbf{x}^{\prime}\in\mathfrak{S}_{1}^{\star\star}$
are \emph{psychologically equal\index{psychological equality}} if
\[
\phi^{\star\star}\left(\mathbf{x},\mathbf{y}\right)=\phi^{\star\star}\left(\mathbf{x}^{\prime},\mathbf{y}\right)
\]
for any $\mathbf{y}\in\mathfrak{S}_{2}^{\star\star}.$ Similarly,
$\mathbf{y},\mathbf{y}^{\prime}\in\mathfrak{S}_{2}^{\star\star}$
are psychologically equal if 
\[
\phi^{\star\star}\left(\mathbf{x},\mathbf{y}\right)=\phi^{\star\star}\left(\mathbf{x},\mathbf{y}^{\prime}\right),
\]
for any $\mathbf{x}\in\mathfrak{S}_{1}^{\star\star}.$ One can always
relabel the elements of the observation areas by assigning identical
labels to all psychologically equal stimuli. For instance, all metameric
colors may be encoded by the same RGB coordinates irrespective of
their spectral composition. Objects of different color but of the
same weight will normally be labeled identically in a task involving
hefting and deciding which of two objects is heavier.

Let us denote by $\mathfrak{S}_{1}^{\star}$ and $\mathfrak{S}_{2}^{\star}$
the observation areas in which psychologically equal stimuli are equal.
The function $\phi^{\star\star}$ is then redefined into 
\begin{equation}
\phi^{\star}:\mathfrak{S}_{1}^{\star}\times\mathfrak{S}_{2}^{\star}\rightarrow R.\label{eq:function}
\end{equation}
We will illustrate this transformation by a toy example. Let the original
function be 
\[
\begin{array}{c|c|c|c|c|c|c|c|}
\phi^{*} & \mathbf{y}_{1} & \mathbf{y}_{2} & \mathbf{y}_{3} & \mathbf{y}_{4} & \mathbf{y}_{5} & \mathbf{y}_{6} & \mathbf{y}_{7}\\
\hline \mathbf{x}_{1} & 0.7 & 0.6 & 0.3 & 0.4 & 0.4 & 0.4 & 0.4\\
\hline \mathbf{x}_{2} & 0.5 & 0.3 & 0.4 & 0.2 & 0.2 & 0.2 & 0.2\\
\hline \mathbf{x}_{3} & 0.5 & 0.3 & 0.4 & 0.2 & 0.2 & 0.2 & 0.2\\
\hline \mathbf{x}_{4} & 0.2 & 0.1 & 0.5 & 0.3 & 0.3 & 0.3 & 0.3\\
\hline \mathbf{x}_{5} & 0.2 & 0.1 & 0.5 & 0.3 & 0.3 & 0.3 & 0.3\\
\hline \mathbf{x}_{6} & 0.1 & 0.3 & 0.8 & 0.6 & 0.6 & 0.6 & 0.6\\
\hline \mathbf{x}_{7} & 0.1 & 0.3 & 0.8 & 0.6 & 0.6 & 0.6 & 0.6
\\\hline \end{array}
\]
The first observation area\index{observation area}, $\mathfrak{S}_{1}^{\star\star}$,
comprises stimuli $\left\{ \mathbf{x}_{1},\ldots,\mathbf{x}_{7}\right\} $
(e.g., weights placed on one's left palm), the second observation
area, $\mathfrak{S}_{2}^{\star\star}$, comprises stimuli $\left\{ \mathbf{y}_{1},\ldots,\mathbf{y}_{7}\right\} $
(weights placed on one's left palm), and the entries in the matrix
above are values of $\phi^{*}\left(\mathbf{x},\mathbf{y}\right)$,
an arbitrary function mapping $\left(\mathbf{x},\mathbf{y}\right)$-pairs
into real numbers (say, the probabilities of deciding that the two
weights differ in heaviness). If two rows (or columns) of the matrix
are identical, then the two corresponding $\mathbf{x}$-stimuli (respectively,
$\mathbf{y}$-stimuli) are psychologically equal, and can be labeled
identically. Thus, the stimuli $\mathbf{x_{2}},\mathbf{x_{3}}$ and
$\mathbf{x_{4}},\mathbf{x_{5}}$ and $\mathbf{x_{6}},\mathbf{x_{7}}$
and $\mathbf{y}_{4},\mathbf{y}_{5},\mathbf{y}_{6},\mathbf{y}_{7}$
are all psychologically equal and they can be replaced by a single
symbol, respectively. The redefined spaces $\mathfrak{S}_{1}^{\star}$
and $\mathfrak{S}_{1}^{\star}$ are then as follows, 
\[
\begin{array}{cccccccc}
\mathfrak{S}_{1}^{\star\star}: & \mathbf{x}_{1} & \mathbf{x}_{2} & \mathbf{x}_{3} & \mathbf{x}_{4} & \mathbf{x}_{5} & \mathbf{x}_{6} & \mathbf{x}_{7}\\
 & \Downarrow & \Downarrow & \Downarrow & \Downarrow & \Downarrow & \Downarrow & \Downarrow\\
\mathfrak{S}_{1}^{\star}: & \mathbf{x}_{a} & \mathbf{x}_{b} & \mathbf{x}_{b} & \mathbf{x}_{c} & \mathbf{x}_{c} & \mathbf{x}_{d} & \mathbf{x}_{d}
\end{array},\begin{array}{cccccccc}
\mathfrak{S}_{2}^{\star\star}: & \mathbf{y}_{1} & \mathbf{y}_{2} & \mathbf{y}_{3} & \mathbf{y}_{4} & \mathbf{y}_{5} & \mathbf{y}_{6} & \mathbf{y}_{7}\\
 & \Downarrow & \Downarrow & \Downarrow & \Downarrow & \Downarrow & \Downarrow & \Downarrow\\
\mathfrak{S}_{2}^{\star:} & \mathbf{y}_{a} & \mathbf{y}_{b} & \mathbf{y}_{c} & \mathbf{y}_{d} & \mathbf{y}_{d} & \mathbf{y}_{d} & \mathbf{y}_{d}
\end{array},
\]
and the function $\phi^{*}$ transforms into $\phi^{\star}$ accordingly,
\[
\begin{array}{c|c|c|c|c|}
\phi^{\star} & \mathbf{y}_{a} & \mathbf{y}_{b} & \mathbf{y}_{c} & \mathbf{y}_{d}\\
\hline \mathbf{x}_{a} & 0.7 & 0.6 & 0.3 & 0.4\\
\hline \mathbf{x}_{b} & 0.5 & 0.3 & 0.4 & 0.2\\
\hline \mathbf{x}_{c} & 0.2 & 0.1 & 0.5 & 0.3\\
\hline \mathbf{x}_{d} & 0.1 & 0.3 & 0.8 & 0.6
\\\hline \end{array}.
\]

As another example, consider the function $\gamma\left(x,y\right)$
of the previous section, and assume that 
\[
\mathfrak{S}_{1}^{\star\star}=[t_{1},u_{1}[,\mathfrak{S}_{2}^{\star\star}=[t_{2},u_{2}[.
\]
Assume that $\gamma\left(x,y\right)$ is strictly increasing in $y$
and strictly decreasing in $x$. Then $\gamma\left(x,y\right)=\gamma\left(x,y'\right)$
implies $y=y'$ and $\gamma\left(x,y\right)=\gamma\left(x',y\right)$
implies $x=x'$, so that in this case 
\[
\mathfrak{S}_{1}^{\star\star}=\mathfrak{S}_{1}^{\star},\mathfrak{S}_{2}^{\star\star}=\mathfrak{S}_{2}^{\star}.
\]

\begin{figure}
\begin{centering}
\includegraphics[scale=0.4]{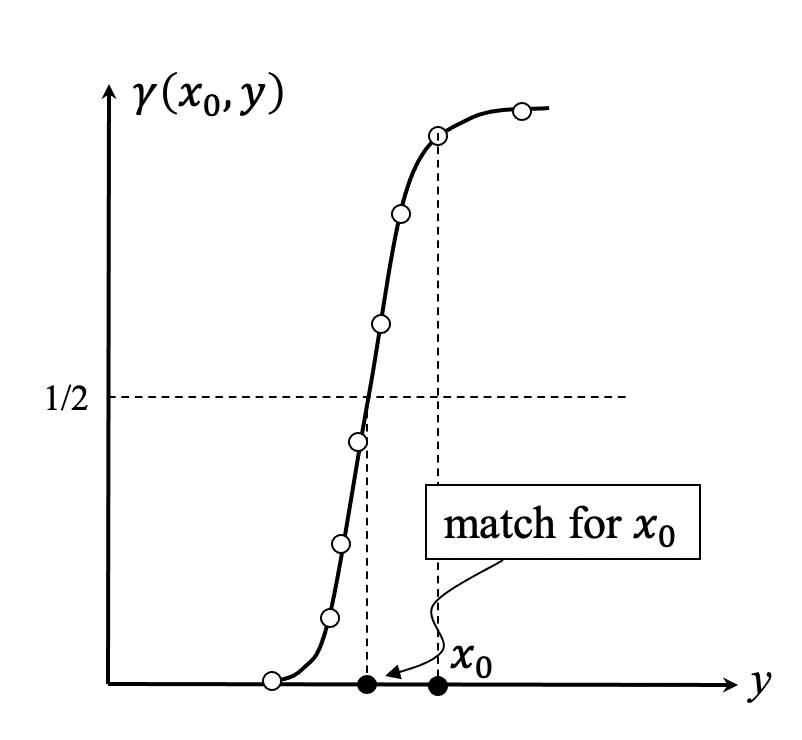} 
\par\end{centering}
\caption{\label{fig:ConstErrorGamma}A ``greater-less'' psychometric function
$y\protect\mapsto\gamma\left(x,y\right)$ defined on an interval of
real numbers.The function shows, for a fixed value of $x=x_{0}$,
the probability with which $y$ is judged to be greater than $x_{0}$
with respect to some designated property. The median value of $y$,
one at which $\gamma\left(x_{0},y\right)=\frac{1}{2}$, is taken to
be a match, or point of subjective equality (PSE) for $x_{0}$, and
the difference between $x_{0}$ and its PSE defines constant error.
(Note that showing $\gamma\left(x,y\right)$ at a fixed value of $x$
does not mean that the value of $x$ was fixed procedurally in an
experiment. The graph is simply a cross-section of $\gamma\left(x,y\right)$
at $x=x_{0}.$)}
\end{figure}

Staying with this example, $\gamma\left(x,y\right)=\nicefrac{1}{2}$
defines here the binary relation ``is matched by'': $x\in\mathfrak{S}_{1}^{\star}$
is matched by $y\in\mathfrak{S}_{2}^{\star}$ if and only if $\gamma\left(x,y\right)=\nicefrac{1}{2}$.
The relation ``$y\in\mathfrak{S}_{2}^{\star}$ is matched by $x\in\mathfrak{S}_{1}^{\star}$''
is defined by the same condition, $\gamma\left(x,y\right)=\nicefrac{1}{2}$.
The traditional psychophysical designation of this relation is that
$y$ is the \emph{point of subjective equality\index{point of subjective equality, PSE}}
(PSE) for $x$ (and then $x$ is the PSE for $y$). The assumptions
(\ref{eq:balanced gamma})-(\ref{eq:balanced gamma 2}) made in the
previous section do not hold generally. In particular, the psychometric
function $\gamma$, as a rule, has a nonzero \emph{constant error},
i.e., $\gamma\left(x,y\right)=\nicefrac{1}{2}$ does not imply $x=y$
(see Figure \ref{fig:ConstErrorGamma}).

\begin{figure}
\begin{centering}
\includegraphics[scale=0.4]{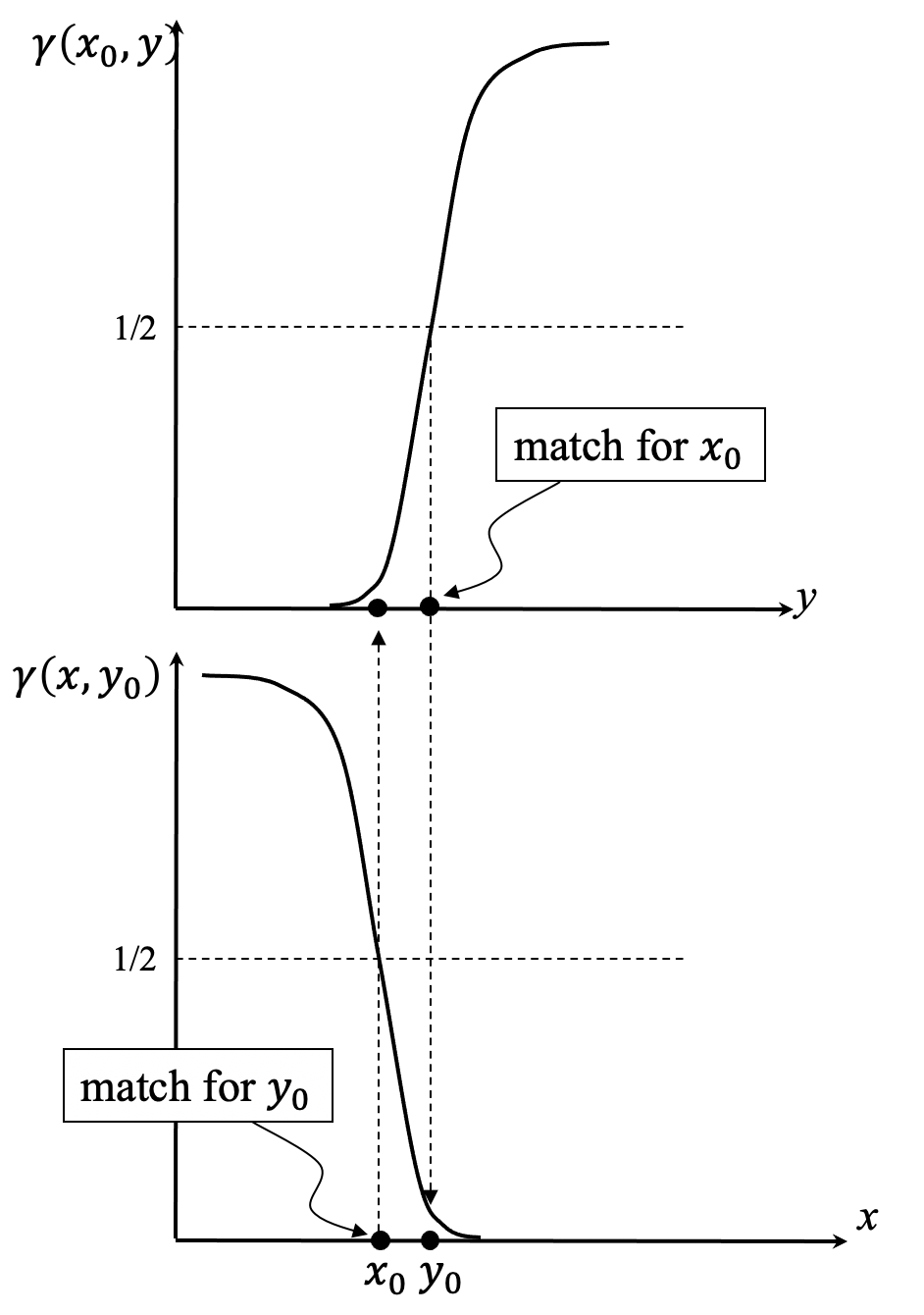} 
\par\end{centering}
\caption{\label{fig:RegMediality}An illustration, for the psychometric function
$\gamma\left(x,y\right)$, of the symmetry of the relation ``to be
a PSE for.'' The upper panel shows the function $y\protect\mapsto\gamma\left(x,y\right)$
at $x=x_{0}$, and $y_{0}$ denotes the PSE for $x_{0}$. The lower
panel shows the function $x\protect\mapsto\gamma\left(x,y\right)$
at $y=y_{0}$, and $x_{0}$ then has to be the PSE for $y_{0}$. Conversely,
if $x_{0}$ denotes the PSE for $y_{0}$ in the lower panel, then
$y_{0}$ has to be the PSE for $x_{0}$ in the upper panel. This follows
from the fact that in both cases the PSE is defined by $\gamma\left(x,y\right)=\frac{1}{2}$,
and the assumption that both $x\protect\mapsto\gamma\left(x,y\right)$
and $y\protect\mapsto\gamma\left(x,y\right)$ are monotone functions
whose range includes the value $\gamma=\frac{1}{2}$ .}
\end{figure}

With the monotonicity assumptions about $\gamma$ made above, if we
also assume that the range of the function $y\mapsto\gamma\left(x,y\right)$
for every $x$ includes the value $\nicefrac{1}{2}$, and that the
same is true for the range of the function $x\mapsto\gamma\left(x,y\right)$
for every $y$, then we have the following properties of the PSE relation
(see Figure \ref{fig:RegMediality}): 
\begin{enumerate}
\item the PSE for every $x\in\mathfrak{S}_{1}^{\star}$ exists and is unique; 
\item the PSE for every $y\in\mathfrak{S}_{2}^{\star}$ exists and is unique; 
\item $y\in\mathfrak{S}_{2}^{\star}$ is a PSE for $x\in\mathfrak{S}_{1}^{\star}$
if and only if $x\in\mathfrak{S}_{1}^{\star}$ is a PSE for $y\in\mathfrak{S}_{2}^{\star}$. 
\end{enumerate}

\index{psychometric function} 

We will assume that these properties generalize to any function $\phi^{\star}$
in (\ref{eq:function}). In other words, we assume that $\phi^{\star}$
is associated with a \emph{bijective} \emph{function} $\mathbf{h:}\mathfrak{S}_{1}^{\star}\longrightarrow\mathfrak{S}_{2}^{\star}$
such that for all $\mathbf{x}\in\mathfrak{S}_{1}^{\star}$ and $\mathbf{y}\in\mathfrak{S}_{2}^{\star}$, 
\begin{description}
\item [{(P1)}] $\mathbf{y}$ is a PSE for $\mathbf{x}$ if and only if
$\mathbf{y}=\mathbf{h}\left(\mathbf{x}\right)$; 
\item [{(P2)}] $\mathbf{x}$ is a PSE for $\mathbf{y}$ if and only if
$\mathbf{x}=\mathbf{h}^{-1}\left(\mathbf{y}\right)$. 
\end{description}
This makes the relation of ``being a PSE of'' or ``being matched
by'' symmetric. As a result, one can always apply to the observation
areas a \emph{canonical transformation} 
\[
\mathbf{f:}\mathfrak{S}_{1}^{\star}\longrightarrow\mathfrak{S},\mathbf{g:}\mathfrak{S}_{2}^{\star}\longrightarrow\mathfrak{S},
\]
with $\mathbf{f}$ and $\mathbf{g}$ arbitrary except for 
\[
\mathbf{h}=\mathbf{g}^{-1}\circ\mathbf{f}.
\]
A canonical transformation redefines the function $\phi^{\star}$
into 
\[
\phi:\mathfrak{S}\times\mathfrak{S}\longrightarrow R,
\]
such that, for any ordered pair $\left(\mathbf{x},\mathbf{y}\right)$,
one of the elements is a PSE for the other element if and only if
$\mathbf{x}=\mathbf{y}$. We say that the stimulus space\index{stimulus space}
and the space-forming function $\phi$ here are in a \emph{canonical
form}. \index{stimulus space!in canonical form}

Let us use the toy example above for an illustration. We assume that
the PSE for any $\mathbf{x}$ is defined here as $\mathbf{y}$ at
which $\mathbf{y}\mapsto\phi^{\star}\left(\mathbf{x},\mathbf{y}\right)$
reaches its minimum; and the PSE for any $\mathbf{y}$ is defined
as $\mathbf{x}$ at which $\mathbf{x}\mapsto\phi^{\star}\left(\mathbf{x},\mathbf{y}\right)$
reaches its minimum. The inspection of the matrix for $\phi^{\star}$
shows that the PSEs are well defined for both $\mathbf{x}$-stimuli
and $\mathbf{y}$-stimuli: 
\[
\begin{array}{c|c|c|c|c|}
\phi^{\star} & \mathbf{y}_{a} & \mathbf{y}_{b} & \mathbf{y}_{c} & \mathbf{y}_{d}\\
\hline \mathbf{x}_{a} & 0.7 & 0.6 & \boxed{0.3} & 0.4\\
\hline \mathbf{x}_{b} & 0.5 & 0.3 & 0.4 & \boxed{0.2}\\
\hline \mathbf{x}_{c} & 0.2 & \boxed{0.1} & 0.5 & 0.3\\
\hline \mathbf{x}_{d} & \boxed{0.1} & 0.3 & 0.8 & 0.6
\\\hline \end{array}.
\]
We also see that in each row the minimal value (shown boxed) is also
minimal in its column. That is, $\mathbf{y}$ is a PSE for $\mathbf{x}$
if and only if $\mathbf{x}$ the PSE for $\mathbf{y}$. The graph
of the bijective $\mathbf{h}$-function in the formulations of the
properties P$1$ and P2 is given by the pairs 
\[
\left\{ \left(\mathbf{x}_{a},\mathbf{y}_{c}\right),\left(\mathbf{x}_{b},\mathbf{y}_{d}\right),\left(\mathbf{x}_{c},\mathbf{y}_{b}\right),\left(\mathbf{x}_{d},\mathbf{y}_{a}\right)\right\} .
\]
Simple relabeling then allows us to have all PSE-pairs on the main
diagonal. Both $\mathfrak{S}_{1}^{\star}$ and $\mathfrak{S}_{1}^{\star}$
can be mapped into one and the same set $\mathfrak{S}$, e.g., as

\[
\begin{array}{ccccc}
\mathfrak{S}_{1}^{\star}: & \mathbf{x}_{a} & \mathbf{x}_{b} & \mathbf{x}_{c} & \mathbf{x}_{d}\\
 & \Downarrow & \Downarrow & \Downarrow & \Downarrow\\
\mathfrak{S:} & \mathbf{a} & \mathbf{b} & \mathbf{c} & \mathbf{d}
\end{array},\begin{array}{ccccc}
\mathfrak{S}_{2}^{\star}: & \mathbf{y}_{c} & \mathbf{y}_{d} & \mathbf{y}_{b} & \mathbf{y}_{a}\\
 & \Downarrow & \Downarrow & \Downarrow & \Downarrow\\
\mathfrak{S:} & \mathbf{a} & \mathbf{b} & \mathbf{c} & \mathbf{d}
\end{array},
\]
and $\phi^{\star}$ transforms into $\phi$ accordingly:

\[
\begin{array}{c|c|c|c|c|}
\phi & \mathbf{a} & \mathbf{b} & \mathbf{c} & \mathbf{d}\\
\hline \mathbf{a} & \boxed{0.3} & 0.4 & 0.6 & 0.7\\
\hline \mathbf{b} & 0.4 & \boxed{0.2} & 0.3 & 0.5\\
\hline \mathbf{c} & 0.5 & 0.3 & \boxed{1} & 0.2\\
\hline \mathbf{d} & 0.8 & 0.6 & 0.3 & \boxed{0.1}
\\\hline \end{array}.
\]

To apply canonical transformation to our second example, the psychometric
function $\gamma\left(x,y\right)$, let us assume that $\gamma\left(x,y\right)=\nicefrac{1}{2}$
holds if and only if $y=h\left(x\right)$ for some homeomorphic mapping
$h$ (i.e., such that both $h$ and $h^{-1}$ are continuous.) Since
$\mathfrak{S}_{1}^{*}=\mathfrak{S}_{1}^{\star}=[t_{1},u_{1}[$ and
$\mathfrak{S}_{2}^{*}=\mathfrak{S}_{2}^{\star}=[t_{2},u_{2}[$, $\mathfrak{S}$
can always be chosen in the form $[t,u[$, by choosing any two homeomorphisms
\[
f:[t_{1},u_{1}[\rightarrow[t,u[,g:[t_{2},u_{2}[\rightarrow[t,u[,
\]
such that $g^{-1}\circ f\equiv h$. Note, however, that this only
ensures compliance with (\ref{eq:balanced gamma 2}), but not with
(\ref{eq:balanced gamma}).

\subsection{\label{sec:Same-different-judgments}Same-different judgments}

\index{same-different judgments} 
\begin{figure}
\begin{centering}
\includegraphics[scale=0.25]{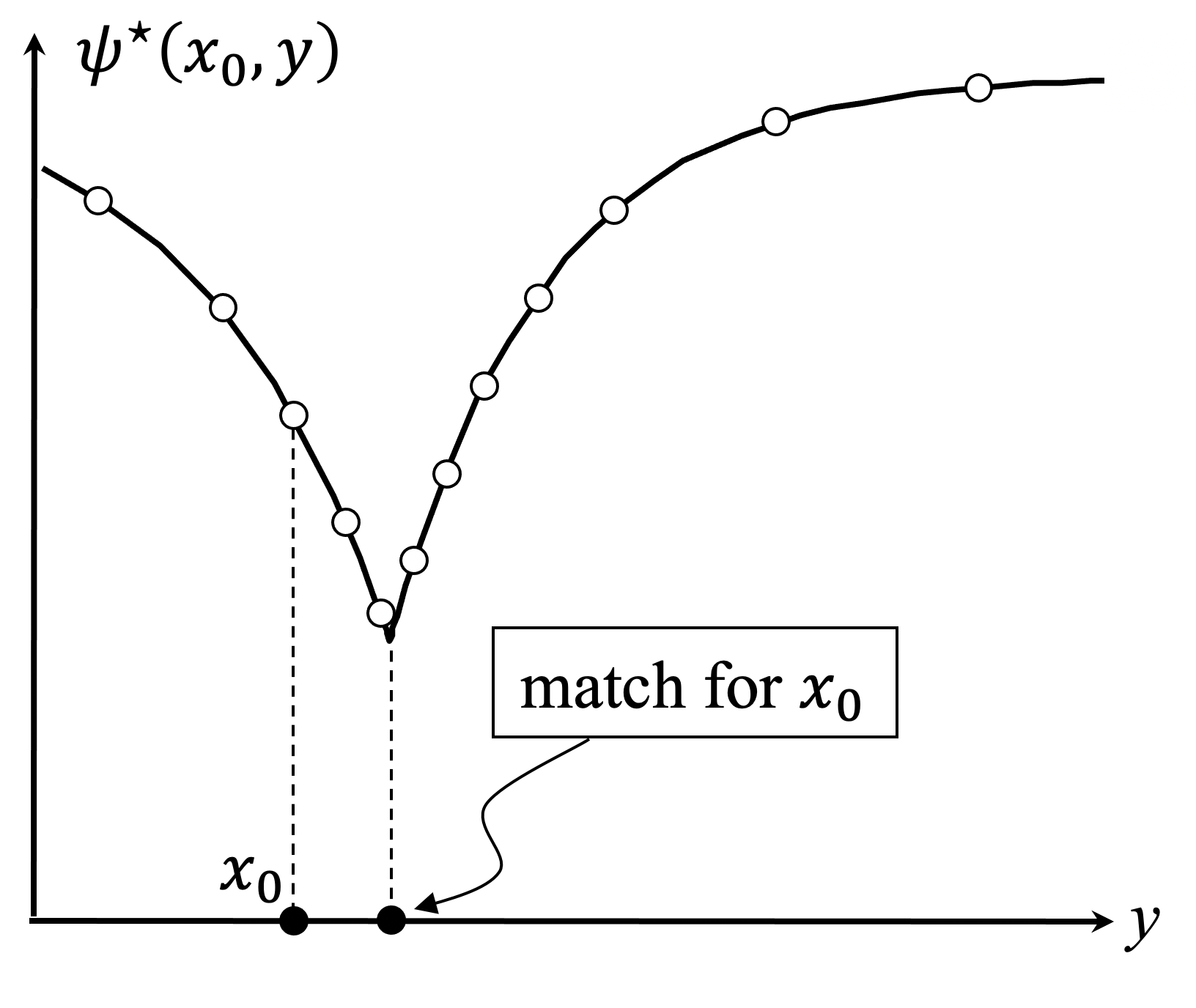} 
\par\end{centering}
\caption{\label{fig:ConstErrorPsi}A ``same-different'' psychometric function
$y\protect\mapsto\psi^{\star}\left(x,y\right)$ defined on an interval
of real numbers. The function shows, for a fixed value of $x=x_{0}$,
the probability with which $y$ is judged to be different from $x_{0}$
(generically or with respect to a designated property). The value
of $y$ at which $\psi^{\star}\left(x_{0},y\right)$ reaches its minimum
is taken to be a match, or point of subjective equality (PSE) for
$x_{0}$.}
\end{figure}

\begin{figure}
\begin{centering}
\includegraphics[scale=0.4]{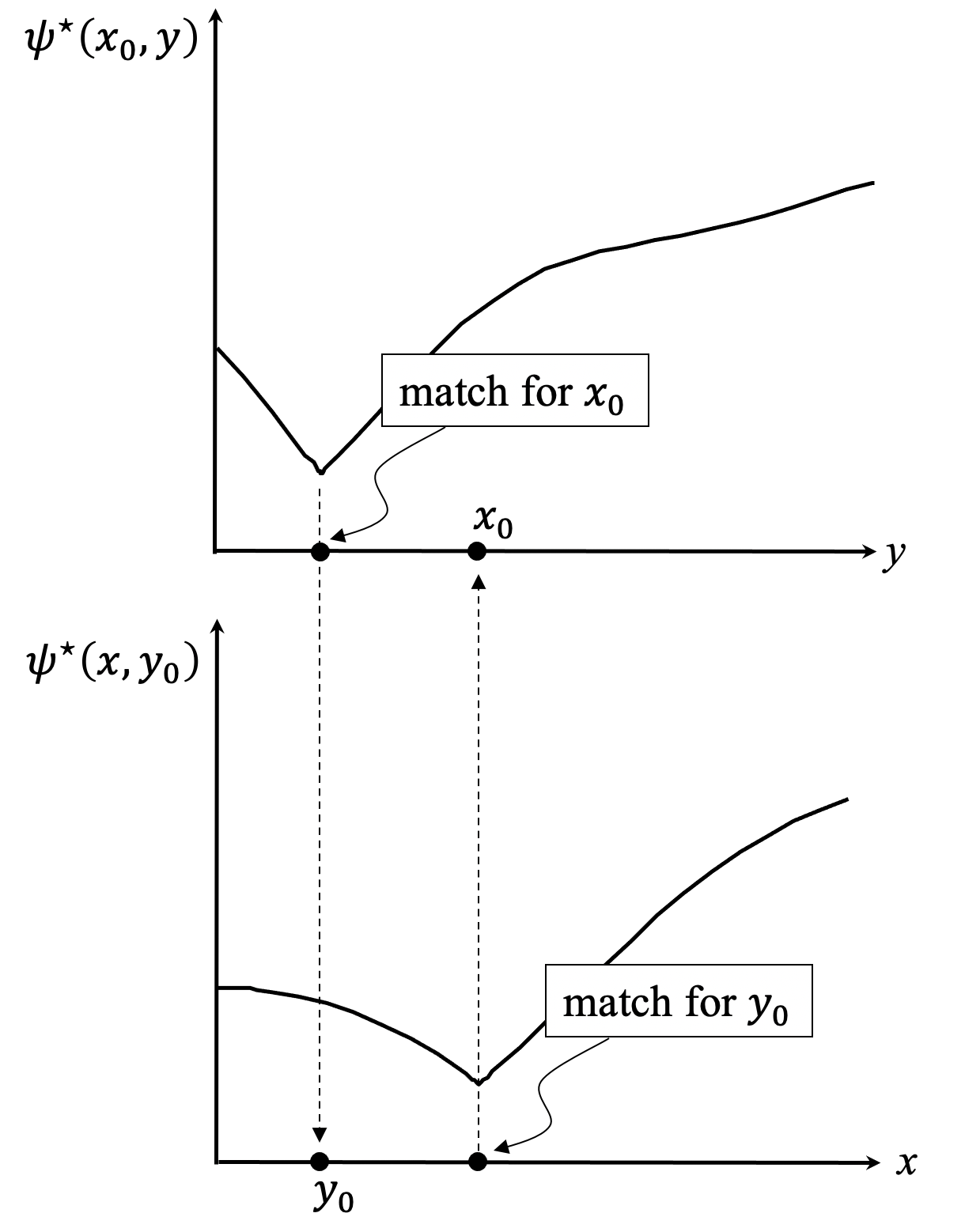} 
\par\end{centering}
\caption{\label{fig:RegMinimality}An illustration, for the psychometric function
$\psi^{\star}\left(x,y\right)$ in Figure \ref{fig:ConstErrorPsi},
of the symmetry of the relation ``to be a PSE for.'' The upper panel
shows the function $y\protect\mapsto\psi^{\star}\left(x,y\right)$
at $x=x_{0}$, and $y_{0}$ denotes the PSE for $x_{0}$. The lower
panel shows the function $x\protect\mapsto\psi^{\star}\left(x,y\right)$
at $y=y_{0}$, and $x_{0}$ is shown to be PSE for $y_{0}$. Conversely,
if $x_{0}$ denotes the PSE for $y_{0}$ in the lower panel, then
$y_{0}$ is shown to be the PSE for $x_{0}$ in the upper panel. Unlike
in the case of the ``great-less'' psychometric function (Figure
\ref{fig:RegMediality}), here the symmetry of the PSE relation is
an assumption rather than a consequence of other properties of the
function $\psi^{\star}$.}
\end{figure}

The greater-than comparisons are possible only with respect to a designated
characteristic, such as loudness or beauty. It is clear, however,
that no such characteristic can reflect all relevant aspects of the
stimuli being compared. Moreover, it is not certain that the characteristic's
values are always comparable in terms of greater-less, given a sufficiently
rich stimulus set. Thus, it may not be clear to an observer which
of two given faces is more beautiful, and even loudness may not be
semantically unidimensional if the sounds are complex. The same-different
comparisons have a greater scope of applicability, and do not have
to make use of designated characteristics. The role of the stimulus-space-defining
function $\phi^{*}$ of the previous section in this case is played
by 
\begin{equation}
\psi^{*}\left(\mathbf{x},\mathbf{y}\right)=\Pr\left[\text{\textbf{y}}\textnormal{ is judged to be different from }\text{\textbf{x}}\right],\label{eq:psi_initial}
\end{equation}
with $\mathbf{x}\in\mathfrak{S}_{1}^{\star\star}$ and $\mathbf{y}\in\mathfrak{S}_{2}^{\star\star}$.
To be different here means to differ in any respect other than the
conspicuous difference between the two observations areas. Thus, if
$\text{\textbf{x}}$ is a visual stimulus always presented to the
left of $\mathbf{y},$this difference in spatial locations does not
enter in the judgments of whether $\text{\textbf{x}}$ and $\text{\textbf{y}}$
are different or the same. Of course, it is also possible to ask whether
the two stimuli differ in a particular respect, such as color or shape.

The reduction of $\left(\psi^{*},\mathfrak{S}_{1}^{\star\star},\mathfrak{S}_{2}^{\star\star}\right)$
to $\left(\psi^{\star},\mathfrak{S}_{1}^{\star},\mathfrak{S}_{2}^{\star}\right)$,
in which psychologically equal stimuli are equal, is effected by assigning
an identical label to any $\mathbf{x},\mathbf{x}^{\prime}\in\mathfrak{S}_{1}^{\star\star}$
such that 
\[
\psi^{\star\star}\left(\mathbf{x},\mathbf{y}\right)=\psi^{\star\star}\left(\mathbf{x}^{\prime},\mathbf{y}\right)
\]
for all $\mathbf{y}\in\mathfrak{S}_{2}^{\star\star},$ and similarly
for the second observation area\index{observation area}.

The PSE relation for the function $\psi^{\star}$ is defined as follows
(see Figure \ref{fig:ConstErrorPsi}): $\mathbf{y}\in\mathfrak{S}_{2}^{\star}$
is a PSE for $\mathbf{x}\in\mathfrak{S}_{1}^{\star}$ if 
\[
\psi^{\star}\left(\mathbf{x},\mathbf{y}\right)<\psi^{\star}\left(\mathbf{x},\mathbf{y'}\right)\textnormal{ for all }\mathbf{y}'\not=\mathbf{y}.
\]
Analogously, $\mathbf{x}\in\mathfrak{S}_{1}^{\star}$ is a PSE for
$\mathbf{y}\in\mathfrak{S}_{2}^{\star}$ if 
\[
\psi^{\star}\left(\mathbf{x},\mathbf{y}\right)<\psi^{\star}\left(\mathbf{x}',\mathbf{y}\right)\textnormal{ for all }\mathbf{x}'\not=\mathbf{x}.
\]
In accordance with the previous section, we assume the existence of
a bijection $\mathbf{h:}\mathfrak{S}_{1}^{\star}\longrightarrow\mathfrak{S}_{2}^{\star}$
such that 
\begin{equation}
\begin{array}{c}
\psi^{\star}\left(\mathbf{x},\mathbf{h}\left(\mathbf{x}\right)\right)<\psi^{\star}\mathbf{\left(\mathbf{x},\mathbf{y}\right)}\text{ for all }\mathbf{y\neq h}\left(\mathbf{x}\right),\\
\psi^{\star}\left(\mathbf{\mathbf{h}}^{-1}\left(\mathbf{y}\right),\mathbf{y}\right)<\psi^{\star}\mathbf{\left(\mathbf{x},\mathbf{y}\right)}\text{ for all }\mathbf{x\neq}\mathbf{\mathbf{h}}^{-1}\left(\mathbf{y}\right).
\end{array}\label{eq:RegMinLaw}
\end{equation}
That is, we assume that the PSEs in the space $\left(\psi^{\star},\mathfrak{S}_{1}^{\star},\mathfrak{S}_{2}^{\star}\right)$
exist, are unique, and that $\mathbf{y}$ is the PSE for $\mathbf{x}$
if and only if $\mathbf{x}$ is the PSE for $\mathbf{y}$. We refer
to this property as the \emph{law of Regular Minimality}.\index{regular minimality}
In this chapter it should be taken as part of the definition of the
functions we are dealing with rather than an empirical claim.

Now, any canonical transformation, as described above, yields a probability
function 
\begin{equation}
\psi:\mathfrak{S}\times\mathfrak{S}\longrightarrow[0,1],\label{eq:psi canonical}
\end{equation}
such that, for any $\mathbf{a},\mathbf{x},\mathbf{y}\in\mathfrak{S}$,
if $\mathbf{x\neq}\mathbf{a}$ and $\mathbf{y\neq}\mathbf{a}$, then
\begin{equation}
\psi\left(\mathbf{a},\mathbf{a}\right)<\left\{ \begin{array}{c}
\psi\left(\mathbf{x},\mathbf{a}\right)\\
\psi\left(\mathbf{a},\mathbf{y}\right)
\end{array}\right..
\end{equation}
We will assume in the following that the discrimination probability
function $\psi$ is presented in this canonical form. This by no means
implies that $\psi\left(\mathbf{x},\mathbf{y}\right)=\psi\left(\mathbf{y},\mathbf{x}\right)$,
the order of the arguments continues to matter. We will continue to
consider the two arguments in $\psi\left(\mathbf{x},\mathbf{y}\right)$
as belonging to the first and second observation areas, respectively.

\section{Notation conventions}

We now introduce notation conventions for the rest of this chapter.
They in part codify and in part modify the notation used in the introductory
section.

Let us agree that from now on real-valued functions of one or several
points of a stimulus set will be indicated by strings without parentheses:
$\psi\mathbf{ab}$ in place of $\psi\left(\mathbf{a},\mathbf{b}\right)$,
$D\mathbf{abc}$ in place of $D\left(\mathbf{a},\mathbf{b},\mathbf{c}\right)$,
etc. Boldface lowercase letters denoting stimuli are merely labels,
with no implied operations between them, so this notation is unambiguous.
(In Section \ref{sec:Finsler}, lowercase boldface letters are also
used to denote direction vectors, in which case the string convention
is not used.) If a stimulus is represented by a real number we may
conveniently confuse the two, and write, e.g., $\gamma\left(x,y\right)$
instead of the more rigorous $\gamma\mathbf{x}\mathbf{y}$ with $\mathbf{x},\mathbf{y}$
represented by (or having values) $x,y$.

A finite sequence (or \emph{chain}) $\left(\mathbf{x}_{1},\ldots,\mathbf{x}_{n}\right)$
of points in stimulus a set will be presented as a string $\mathbf{x}_{1}\ldots\mathbf{x}_{n}$.
If a chain of stimuli is to be referred to without indicating its
elements, then it is indicated by uppercase boldface letters. Thus
$\mathbf{X}$ may stand for $\mathbf{abc}$, $\mathbf{Y}$ stand for
$\mathbf{y}_{1}\ldots\mathbf{y}_{n}$, etc. If $\mathbf{X}=\mathbf{x}_{1}...\mathbf{x}_{k}$and
$\mathbf{Y}=\mathbf{y}_{1}...\mathbf{y}_{l}$ are two chains, then
\[
\begin{array}{c}
\mathbf{XY}=\mathbf{x}_{1}...\mathbf{x}_{k}\mathbf{y}_{1}...\mathbf{y}_{l},\\
\mathbf{aXb}=\mathbf{a}\mathbf{x}_{1}...\mathbf{x}_{k}\mathbf{b},\\
\mathbf{aXbYa}=\mathbf{a}\mathbf{x}_{1}...\mathbf{x}_{k}\mathbf{b}\mathbf{y}_{1}...\mathbf{y}_{l}\mathbf{a},\\
\textnormal{etc.}
\end{array}
\]
The number of elements in a chain $\mathbf{X}$ is its cardinality
$\left|\mathbf{X}\right|$. \index{chain} Infinite sequences $\left\{ x_{1},\ldots,x_{n},\ldots\right\} $,
$\left\{ \mathbf{x}_{1},\ldots,\mathbf{x}_{n},\ldots\right\} $, $\left\{ \mathbf{X}_{1},\ldots,\mathbf{X}_{n},\ldots\right\} ,$
etc., are almost always indicated by their generic elements: numerical
sequence $\left\{ x_{n}\right\} ,$ stimulus sequence $\left\{ \mathbf{x}_{n}\right\} $,
sequence of chains $\left\{ \mathbf{X}_{n}\right\} $, etc. Convergence
of a sequence, such as $\mathbf{x}_{n}\rightarrow\text{\textbf{x}}$,
is understood as conditioned on $n\rightarrow\infty.$ In a sequence
of chains, the cardinality $\left|\mathbf{X}_{n}\right|$ is generally
changing.

As mentioned earlier, we indicate intervals of reals (closed, open
and half-open) by square-brackets: $\left[a,b\right],$ $\left[a,b\right[,$
$\left]a,b\right],$ and $\left]a,b\right[.$ Round-bracketed pairs
of numbers of stimuli, $\left(a,b\right)$ or $\left(\mathbf{a},\mathbf{b}\right)$,
always indicate an ordered pair.

Sets of stimuli are denoted by Gothic letters, $\mathfrak{S}$, $\mathfrak{S}_{1}^{\star\star}$,
$\mathfrak{s}$, etc. For sets of chains and paths in stimulus spaces
we use script letters, $\mathcal{C},$$\mathcal{P}_{a}^{b}$, etc.
For other types of sets we use blackboard and sans serif fonts on
an ad hoc basis. The set of reals is denoted as usual $\mathbb{R}$.

\section{\label{sec:UFS}Basics of Fechnerian Scaling}

\index{Fechnerian Scaling} Using our new notation, and considering
an at least two-element stimulus space $\mathfrak{S}$ in a canonical
form, we have, for any distinct points $\mathbf{x}$ and $\mathbf{y}$
in $\mathfrak{S}$, 
\begin{equation}
\begin{array}{c}
\Psi^{\left(1\right)}\mathbf{xy}=\psi\mathbf{xy}-\psi\mathbf{xx}>0,\\
\Psi^{\left(2\right)}\mathbf{xy}=\psi\mathbf{yx}-\psi\mathbf{xx}>0.
\end{array}
\end{equation}
We call the quantities $\Psi^{\left(1\right)}\mathbf{xy}$ and $\Psi^{\left(2\right)}\mathbf{xy}$
\emph{psychometric increments} of the first and second kind, respectively.
Both can be interpreted as ways of quantifying the intuition of a
dissimilarity of $\mathbf{y}$ from $\mathbf{x}$. The order ``from-to''
is important here, as $\Psi^{\left(i\right)}\mathbf{yx}\not=\Psi^{\left(i\right)}\mathbf{yx}$
($i=1,2$).

In Fechnerian Scaling we use the psychometric increments to compute
subjective distances in the spirit of Fechner's idea of cumulation
of small dissimilarities. We will see that this cumulation can assume
different forms, depending on the properties of a stimulus space\index{stimulus space}.
However, the general construction, applicable to all spaces, is as
follows.

\subsection{Step 1}

First, we assume that both $\Psi^{\left(1\right)}$ or $\Psi^{\left(2\right)}$
are \emph{dissimilarity functions}, in accordance with the following
definition (to be explained and elaborated later on). 
\begin{defn}
\label{def:dissimilarity}We say that $D:\mathfrak{S}\times\mathfrak{S}\rightarrow\mathbb{R}$
is a dissimilarity function if it has the following properties: \index{dissimilarity function} 

$\mathcal{D}1$(\emph{positivity}) $D\mathbf{ab}>0$ for any distinct
$\mathbf{a},\mathbf{b}\in\mathfrak{S}$;

$\mathcal{D}2$ (\emph{zero property}) $D\mathbf{aa}=0$ for any $\mathbf{a}\in\mathfrak{S}$;

$\mathcal{D}3$ (\emph{uniform continuity}) for any $\varepsilon>0$
one can find a $\delta>0$ such that, for any $\mathbf{a},\mathbf{b},\mathbf{a'},\mathbf{b'}\in\mathfrak{S}$,
\[
\textnormal{if }D\mathbf{aa}^{\prime}<\delta\textnormal{ and }D\mathbf{bb}^{\prime}<\delta,\mathbf{\textnormal{ then }}\left\vert D\mathbf{a}^{\prime}\mathbf{b}^{\prime}-D\mathbf{ab}\right\vert <\varepsilon;
\]

$\mathcal{D}4$ (\emph{chain property}) for any $\varepsilon>0$ one
can find a $\delta>0$ such that for any chain $\mathbf{aXb}$, 
\[
\textnormal{if }D\mathbf{aXb}<\delta,\textnormal{ then }D\mathbf{ab}<\varepsilon.
\]
\end{defn}

For the chain property, we need to define $D\mathbf{aXb}$. 
\begin{defn}
Given a chain $\mathbf{X}=\mathbf{x}_{1}...\mathbf{x}_{k}$ in $\mathfrak{S}$,
its\emph{ D-length} (or just \emph{length} once $D$ is specified)
is defined as 
\[
D\mathbf{X}=\left\{ \begin{array}{cc}
D\mathbf{x}_{1}\mathbf{x}_{2}+...+D\mathbf{x}_{k-1}\mathbf{x}_{k} & \textnormal{if }\left|\mathbf{X}\right|>1\\
0 & \textnormal{if }\left|\mathbf{X}\right|\leq1
\end{array}\right..
\]
\end{defn}

Then, for a given pair of points $\mathbf{a},\mathbf{b}$, the length
of $\mathbf{aXb}$ is 
\[
D\mathbf{aXb}=\left\{ \begin{array}{cc}
D\mathbf{a}\mathbf{x}_{1}+D\mathbf{X}+D\mathbf{x}_{k}\mathbf{b} & \textnormal{if }\left|\mathbf{X}\right|>0\\
D\mathbf{ab} & \textnormal{if }\left|\mathbf{X}\right|=0
\end{array}\right..
\]
\index{chain!length of}

\subsection{Step 2}

Next, we consider the set $\mathcal{C}$ of all (finite) chains in
$\mathfrak{S}$, 
\[
\mathcal{C}=\bigcup_{k=0}^{\infty}\mathfrak{S}^{k},
\]
and define 
\begin{equation}
G\mathbf{ab}=\inf_{\mathbf{X}\in\mathcal{C}}D\mathbf{aXb}.\label{eq:Gdefined}
\end{equation}
We will see below that the function $G:\mathfrak{S}\times\mathfrak{S}\rightarrow\mathbb{R}$
is a \emph{quasimetric dissimilarity}, in accordance with the following
definition. 
\begin{defn}
\label{def:quasimetric dissimilarity}Function $M:\mathfrak{S}\times\mathfrak{S}\rightarrow\mathbb{R}$
is a quasimetric dissimilarity function if it has the following properties:
\index{dissimilarity function!quasimetric}

$\mathcal{QM}1$ (\emph{positivity}) $M\mathbf{ab}>0$ for any distinct
$\mathbf{a},\mathbf{b}\in\mathfrak{S}$;

$Q\mathcal{M}2$ (\emph{zero property}) $M\mathbf{aa}=0$ for any
$\mathbf{a}\in\mathfrak{S}$;

$\mathcal{QM}3$ (\emph{triangle inequality})\index{inequality!triangle}
$M\mathbf{ab}+M\mathbf{bc}\geq M\mathbf{ac}$ for all $\mathbf{a},\mathbf{b},\mathbf{c}\in\mathfrak{S}$.

$\mathcal{\mathcal{QM}}4$ (\emph{symmetry in the small}) for any
$\varepsilon>0$ one can find a $\delta>0$ such that $M\mathbf{ab}<\delta$
implies $M\mathbf{ba}<\varepsilon$, for any $\mathbf{a},\mathbf{b}\in\mathfrak{S}$. 
\end{defn}

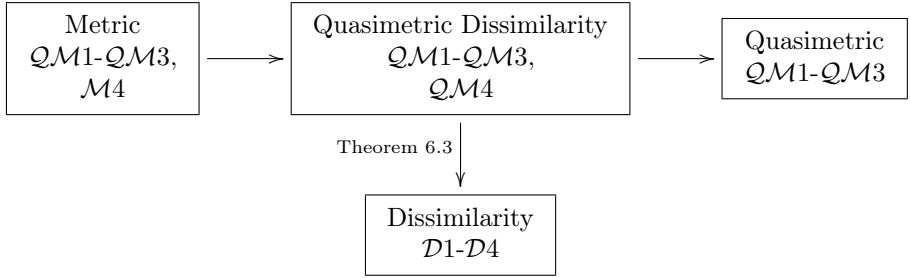
\begin{figure}[h]
\begin{centering}
\[
\vcenter{\xymatrix@C=1cm{\boxed{\begin{array}{c}
\textnormal{Metric}\\
\mathcal{QM}1\textnormal{-}\mathcal{QM}3,\\
\mathcal{\mathcal{M}}4
\end{array}}\ar@{->}[r] & \boxed{\begin{array}{c}
\textnormal{Quasimetric Dissimilarity}\\
\mathcal{QM}1\textnormal{-}\mathcal{QM}3,\\
\mathcal{\mathcal{QM}}4
\end{array}}\ar@{->}[d]_{\textnormal{Theorem }\ref{thm:Any-quasimetric-dissimilarity}}\ar@{->}[r] & \boxed{\begin{array}{c}
\textnormal{Quasimetric}\\
\mathcal{QM}1\textnormal{-}\mathcal{QM}3
\end{array}}\\
 & \boxed{\begin{array}{c}
\textnormal{Dissimilarity}\\
\mathcal{D}1\textnormal{-}\mathcal{\mathcal{D}}4
\end{array}}
}
}
\]
 
\par\end{centering}
\caption{\label{fig:interrelations}Interrelations between metric-like concepts.
Arrows between the boxes stand for ``is a special case of.''}
\end{figure}

To relate quasimetric dissimilarity to two familiar terms, a function
satisfying $\mathcal{QM}1$-$\mathcal{QM}3$ is called a \emph{quasimetric\index{quasimetric}},
and a quasimetric is called a \emph{metric}\index{metric} if it satisfies
the property 
\begin{quote}
$\mathcal{\mathcal{M}}4$ (symmetry) $M\mathbf{ab}=M\mathbf{ba}$,
for any $\mathbf{a},\mathbf{b}\in\mathfrak{S}$. 
\end{quote}
Quasimetric dissimilarity therefore can be viewed as a concept intermediate
between quasimetric and metric. More importantly, however, a quasimetric
dissimilarity (hence also a metric), as shown below, is a special
form of dissimilarity, whereas quasimetric generally is not (see Figure~\ref{fig:interrelations}).

\subsection{Step 3}

The quasimetric dissimilarities 
\[
G^{\left(1\right)}\mathbf{ab}=\inf_{\mathbf{X}\in\mathcal{C}}\Psi^{\left(1\right)}\mathbf{aXb}
\]
and 
\[
G^{\left(2\right)}\mathbf{ab}=\inf_{\mathbf{X}\in\mathcal{C}}\Psi^{\left(2\right)}\mathbf{aXb}
\]
are generally different. However, we will see below that 
\begin{equation}
G^{\left(1\right)}\mathbf{ab}+G^{\left(1\right)}\mathbf{ba}=G^{\left(2\right)}\mathbf{ab}+G^{\left(2\right)}\mathbf{ba},\label{eq:G1G2sums}
\end{equation}
and this quantity is clearly a metric. We will denote it $\overleftrightarrow{G}\mathbf{ab}$,
and interpret it as the \emph{Fechnerian distance\index{Fechnerian distance}}
between $\mathbf{a}$ and $\mathbf{b}$ in the canonical stimulus
space\index{stimulus space!in canonical form} $\mathfrak{S}$. The
double-arrow in $\overleftrightarrow{G}$ is suggestive of the following
way of presenting this quantity: 
\begin{equation}
\overleftrightarrow{G}\mathbf{ab}=\inf_{\mathbf{X},\mathbf{Y}\in\mathcal{C}}\Psi^{\left(1\right)}\mathbf{aXbYa}=\inf_{\mathbf{X},\mathbf{Y}\in\mathcal{C}}\Psi^{\left(2\right)}\mathbf{aXbYa},\label{eq:symmetrizedG}
\end{equation}
the $\mathbf{aXbYa}$ (equivalently, $\mathbf{bYaXb}$) being a closed
chain containing the points $\mathbf{a}$ and $\mathbf{b}$.

\subsection{Subsequent development}

The function $\overleftrightarrow{G}$ is, in a sense, the ultimate
goal of Fechnerian Scaling\index{Fechnerian Scaling}. However, the
metric structure of a space is part of its geometry, and this is what
a full theory of Fechnerian Scaling deals with. In discrete spaces,
consisting of isolated points, the general definition of $\overleftrightarrow{G}$
provides the algorithm for computing it. In more structured spaces,
however, the Fechnerian metric\index{metric!Fechnerian} may be computed
in specialized ways. Rather than considering all possible chains,
in some spaces one integrates infinitesimal dissimilarities along
continuous paths and seeks the shortest paths. In still more structured
spaces this leads to a generalized form of Finsler geometry, where
computations of distances are based on indicatrices or submetric functions.

\begin{figure}
\begin{centering}
\includegraphics[scale=0.4]{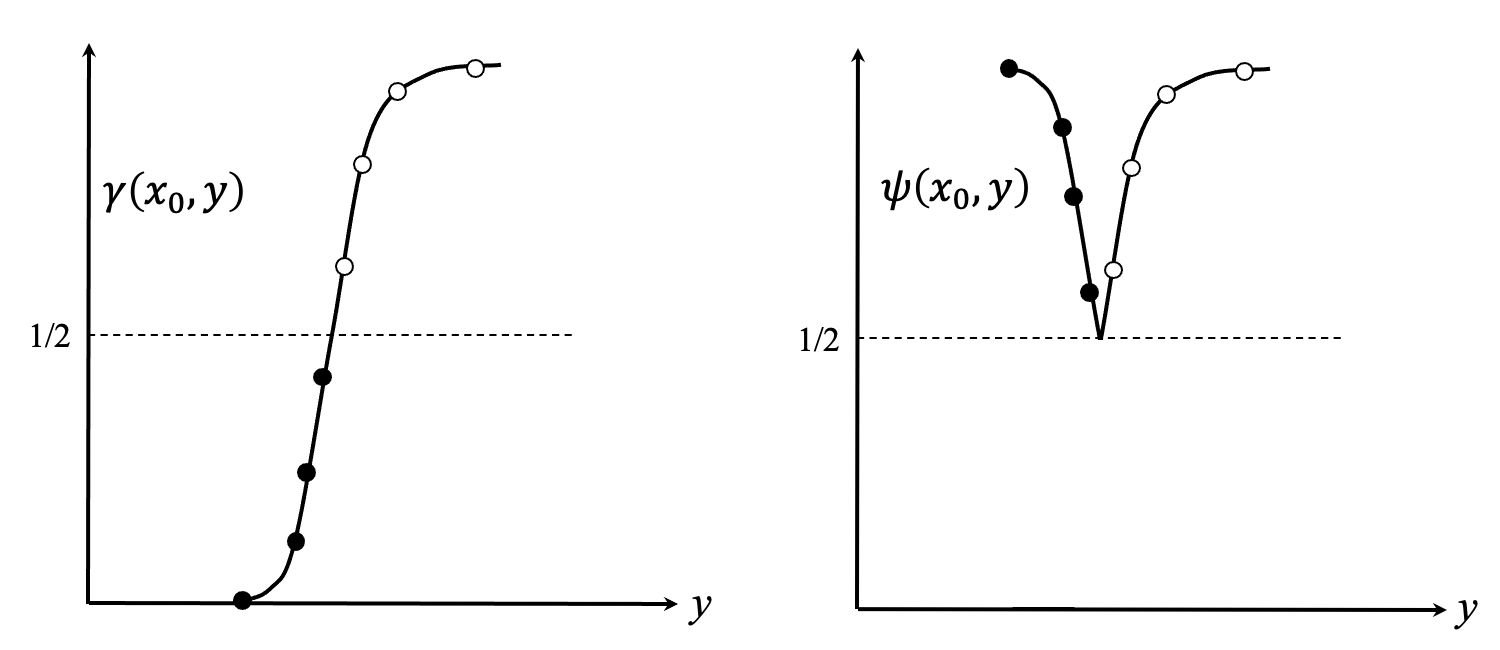} 
\par\end{centering}
\caption{\label{fig:gamma_psi}An illustration of how a ``greater-less''
discrimination probability function (on the left) can be redefined
into a ``same-different''-type discrimination probability function.}
\end{figure}

The psychometric increments $\Psi^{\left(1\right)}$ and $\Psi^{\left(2\right)}$
are at the foundation of Fechnerian Scaling. In this chapter they
are defined through the psychometric function\index{psychometric function}
$\psi$ in (\ref{eq:psi_initial}), which is usually associated with
the same-different version of the \emph{method of constant stimuli}.
In this method, same-different judgements are recorded for repeatedly
presented multiple pairs of stimuli, as indicated, e.g., by the open
circles in Figure \ref{fig:ConstErrorPsi}. However, virtually any
pairwise comparison procedure can be, in principle, used to define
analogues of $\Psi^{\left(1\right)}$ and $\Psi^{\left(2\right)}$.
For instance, if the observer judges pairs of stimuli in terms of
``greater-less'' with respect to some property, the psychometric
function $\gamma$ of Figure \ref{fig:ConstErrorGamma} (assuming
it is in a canonical form) can be converted into a $\psi$-like function
by putting 
\[
\psi\mathbf{xy}=\left\{ \begin{array}{cc}
\gamma\mathbf{xy} & \textnormal{if }\gamma\mathbf{xy}\geq\frac{1}{2}\\
1-\gamma\mathbf{xy} & \textnormal{if }\gamma\mathbf{xy}<\frac{1}{2}
\end{array}\right..
\]
This is illustrated in Figure \ref{fig:gamma_psi} for the case $\mathfrak{S}$
is an interval of real numbers. The psychometric increments then are
defined as 
\[
\Psi^{\left(1\right)}\mathbf{xy}=\left|\gamma\mathbf{xy}-\frac{1}{2}\right|,\Psi^{\left(2\right)}\mathbf{xy}=\left|\gamma\mathbf{yx}-\frac{1}{2}\right|.
\]

Some experimental procedures may yield dissimilarity values $D\mathbf{ab}$
``directly.'' Thus, in one of the procedures of \emph{Multidimensional
Scaling} (MDS), observers are presented pairs of stimuli and asked
to numerically estimate ``how different they are.'' Then, for every
pair of stimuli $\mathbf{a},\mathbf{b}$, some measure of central
tendency of these numerical estimates can be hypothesized to be an
efficient estimator of a dissimilarity function\index{dissimilarity function}
\[
\Psi^{\left(1\right)}\mathbf{ab}=\Psi^{\left(2\right)}\mathbf{ba}=D\mathbf{ab}.
\]
If one can establish that $D\mathbf{aa}=0$ for all stimuli and that
$D\mathbf{ab}>0$ for distinct $\mathbf{a},\mathbf{b}$, then the
stimulus space\index{stimulus space!in canonical form} is in a canonical
form, and the hypothesis that $D$ is a dissimilarity function cannot
be falsified on any finite set of data. However, given sufficient
amount of data, one can usually falsify the hypothesis that $D\mathbf{ab}$
is a quasimetric, by establishing that $D\mathbf{ab}$ violates the
triangle inequality\index{inequality!triangle}. In such situations,
MDS seeks a monotone transformation $g\circ D$ that would yield a
quasimetric. Dissimilarity cumulation offers an alternative approach,
to use $D$ to compute by (\ref{eq:Gdefined}) a quasimetric dissimilarity
$G$ and then symmetrize it by (\ref{eq:symmetrizedG}). We will return
to this situation in Section \ref{sec:Data-analytic}.

\section{Dissimilarity function\index{dissimilarity function}}

The properties $\mathcal{D}3$ and $\mathcal{D}4$ of Definition \ref{def:dissimilarity}
are more conveniently presented in terms of convergence of sequences.
Let us introduce convergence in a stimulus space.\index{stimulus space} 
\begin{defn}
Given two sequences of points in $\mathfrak{S}$, $\left\{ \mathbf{a}_{n}\right\} $
and $\left\{ \mathbf{b}_{n}\right\} $, we say that $\mathbf{a}_{n}$
and $\mathbf{b}_{n}$ \emph{converge to each other, and write this
}$\mathbf{a}_{n}\leftrightarrow\mathbf{b}_{n}$, if $D\mathbf{a}_{n}\mathbf{b}_{n}\rightarrow0$.
In the special case $\mathbf{b}_{n}\equiv\mathbf{b}$, we say that
$\mathbf{a}_{n}$ \emph{converges to} $\mathbf{b},$and write $\mathbf{a}_{n}\rightarrow\mathbf{b}$. 
\end{defn}

\begin{figure}
\begin{centering}
\includegraphics[scale=0.2]{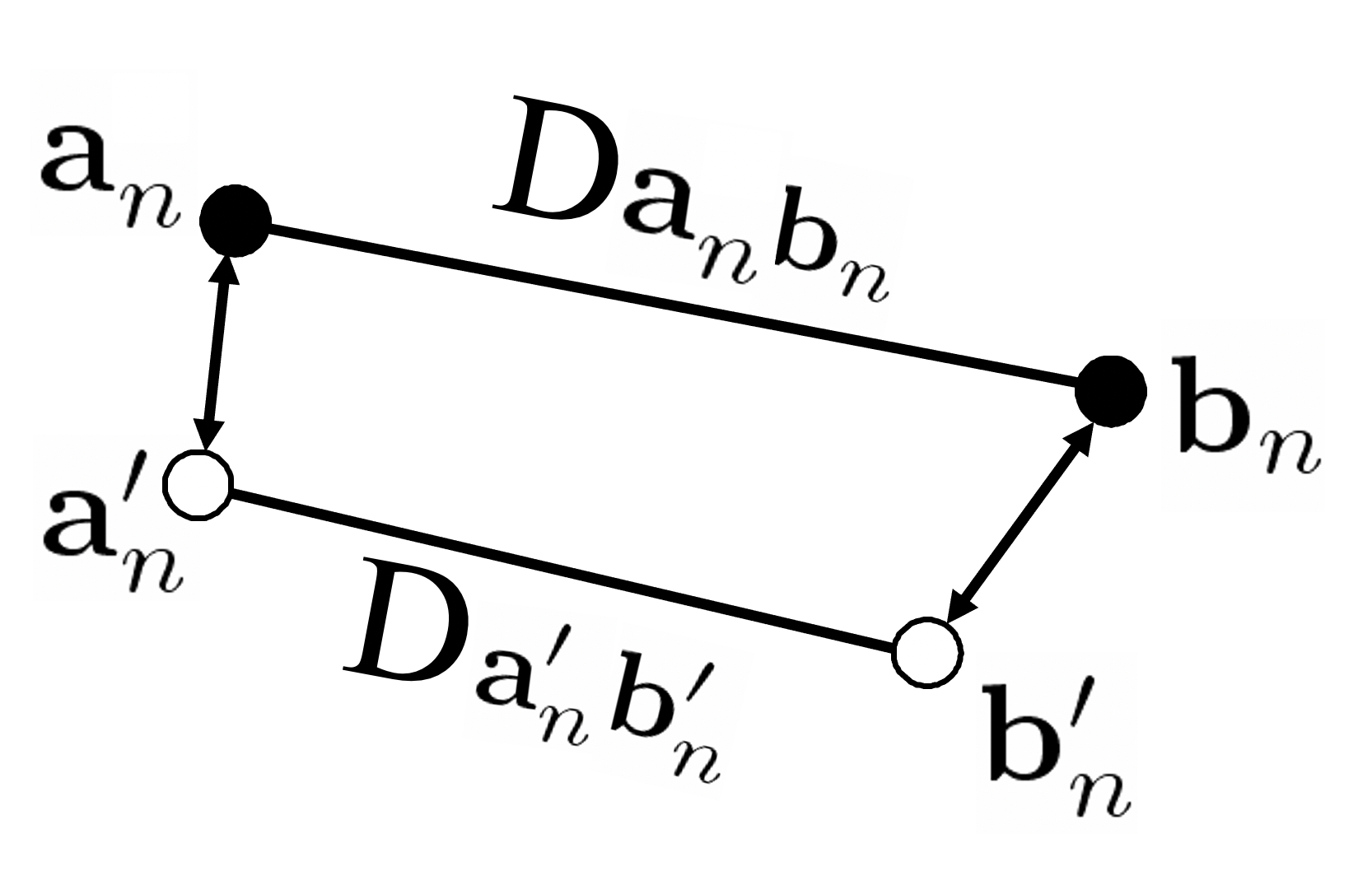} 
\par\end{centering}
\caption{\label{fig:uniform_cont}Illustration of the uniform continuity of
$D$. The dissimilarities $D\mathbf{a}_{n}\mathbf{b}_{n}$ and $D\mathbf{a}'_{n}\mathbf{b}'_{n}$
converge to each other as $\mathbf{a}_{n}$ with $\mathbf{a}'_{n}$
converge to each other and $\mathbf{b}_{n}$ with $\mathbf{b}'_{n}$
converge to each other.}
\end{figure}

The property $\mathcal{D}3$ (uniform continuity)\index{uniform continuity}
then can be presented as follows: 
\[
\textnormal{if }\mathbf{a}_{n}\leftrightarrow\mathbf{a^{\prime}}_{n}\textnormal{ and }\mathbf{b}_{n}\leftrightarrow\mathbf{b^{\prime}}_{n},\mathbf{\textnormal{ then }}D\mathbf{a}_{n}^{\prime}\mathbf{b}_{n}^{\prime}-D\mathbf{a}_{n}\mathbf{b}_{n}\rightarrow0.
\]
In other words, $D$ is a uniformly continuous function (Figure \ref{fig:uniform_cont}).

It is clear that $\mathbf{a}_{n}\leftrightarrow\mathbf{a}_{n}$ is
true for any sequence $\left\{ \mathbf{a}_{n}\right\} $ (because
$D\mathbf{a}_{n}\mathbf{a}_{n}=0$). Assuming that $\mathbf{a}_{n}\leftrightarrow\mathbf{b}_{n}$,
we can use $\mathcal{D}3$ to observe that 
\[
\mathbf{a}_{n}\leftrightarrow\mathbf{b}_{n}\textnormal{ and }\mathbf{a}_{n}\leftrightarrow\mathbf{a}_{n}\Longrightarrow D\mathbf{a}_{n}\mathbf{a}_{n}-D\mathbf{b}_{n}\mathbf{a}_{n}\rightarrow0\Longleftrightarrow D\mathbf{b}_{n}\mathbf{a}_{n}\rightarrow0.
\]
But $D\mathbf{b}_{n}\mathbf{a}_{n}\rightarrow0$ means $\mathbf{b}_{n}\leftrightarrow\mathbf{a}_{n}$,
and we obtain the following proposition. 
\begin{thm}[\emph{symmetry in the small}]
\index{symmetry in the small} For any $\left\{ \mathbf{a}_{n}\right\} ,\left\{ \mathbf{b}_{n}\right\} $,
\[
\mathbf{a}_{n}\leftrightarrow\mathbf{b}_{n}\textnormal{ iff }\mathbf{b}_{n}\leftrightarrow\mathbf{a}_{n}.
\]
\end{thm}

This justifies the terminology (convergence to each other) and notation
in the definition of $\mathbf{a}_{n}\leftrightarrow\mathbf{b}_{n}$.

\begin{figure}
\begin{centering}
\includegraphics[scale=0.3]{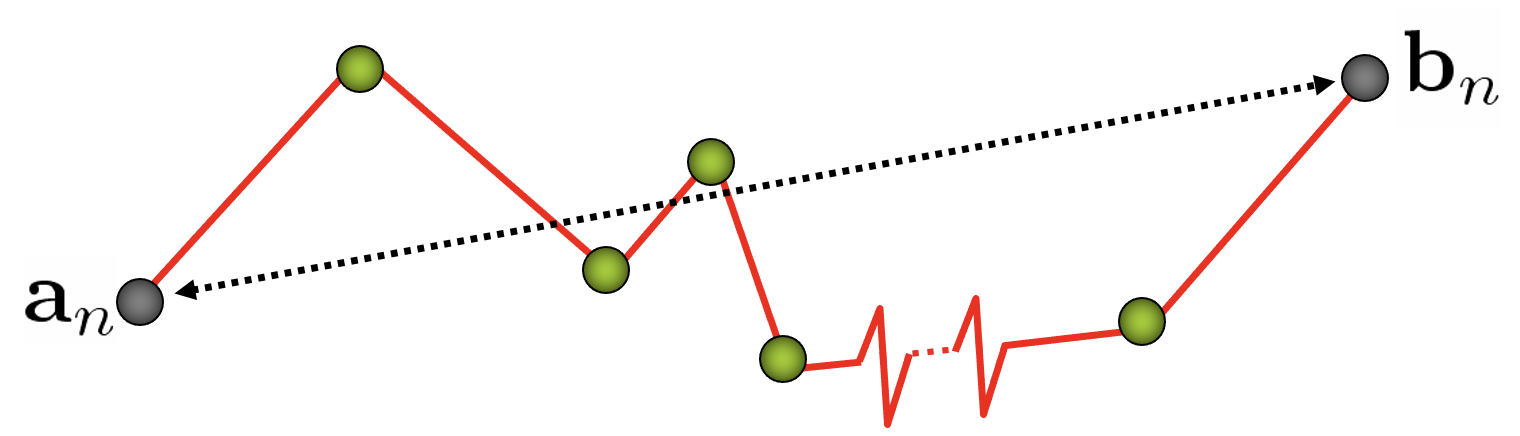} 
\par\end{centering}
\caption{\label{fig:Figchain_prop}Illustration of the chain property of $D$.
If the overall length of the chains $\mathbf{X}_{n}$ connecting $\mathbf{b}_{n}$
to $\mathbf{a}_{n}$ tends to zero, then $\mathbf{a}_{n}$ and $\mathbf{b}_{n}$
converge to each other. This property is nontrivial only if $\left|\mathbf{X}_{n}\right|$,
the number of elements in the chains, tends to infinity. If it is
bounded, $\mathbf{a}_{n}\leftrightarrow\mathbf{b}_{n}$ is a consequence
of the transitivity of the $\leftrightarrow$ relation (not discussed
in the text, but easily established).}
\end{figure}

Property $\mathcal{D}4$ (chain property) can be presented as follows:
for any sequences $\left\{ \mathbf{a}_{n}\right\} ,\left\{ \mathbf{b}_{n}\right\} $
in $\mathfrak{S}$ and $\left\{ \mathbf{X}_{n}\right\} $ in $\mathcal{C}$
(the set of chains), 
\begin{equation}
\textnormal{if }D\mathbf{a}_{n}\mathbf{X}_{n}\mathbf{b}_{n}\rightarrow0,\textnormal{ then }\mathbf{a}_{n}\leftrightarrow\mathbf{b}_{n}.
\end{equation}
Figures \ref{fig:Figchain_prop} provides an illustration.

The properties $\mathcal{D}1$-$\mathcal{D}4$ are logically independent:
none of them is a consequence of the remaining three. This is proved
by constructing examples, for each of these properties, that violate
this property while conforming to the others. For example, to prove
the independence of $\mathcal{D}4$, consider $\mathfrak{S}=\mathbb{R}$,
and let $D\mathbf{xy}=\left(x-y\right)^{2}$ (where $x,y$ are the
numerical values representing $\mathbf{x},\mathbf{y}$, respectively).
The function $D$ clearly satisfies $\mathcal{D}1$-$\mathcal{D}3$.
However, for any points $\mathbf{a},\mathbf{b}$, if the elements
of a chain $\mathbf{X}_{n}$ subdivide $\left[a,b\right]$ into $n$
equal parts, then 
\[
D\mathbf{a}\mathbf{X}_{n}\mathbf{b}=n\left(\frac{b-a}{n}\right)^{2}\rightarrow0,
\]
while the value of $D\mathbf{ab}$ remains equal to $\left(b-a\right)^{2}.$

\section{Quasimetric dissimilarity}

\index{dissimilarity function!quasimetric} We begin by establishing
an important fact: the function $G$ defined by \ref{eq:Gdefined}
and the dissimilarity $D$ are \emph{equivalent in the small}. 
\begin{thm}
\label{thm:GandDinsmall}For any $\left\{ \mathbf{a}_{n}\right\} ,\left\{ \mathbf{b}_{n}\right\} $,
\[
\mathbf{a}_{n}\leftrightarrow\mathbf{b}_{n}\textnormal{ iff }G\mathbf{a}_{n}\mathbf{b}_{n}\rightarrow0.
\]
\end{thm}

To prove this, we first observe that $G\mathbf{a}\mathbf{b}\geq0$,
as the infimum of nonnegative $D\mathbf{a}\mathbf{X}\mathbf{b}$.
If $D\mathbf{a}_{n}\mathbf{b}_{n}\rightarrow0$, we have 
\[
0\leq G\mathbf{a}_{n}\mathbf{b}_{n}=\inf_{\mathbf{X}\in\mathcal{C}}D\mathbf{a}_{n}\mathbf{X}\mathbf{b}_{n}\leq D\mathbf{a}_{n}\mathbf{b}_{n}\rightarrow0,
\]
and this implies $G\mathbf{a}_{n}\mathbf{b}_{n}\rightarrow0$. Conversely,
$\inf_{\mathbf{X}\in\mathcal{C}}D\mathbf{a}_{n}\mathbf{X}\mathbf{b}_{n}\rightarrow0$
means that for some sequence of chains $\left\{ \mathbf{X}_{n}\right\} $,
$D\mathbf{a}_{n}\mathbf{X}_{n}\mathbf{b}_{n}\rightarrow0$. By the
chain property then, $D\mathbf{a}_{n}\mathbf{b}_{n}\rightarrow0$.

Let us now see if $G$ satisfies the properties defining a quasimetric
dissimilarity, $\mathcal{QM}1$-$\mathcal{QM}4$ We immediately see
that it satisfies the triangle inequality\index{inequality!triangle}
($\mathcal{QM}3$): 
\[
G\mathbf{ab}\leq G\mathbf{ac}+G\mathbf{cb},
\]
for any $\mathbf{a},\mathbf{b},\mathbf{c}\in\mathfrak{S}$. Indeed,
\[
G\mathbf{ac}+G\mathbf{cb}=\inf_{\mathbf{X}\in\mathcal{C}}D\mathbf{aXc}+\inf_{\mathbf{Y}\in\mathcal{C}}D\mathbf{cYb}=\inf_{\mathbf{X},\mathbf{Y}\in\mathcal{C}}D\mathbf{aXcYb},
\]
and the set of all possible $\mathbf{aXb}$ contains the set of all
possible $\mathbf{aXcYb}$ chains. It is also easy to see that the
function $G$ is symmetric in the small ($\mathcal{QM}4$). Written
in convergence terms, the property is 
\[
\textnormal{if }G\mathbf{a}_{n}\mathbf{b}_{n}\rightarrow0\textnormal{ then }G\mathbf{b}_{n}\mathbf{a}_{n}\rightarrow0.
\]
It is proved by observing that, by the previous theorem, if $G\mathbf{a}_{n}\mathbf{b}_{n}\rightarrow0$
then $\mathbf{a}_{n}\leftrightarrow\mathbf{b}_{n}$, and then $G\mathbf{b}_{n}\mathbf{a}_{n}\rightarrow0$.
Because we know that $G\mathbf{ab}$ is nonnegative, the properties
$\mathcal{QM}1$ and $\mathcal{QM}2$ follow from 
\[
G\mathbf{ab}=\inf_{\mathbf{X}\in\mathcal{C}}D\mathbf{a}\mathbf{X}\mathbf{b}=0\Longrightarrow D\mathbf{a}\mathbf{X}_{n}\mathbf{b}\rightarrow0,
\]
for some sequence of chains $\left\{ \mathbf{X}_{n}\right\} $. But
this means, by the chain property, $D\mathbf{a}\mathbf{b}=0$, which
is true if and only if $\mathbf{a}=\mathbf{b}$. We have established
therefore 
\begin{thm}
\label{thm:GisQMdissimilarity}The function $G$ is a quasimetric
dissimilarity. 
\end{thm}

It is instructive to see why, as mentioned earlier and as its name
suggests, any quasimetric dissimilarity, and $G$ in particular, is
a dissimilarity function\index{dissimilarity function}. Let $M$
satisfy the properties $\mathcal{QM}1$-$\mathcal{QM}4$. Then $\mathcal{D}1$
and $\mathcal{D}2$ are satisfied trivially. The property $\mathcal{D}3$
(uniform continuity) follows from the fact that, by the triangle inequality\index{inequality!triangle},
\[
\begin{cases}
M\mathbf{aa'}+M\mathbf{b'b} & \geq M\mathbf{ab}-M\mathbf{a'b'},\\
M\mathbf{a'a}+M\mathbf{bb'} & \geq M\mathbf{a'b'}-M\mathbf{ab}.
\end{cases}
\]
By the symmetry in the small property, 
\[
\begin{array}{c}
M\mathbf{a}_{n}\mathbf{a'}_{n}\rightarrow0\Longleftrightarrow M\mathbf{a'}_{n}\mathbf{a}_{n}\rightarrow0,\\
M\mathbf{b'}_{n}\mathbf{b}_{n}\rightarrow0\Longleftrightarrow M\mathbf{b}_{n}\mathbf{b'}_{n}\rightarrow0,
\end{array}
\]
so these convergences imply 
\[
\left|M\mathbf{ab}-M\mathbf{a'b'}\right|\rightarrow0.
\]
The chain property, $\mathcal{D}4$, follows from $M\mathbf{aXb}\geq M\mathbf{ab}$,
by the triangle inequality\index{inequality!triangle}. We have established
therefore 
\begin{thm}
\label{thm:Any-quasimetric-dissimilarity}Any quasimetric dissimilarity
(hence also any metric) is a dissimilarity function. 
\end{thm}

\begin{figure}
\begin{centering}
\includegraphics[scale=0.4]{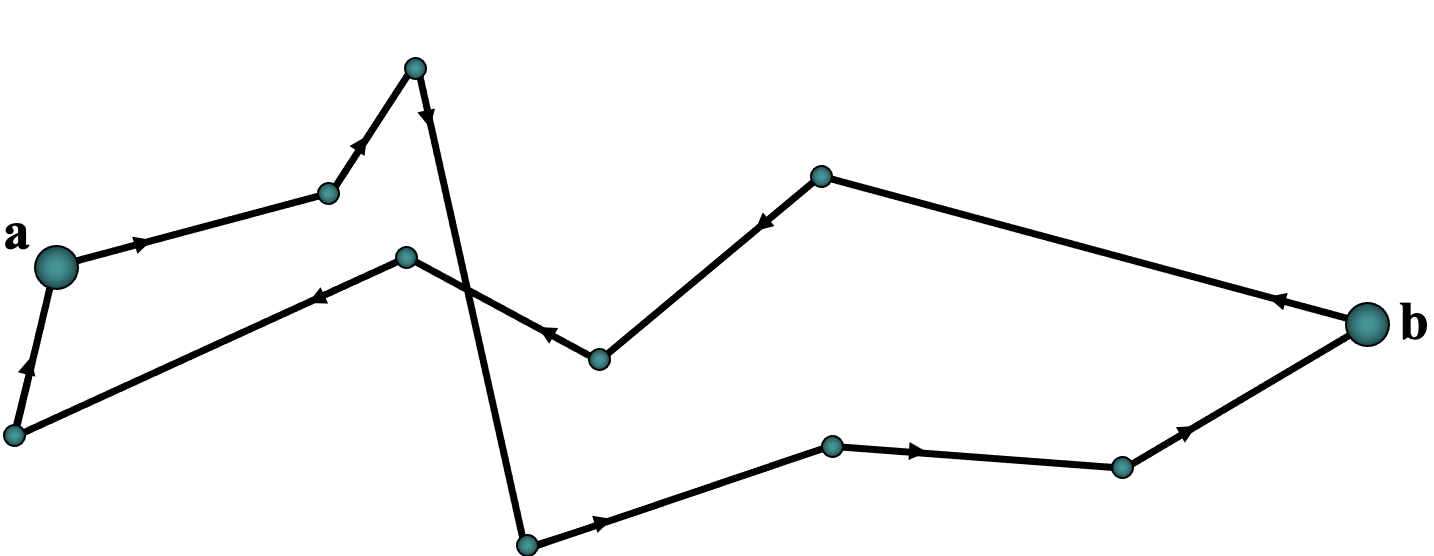} 
\par\end{centering}
\caption{\label{fig:Loop}For any closed chain $\mathbf{X}$ containing points
$\mathbf{a},\mathbf{b}$, the value of $\Psi^{\left(1\right)}\mathbf{X}$
is the same as the value of $\Psi^{\left(2\right)}\mathbf{X}^{\dagger}$,
the same chain traversed in the opposite direction.}
\end{figure}

Let us now return to the to the definition of $G^{\left(1\right)}$,
$G^{\left(2\right)}$, and $\overleftrightarrow{G}$. We need to establish
(\ref{eq:symmetrizedG}), from which (\ref{eq:G1G2sums}) follows.
Given a chain $\mathbf{X}=\mathbf{x}_{1}\mathbf{x}_{2}...\mathbf{x}_{k}$,
let us define the opposite chain $\mathbf{X}^{\dagger}$ as $\mathbf{\mathbf{\mathbf{x}}}_{k}\mathbf{\mathbf{\mathbf{x}}}_{k-1}...\mathbf{x}_{1}$.
By straightforward algebra, 
\[
\Psi^{\left(1\right)}\mathbf{X}=\sum_{i=1}^{k-1}\Psi^{\left(1\right)}\mathbf{\mathbf{x}}_{i}\mathbf{\mathbf{x}}_{i+1}=\sum_{i=1}^{k-1}\left(\psi\mathbf{\mathbf{x}}_{i}\mathbf{\mathbf{x}}_{i+1}-\psi\mathbf{\mathbf{x}}_{i}\mathbf{\mathbf{x}}_{i}\right),
\]
\[
\Psi^{\left(2\right)}\mathbf{X^{\dagger}}=\sum_{i=1}^{k-1}\Psi^{\left(2\right)}\mathbf{\mathbf{x}}_{i+1}\mathbf{\mathbf{x}}_{i}=\sum_{i=1}^{k-1}\left(\psi\mathbf{\mathbf{x}}_{i}\mathbf{\mathbf{x}}_{i+1}-\psi\mathbf{\mathbf{x}}_{i+1}\mathbf{\mathbf{x}}_{i+1}\right).
\]
It follows that 
\[
\Psi^{\left(1\right)}\mathbf{X}-\Psi^{\left(2\right)}\mathbf{X^{\dagger}}=\psi\mathbf{\mathbf{\mathbf{x}}}_{k}\mathbf{\mathbf{\mathbf{x}}}_{k}-\psi\mathbf{\mathbf{\mathbf{x}}}_{1}\mathbf{\mathbf{\mathbf{x}}}_{1}.
\]
In particular, if the chain is closed,$\mathbf{\mathbf{\mathbf{x}}}_{k}=\mathbf{x}_{1}$,
we have 
\[
\Psi^{\left(1\right)}\mathbf{X}=\Psi^{\left(2\right)}\mathbf{X^{\dagger}}.
\]
That is, the $\Psi^{\left(1\right)}$-length of a closed chain equals
the $\Psi^{\left(2\right)}$-length of the same chain traversed in
the opposite direction (see Figure \ref{fig:Loop}). Applying this
to a chain $\mathbf{aXbYa}$, 
\[
\Psi^{\left(1\right)}\mathbf{aXbYa}\mathbf{=}\Psi^{\left(2\right)}\mathbf{aY^{\dagger}b}\mathbf{X}^{\dagger}\mathbf{a},
\]
whence 
\[
\inf_{\mathbf{X},\mathbf{Y}\in\mathcal{C}}\Psi^{\left(1\right)}\mathbf{aXbYa}\mathbf{=}\inf_{\mathbf{Y^{\dagger}},\mathbf{X^{\dagger}}\in\mathcal{C}}\Psi^{\left(2\right)}\mathbf{\mathbf{aY^{\dagger}b}\mathbf{X}^{\dagger}\mathbf{a}}.
\]
Clearly, the set of all possible pairs of chains $\left(\mathbf{X},\mathbf{Y}\right)$
is the same as the set of all pairs $\left(\mathbf{Y^{\dagger}},\mathbf{X^{\dagger}}\right)$,
and by simple renaming, 
\[
\inf_{\mathbf{X},\mathbf{Y}\in\mathcal{C}}\Psi^{\left(1\right)}\mathbf{aXbYa}\mathbf{=}\inf_{\mathbf{X},\mathbf{Y}\in\mathcal{C}}\Psi^{\left(2\right)}\mathbf{aXbYa}.
\]
This proves the following 
\begin{thm}
For any $\mathbf{a},\mathbf{b\in}\mathfrak{S}$, 
\[
G^{\left(1\right)}\mathbf{ab}+G^{\left(1\right)}\mathbf{ba}=G^{\left(2\right)}\mathbf{ab}+G^{\left(2\right)}\mathbf{ba}=\overleftrightarrow{G}\mathbf{ab}.
\]
The function $\overleftrightarrow{G}$ is a metric. 
\end{thm}

The last statement is an immediate corollary of Theorem \ref{thm:GisQMdissimilarity}.

One can think of other ways of combining quasimetric dissimilarities
$G^{\left(1\right)}\mathbf{ab}$ and $G^{\left(1\right)}\mathbf{ba}$
into a metric, such as 
\[
\max\left(G^{\left(1\right)}\mathbf{ab},G^{\left(1\right)}\mathbf{ba}\right),\sqrt{G^{\left(1\right)}\mathbf{ab}+G^{\left(1\right)}\mathbf{ba}},\textnormal{etc.}
\]
Denoting a combination like this $f\left(G^{\left(1\right)}\mathbf{ab},G^{\left(1\right)}\mathbf{ba}\right)$,
the natural requirements are that 
\begin{description}
\item [{(i)}] it should equal $f\left(G^{\left(2\right)}\mathbf{ab},G^{\left(2\right)}\mathbf{ba}\right)$,
and 
\item [{(ii)}] $f\left(x,x\right)\propto x$. 
\end{description}
The latter requirement ensures that if $G^{\left(1\right)}\mathbf{ab}$
always equals $G^{\left(1\right)}\mathbf{ba}$ (i.e., it is already
a metric), then $f\left(G^{\left(1\right)}\mathbf{ab},G^{\left(1\right)}\mathbf{ab}\right)$
is just a multiple of $G^{\left(1\right)}\mathbf{ab}$. Clearly, function
$\overleftrightarrow{G}$ satisfies these requirements. In fact, up
to a scaling coefficient, it is the only such function. 
\begin{thm}
\label{thm:The-only-function}Function $f\left(x,y\right)$ satisfies
\textbf{(i)} and \textbf{(ii)} above for all stimulus spaces if and
only if $f\left(x,y\right)=k\left(x+y\right)$. 
\end{thm}

For a proof, consider a canonical space $\left(\psi,\mathfrak{S}\right)$
with $\mathfrak{S}=\left\{ \mathbf{a,b}\right\} $. It is easy to
see that for any $s,z\in\left(0,1\right]$ one can find probabilities
$\psi\mathbf{aa,}$ $\psi\mathbf{ab,}$ $\psi\mathbf{ba,}$ $\psi\mathbf{bb}$
satisfying 
\[
\begin{array}{l}
G_{1}\mathbf{ab}=\psi\mathbf{ab}-\psi\mathbf{aa}=s\\
G_{1}\mathbf{ba}=\psi\mathbf{ba}-\psi\mathbf{bb}=s\\
G_{2}\mathbf{ab}=\psi\mathbf{ba}-\psi\mathbf{aa}=2s-z\\
G_{2}\mathbf{ba}=\psi\mathbf{ab}-\psi\mathbf{bb}=z
\end{array}.
\]
Then the requirement \textbf{(i)} means that
\[
f\left(s,s\right)=f\left(2s-z,z\right)
\]
should hold for all $s,z\in\left(0,1\right]$. That is, $f\left(2s-z,z\right)$
depends on $s$ only, and we have
\[
f\left(x,y\right)=g\left(x+y\right).
\]
Putting $x=y=\frac{u}{2}$, it follows from the requirement \textbf{(ii)}
that
\[
g\left(x+y\right)=g\left(u\right)=ku,
\]
for some $k>0$. So, our definition of $\overleftrightarrow{G}$ is
not arbitrary, except for choosing $k=1$. 

\section{\label{sec:Discrete}Dissimilarity cumulation in discrete spaces}

\index{dissimilarity cumulation!in discrete spaces} 

\subsection{Direct computation of distances}

A discrete stimulus space\index{stimulus space!discrete} $\left(\mathfrak{S},D\right)$
consists of isolated points, i.e., for every $\mathbf{x}\in\mathfrak{S}$,
\begin{equation}
\inf_{\text{\textbf{y}\ensuremath{\in}\ensuremath{\mathfrak{S}},}\textbf{y}\not=\mathbf{x}}\text{\ensuremath{D\mathbf{x}\mathbf{y}}}>0.
\end{equation}
Although genuinely discrete and even finite stimulus spaces exist
(e.g., the Morse codes of letters and digits studied for their confusability),
this special case is important not so much in its own right as because
any set of empirical data forms a discrete (in fact, finite) space.
This means, e.g., that even if an observer is asked to compare colors
or sounds, the data will form a finite set of pairs associated with
some estimate of discriminability. If the data are sufficiently representative,
the results of applying to them Fechnerian Scaling\index{Fechnerian Scaling}
of discrete spaces should provide a good approximation to the theoretical
Fechnerian Scaling using dissimilarity cumulation along continuous
or smooth paths, as described later in this chapter.

\begin{figure}
\begin{centering}
\includegraphics[scale=0.4]{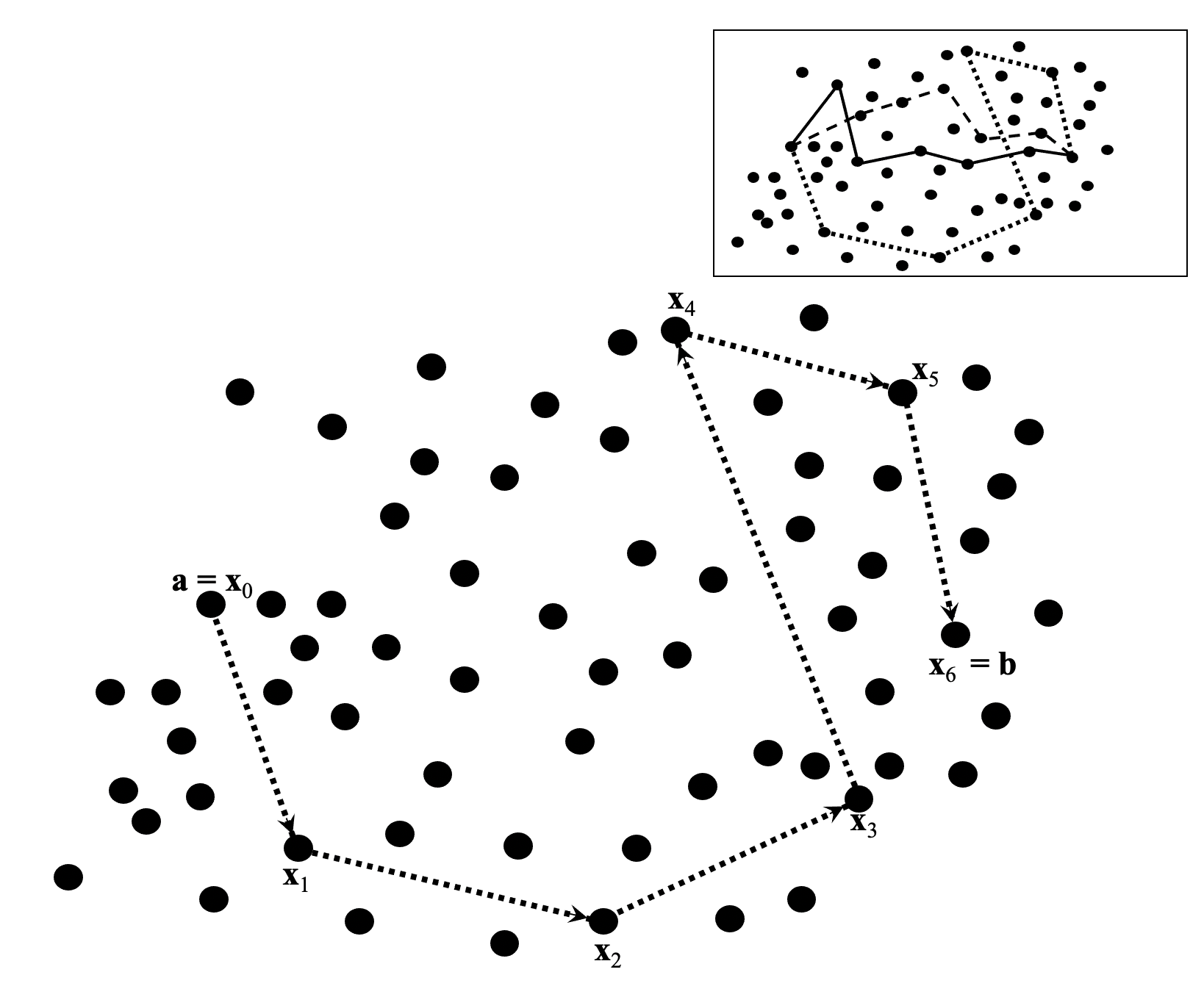} 
\par\end{centering}
\caption{\label{fig:DCdiscrete}Dissimilarity cumulation is discrete spaces.
One considers all possible chains connecting a point $\mathbf{a}$
to a point $\mathbf{b}$ and seeks the infimum of their $D$-lengths.
In a finite space this infimum is the smallest among the $D$-lengths,
and it may be attained by more than one chain. }
\end{figure}

As mentioned earlier, in discrete spaces the general definition of
a Fechnerian distance directly determines the algorithm of computing
them: one tries all possible chains leading from one point to another
(with some obvious heuristics shrinking this set), and finds their
infimum or, in special cases, minimum. This is illustrated in Figure
\ref{fig:DCdiscrete}.

Let us return to the toy example presented in Section \ref{sec:Observation-areas},
and assume that the function $\phi$ there is in fact the discrimination
probability function $\psi$. The canonical space $\left(\mathfrak{S}=\left\{ \mathbf{a},\mathbf{b},\mathbf{c},\mathbf{d}\right\} ,\psi\right)$
is represented by the matrix that we reproduce here for convenience,
\[
\begin{array}{c|c|c|c|c|}
\psi & \mathbf{a} & \mathbf{b} & \mathbf{c} & \mathbf{d}\\
\hline \mathbf{a} & 0.3 & 0.4 & 0.6 & 0.7\\
\hline \mathbf{b} & 0.4 & 0.2 & 0.3 & 0.5\\
\hline \mathbf{c} & 0.5 & 0.3 & 0.1 & 0.2\\
\hline \mathbf{d} & 0.8 & 0.6 & 0.3 & 0.1
\\\hline \end{array}.
\]
We know that all computations can be performed with either $\Psi^{\left(1\right)}$
or $\Psi^{\left(2\right)}$, the final result will be the same. Let
us therefore compute $\Psi^{\left(1\right)}\mathbf{xy}$ by subtracting
from each entry $\psi\mathbf{xy}$ the diagonal value in the same
row, $\psi\mathbf{xx}$ (because the row labels are representing the
stimuli in the first observation area)\index{observation area}. The
result is 
\begin{equation}
\begin{array}{c|c|c|c|c|}
\Psi^{\left(1\right)} & \mathbf{a} & \mathbf{b} & \mathbf{c} & \mathbf{d}\\
\hline \mathbf{a} & 0 & 0.1 & 0.3 & 0.4\\
\hline \mathbf{b} & 0.2 & 0 & 0.1 & 0.3\\
\hline \mathbf{c} & 0.4 & 0.2 & 0 & 0.1\\
\hline \mathbf{d} & 0.7 & 0.5 & 0.2 & 0
\\\hline \end{array}.\label{eq:matrix D toy}
\end{equation}
Let us, e.g., consider next all chains leading from $\mathbf{a}$
to $\mathbf{d}$, and from $\mathbf{d}$ to $\mathbf{a}$. We obviously
need not consider chains with loops in them (such as $\mathbf{adcacb}$,
containing loops $\mathbf{cac}$ and $\mathbf{adca}$). 
\[
\begin{array}{c|c}
\textnormal{from }\mathbf{a}\textnormal{ to }\mathbf{d} & \textnormal{\ensuremath{\Psi^{\left(1\right)}}-ength}\\
\hline \mathbf{ad} & 0.4\\
\mathbf{abd} & 0.1+0.3\\
\mathbf{acd} & 0.3+0.1\\
\mathbf{abcd} & 0.1+0.1+0.1\\
\mathbf{acbd} & 0.3+0.2+0.3
\end{array},\quad\begin{array}{c|c}
\textnormal{from }\mathbf{d}\textnormal{ to }\mathbf{a} & \textnormal{\ensuremath{\Psi^{\left(1\right)}}-ength}\\
\hline \mathbf{da} & 0.7\\
\mathbf{dba} & 0.5+0.2\\
\mathbf{dca} & 0.2+0.4\\
\mathbf{dcba} & 0.2+0.2+0.2\\
\mathbf{dbca} & 0.5+0.1+0.4
\end{array}.
\]
The shortest chains here are $\mathbf{abcd}$ and either of $\mathbf{dca}$
and $\mathbf{dcba}$, their $\Psi^{\left(1\right)}$-lengths being,
respectively, 
\[
G^{\left(1\right)}\mathbf{ad}=0.3,G^{\left(1\right)}\mathbf{da}=0.6.
\]
Thence 
\[
\overleftrightarrow{G}\mathbf{ab}=0.3+0.6=0.9.
\]
Repeating this procedure for each other pair of stimuli, we obtain
the following complete set of $G^{\left(1\right)}$-distances, 
\begin{equation}
\begin{array}{c|c|c|c|c|}
G^{\left(1\right)} & \mathbf{a} & \mathbf{b} & \mathbf{c} & \mathbf{d}\\
\hline \mathbf{a} & 0 & 0.1 & 0.2 & 0.3\\
\hline \mathbf{b} & 0.2 & 0 & 0.1 & 0.2\\
\hline \mathbf{c} & 0.4 & 0.2 & 0 & 0.1\\
\hline \mathbf{d} & 0.6 & 0.4 & 0.2 & 0
\\\hline \end{array},\label{eq:matrix G1 toy}
\end{equation}
and, by symmetrization, the complete set of Fechnerian distances,
\begin{equation}
\begin{array}{c|c|c|c|c|}
\overleftrightarrow{G} & \mathbf{a} & \mathbf{b} & \mathbf{c} & \mathbf{d}\\
\hline \mathbf{a} & 0 & 0.3 & 0.6 & 0.9\\
\hline \mathbf{b} & 0.3 & 0 & 0.3 & 0.6\\
\hline \mathbf{c} & 0.6 & 0.3 & 0 & 0.3\\
\hline \mathbf{d} & 0.9 & 0.6 & 0.3 & 0
\\\hline \end{array}.\label{eq:matrix G toy}
\end{equation}
The shortest chains are not generally unique, as we have seen in our
toy example. However, their infimum for any given pair of points (in
the case of finite sets, minimum) is always determined uniquely. (Note
that it is only a numerical accident that all $\overleftrightarrow{G}$
in our example are below 1, there is no general upper bound for $\overleftrightarrow{G}$
computed from probability values.)

Recall that a label in the canonical stimulus space\index{stimulus space!in canonical form},
say, $\mathbf{a}$, is a representations of two different stimuli
in the two observation areas. If one goes back to the original stimulus
spaces, the Fechnerian distance 0.6 between points $\mathbf{b}$ and
$\mathbf{d}$ in the canonical space $\mathfrak{S}$, is in fact both 
\begin{description}
\item [{(i)}] the distance between either of the stimuli $\mathbf{x}_{2},\mathbf{x}_{3}$
and either of the stimuli $\mathbf{x}_{6},\mathbf{x}_{7}$ in the
stimulus space\index{stimulus space} $\mathfrak{S}_{1}^{*}$ (first
observation area); and 
\item [{(ii)}] the distance between any of the stimuli $\mathbf{y}_{4},\mathbf{y}_{5},\mathbf{y}_{6},\mathbf{y}_{7}$
and the stimulus $\mathbf{y}_{1}$ in the stimulus space $\mathfrak{S}_{2}^{*}$
(second observation area).\index{observation area} 
\end{description}
Indeed, any of the stimuli $\mathbf{y}_{4},\mathbf{y}_{5},\mathbf{y}_{6},\mathbf{y}_{7}$
and either of $\mathbf{x}_{2},\mathbf{x}_{3}$ are each other's PSEs,
mapped into $\mathbf{b}$ in the canonical representation. Similarly,
either of the stimuli $\mathbf{x}_{6},\mathbf{x}_{7}$ and $\mathbf{y}_{1}$
are each other's PSEs, mapped into $\mathbf{d}$.

Let us emphasize that Fechnerian distances are always defined \emph{within}
observation areas rather than across them. This is the reason Fechnerian
distance $\overleftrightarrow{G}$ is a true metric, with the symmetry
property\index{metric}. Within a single observation area the order
of two stimuli has no operational meaning, so $\overleftrightarrow{G}\mathbf{xy}$
cannot be different from $\overleftrightarrow{G}\mathbf{yx}$. The
situation is different when we consider a discrimination probability
function $\psi$ or a dissimilarity function\index{dissimilarity function}
$D$ (e.g., $\Psi^{\left(1\right)}$ or $\Psi^{\left(2\right)}$).
In $\psi\mathbf{xy}$ and $D\mathbf{xy}$ the first and second stimuli
belong to, respectively, the first and second observation areas, making
them meaningfully asymmetric.

The quasimetric dissimilarity $G$ (e.g., $G^{\left(1\right)}$or
$G^{\left(2\right)}$) from which $\overleftrightarrow{G}$ is computed,
strictly speaking, is not interpretable before it is symmetrized.
$G\mathbf{xy}$ is merely a component of $\overleftrightarrow{G}\mathbf{xy}$,
the other component being $G\mathbf{yx}$. However, in the rest of
this paper we are focusing on $G$ rather than $\overleftrightarrow{G}$
because the computation of $G$ from $D$ is the nontrivial part of
Fechnerian Scaling, leaving one only the trivial step of adding $G\mathbf{yx}$
to $G\mathbf{xy}$.

\subsection{Recursive corrections for violations of the triangle inequality\index{inequality!triangle}}

\label{sec: recurs} The procedure described in this section is not
the only way to compute $G$ from $D$. Another way, known as the
Floyd-Warshall algorithm\index{Floyd-Warshall algorithm}, is based
on the following logic. If one considers in $\mathfrak{S}$ all possible
ordered triples $\mathbf{xyz}$ with pairwise distinct elements, and
finds out that all of them satisfy the triangle inequality, 
\[
D\mathbf{xz}\le D\mathbf{xy}+D\mathbf{yz},
\]
then $D$ simply coincides with $G$. If therefore, in the general
case, one could ``correct'' all ordered triples $\mathbf{xyz}$
for violations of the triangle inequality, one would transform $D$
into $G$. The following is how this can be done for any finite stimulus
space (a generalization to be discussed in Section \ref{subsec:DzhDzh}).

Let $\mathfrak{S}$ contains $k$ points, and let $\mathfrak{S}_{3}$
denote the set of $t=k\left(k-1\right)\left(k-2\right)$ ordered triples
of pairwise distinct points of $\mathfrak{S}$. We will call the elements
of $\mathfrak{S}_{3}$ \emph{triangles}. For $n=0,1,\ldots$, let
$\mathbf{T}^{\left(n\right)}$ denote a sequence of the $t$ triangles
in $\mathfrak{S}_{3}$ (in an arbitrary order, as its choice will
be shown to be immaterial for the end result). For each $n$, we index
the triangles in $\mathbf{T}^{\left(n\right)}$ by double indices
$\left(n,1\right),\left(n,2\right),\ldots,\left(n,t\right)$, and
we order all such pairs lexicographically: the successor $\left(n,i\right)'$
of $\left(n,i\right)$ is $\left(n,i+1\right)$ if $i<t$ and $\left(n,t\right)'=\left(n+1,1\right)$.
So the triangle indexed $\left(n,i\right)'$ is in $\mathbf{T}^{\left(n\right)}$,
while the triangle indexed $\left(n,t\right)'$ is the first one in
$\mathbf{T}^{\left(n+1\right)}$. 
\begin{defn}
\label{def:cor}Given a finite space $\left(\mathfrak{S},D\right)$
and the triangle sequences $\mathbf{T}^{\left(0\right)},\mathbf{T}^{\left(1\right)},\ldots$,
the dissimilarity function\index{dissimilarity function} $M^{\left(n,i\right)}$
for $n=0,1,\ldots$ and $i=1,2,\ldots,t$ is defined by induction
as follows.

(i) $M^{\left(0,i\right)}\equiv D$ for $i=1,2,\ldots,t$.

(ii) Let $M^{\left(n,i\right)}$ be defined for some $\left(n,i\right)\geq\left(0,t\right)$,
and let $\mathbf{abc}$ be the triangle indexed by $\left(n,i\right)'$.
Then $M^{\left(n,i\right)'}\mathbf{xy}=M^{\left(n,i\right)}\mathbf{xy}$
for all $\mathbf{x},\mathbf{y}\in\mathfrak{S}$ except, possibly,
for $M^{\left(n,i\right)'}\mathbf{ac}$, defined as 
\[
M^{\left(n,i\right)'}\mathbf{ac}=\min\left(M^{\left(n,i\right)}\mathbf{ac},M^{\left(n,i\right)}\mathbf{ab}+M^{\left(n,i\right)}\mathbf{bc}\right).
\]
\end{defn}

(Note that in every triangle $\mathbf{xyz}$ the triangle inequality\index{inequality!triangle}
is tested only in the form $D\mathbf{xz}\leq D\mathbf{xy}+D\mathbf{yz}$,
irrespective of whether any of the remaining five triangles inequalities
is violated, $D\mathbf{xy}\leq D\mathbf{xz}+D\mathbf{zy}$, $D\mathbf{zy}\leq D\mathbf{zx}+D\mathbf{xy},$
etc.)

The function $M^{\left(n,i\right)}$ for every $\left(n,i\right)$
is clearly a dissimilarity function\index{dissimilarity function},
and it is referred as the \emph{corrected dissimilarity function}\index{dissimilarity function!corrected}.
If, at some $\left(n,i\right)$, the function $M^{\left(n,i\right)}$
is a quasimetric dissimilarity, it is called the \emph{terminal corrected
dissimilarity function}.

It follows from Definition \ref{def:cor} that if $\left(m,j\right)\geq\left(n,i\right)$,
then $M^{\left(m,j\right)}\mathbf{xy}\leq M^{\left(n,i\right)}\mathbf{xy}$
for all $\mathbf{x},\mathbf{y}\in\mathfrak{S}$. Therefore, if, for
some $n$, $M^{\left(n+1,t\right)}\equiv M^{\left(n,t\right)}$, then
$M^{\left(n+1,1\right)}\equiv M^{\left(n,t\right)}$, implying that
$M^{\left(n,t\right)}$ is the terminal dissimilarity function. The
converse being obvious, we have 
\begin{lem}
\label{LM:add}$M^{\left(n,i\right)}$ is the terminal corrected dissimilarity
function if and only if $M^{\left(n+1,t\right)}\equiv M^{\left(n,t\right)}$. 
\end{lem}

The next lemma provides a link between the algorithm being considered
and the use of chains in the definition of $G$. Recall that $\mathcal{C}$
denotes the set of all chains in $\mathfrak{S}$. 
\begin{lem}
\label{LM}For any $n=0,1,\ldots$, any $i=1,2,\ldots,t$, and any
$\mathbf{a},\mathbf{b}\in\mathfrak{S}$, there is a chain $\mathbf{X}\in\mathcal{C}$
such that 
\[
M^{\left(n,i\right)}\mathbf{ab}=D\mathbf{aXb}.
\]
\end{lem}

The proof obtains by induction on the lexicographically ordered $\left(n,i\right)$.
The statement holds for $n=0$, with $\mathbf{X}$ an empty chain.
Let it hold for all double indices up to and including $\left(n,i\right)\geq\left(0,t\right)$,
and let $\mathbf{abc}$ be the triangle indexed $\left(n,i\right)'$.
Then the statement is clearly true for $M^{\left(n,i\right)'}\mathbf{ac}$
whether it equals $M^{\left(n,i\right)}\mathbf{ac}$ or $M^{\left(n,i\right)}\mathbf{ab}+M^{\left(n,i\right)}\mathbf{bc}$,
and it is true for all other $\mathbf{xy}$ because then $M^{\left(n,i\right)'}\mathbf{xy}=M^{\left(n,i\right)}\mathbf{xy}$.

Does a terminal dissimilarity function\index{dissimilarity function}
necessarily exist? Let us assume it does not. Then, by Lemma \ref{LM:add},
$M^{\left(n+1,t\right)}$ and $M^{\left(n,t\right)}$ do not coincide
for all $n=0,1,\ldots$. Since $\mathfrak{S}\times\mathfrak{S}$ is
finite, there should exist distinct points $\mathbf{a},\mathbf{b}\in\mathfrak{S}$
and an infinite sequence of positive integers $n_{1}<n_{2}<\ldots$
for which 
\[
D\mathbf{ab}\neq M^{\left(n_{1},t\right)}\mathbf{ab}\neq M^{\left(n_{2},t\right)}\mathbf{ab}\neq\ldots.
\]
From Definition \ref{def:cor} it follows then that 
\[
D\mathbf{ab}>M^{\left(n_{1},t\right)}\mathbf{ab}>M^{\left(n_{2},t\right)}\mathbf{ab}>\ldots.
\]
By Lemma \ref{LM}, for every $\left(n_{i},t\right)$ there should
exist a chain $\mathbf{X}_{i}$ such that 
\[
M^{\left(n_{i},t\right)}\mathbf{ab}=D\mathbf{aX}_{i}\mathbf{b},\;i=1,2,\ldots.
\]
But a sequence of inequalities 
\[
D\mathbf{a}\mathbf{b}>D\mathbf{aX}_{n_{1}}\mathbf{b}>D\mathbf{aX}_{n_{2}}\mathbf{b}>\ldots
\]
is impossible in a finite set, because the set of chains with lengths
below a given value is finite. This contradiction proves the existence
of a terminal dissimilarity function\index{dissimilarity function}.
Let us denote it by $M$. Observe that for any $\mathbf{a},\mathbf{b}\in\mathfrak{S}$
and any chain $\mathbf{X\in\mathcal{C}}$, 
\[
D\mathbf{aXb}\geq M\mathbf{aXb}.
\]
But $M$ satisfies the triangle inequality\index{inequality!triangle},
whence 
\[
M\mathbf{aXb}\geq M\mathbf{ab},
\]
whence

\[
M\mathbf{ab}\leq D\mathbf{aXb}.
\]
By Lemma \ref{LM}, this implies 
\[
M\mathbf{ab}=\min_{\mathbf{X}\in\mathcal{C}}D\mathbf{aXb},
\]
which equals $G\mathbf{ab}$ by definition. We have established therefore 
\begin{thm}
A terminal corrected dissimilarity function\index{dissimilarity function}
exists, and it coincides with the quasimetric dissimilarity $G$ induced
by the initial dissimilarity function $D$. 
\end{thm}

It is worthwhile to emphasize that nowhere in the proof we have used
a specific order of the triangles in $\mathbf{T}^{\left(n\right)}$.

We see that dissimilarities on finite sets can be viewed as ``imperfect''
quasimetric dissimilarities, and the dissimilarity cumulation procedure
can be recast as a series of recursive corrections of the dissimilarities
for the violations of the triangle inequality\index{inequality!triangle}.

Let us illustrate the procedure on our toy example, starting with
the matrix of dissimilarities 
\[
\begin{array}{c|c|c|c|c|}
\Psi^{\left(1\right)}=D & \mathbf{a} & \mathbf{b} & \mathbf{c} & \mathbf{d}\\
\hline \mathbf{a} & 0 & 0.1 & 0.3 & 0.4\\
\hline \mathbf{b} & 0.2 & 0 & 0.1 & 0.3\\
\hline \mathbf{c} & 0.4 & 0.2 & 0 & 0.1\\
\hline \mathbf{d} & 0.7 & 0.5 & 0.2 & 0
\\\hline \end{array}
\]
and using, for each $\mathbf{T}^{\left(n\right)}$ the same sequence
of $t=24$ triangles 
\begin{equation}
\begin{array}{rccccc}
i=1 & 2 & 3 & \ldots & 23 & 24\\
\mathbf{acb} & \mathbf{abc} & \mathbf{adc} & \ldots & \mathbf{dac} & \mathbf{dbc}
\end{array}.\label{eq:sequence toy}
\end{equation}
It is obtained by cycling through the first element (4 values), subcycling
through the last element (3 values), and sub-subcycling through the
middle element (2 values), in the alphabetic order.

Testing the triangles in $\mathbf{T}^{\left(1\right)}$ one by one,
$M^{\left(1,1\right)}$coincides with $D$ because the triangle indexed
$\left(1,1\right)$ is $\mathbf{acb}$, and the triangle inequality
in it is not violated. Similarly, $M^{\left(1,2\right)}\equiv M^{\left(1,1\right)}$
and $M^{\left(1,3\right)}\equiv M^{\left(1,2\right)}$ because the
triangle inequality is not violated in the triangles labeled $\left(1,2\right)$
and $\left(1,3\right)$. The first violation of the triangle inequality
occurs in the triangle indexed $\left(1,3\right)$, $\mathbf{abc}$:
\[
0.3=D\mathbf{ac}>D\mathbf{ab}+D\mathbf{bc}=0.1+0.1.
\]
We ``correct'' the value of $D\mathbf{ac}$ therefore by replacing
0.3 with 0.2 (shown in parentheses in matrix $M^{\left(1,3\right)}$
below): 
\[
\begin{array}{c|c|c|c|c|}
D & \mathbf{a} & \mathbf{b} & \mathbf{c} & \mathbf{d}\\
\hline \mathbf{a} & 0 & 0.1 & 0.3 & 0.4\\
\hline \mathbf{b} & 0.2 & 0 & 0.1 & 0.3\\
\hline \mathbf{c} & 0.4 & 0.2 & 0 & 0.1\\
\hline \mathbf{d} & 0.7 & 0.5 & 0.2 & 0
\\\hline \end{array}\Rightarrow\begin{array}{c|c|c|c|c|}
M^{\left(1,3\right)} & \mathbf{a} & \mathbf{b} & \mathbf{c} & \mathbf{d}\\
\hline \mathbf{a} & 0 & 0.1 & (0.2) & 0.4\\
\hline \mathbf{b} & 0.2 & 0 & 0.1 & 0.3\\
\hline \mathbf{c} & 0.4 & 0.2 & 0.0 & 0.1\\
\hline \mathbf{d} & 0.7 & 0.5 & 0.2 & 0.0
\\\hline \end{array}
\]
No violations occur until we reach the triangle indexed $\left(1,20\right)$,
so $M^{\left(1,19\right)}\equiv M^{\left(1,18\right)}\equiv\ldots\equiv M^{\left(1,3\right)}$.
In $M^{\left(1,19\right)},$however, we have, for the triangle $\mathbf{dca}$:
\[
0.7=M^{\left(1,19\right)}\mathbf{da}>M^{\left(1,19\right)}\mathbf{dc}+M^{\left(1,19\right)}\mathbf{ca}=0.2+0.4,
\]
We correct $M^{\left(1,19\right)}\mathbf{da}$ from 0.7 to 0.6, as
shown in the parentheses in matrix $M^{\left(1,20\right)}$. 
\[
\begin{array}{c|c|c|c|c|}
M^{\left(1,19\right)} & \mathbf{a} & \mathbf{b} & \mathbf{c} & \mathbf{d}\\
\hline \mathbf{a} & 0 & 0.1 & 0.2 & 0.4\\
\hline \mathbf{b} & 0.2 & 0 & 0.1 & 0.3\\
\hline \mathbf{c} & 0.4 & 0.2 & 0 & 0.1\\
\hline \mathbf{d} & 0.7 & 0.5 & 0.2 & 0
\\\hline \end{array}\Rightarrow\begin{array}{c|c|c|c|c|}
M^{\left(1,20\right)} & \mathbf{a} & \mathbf{b} & \mathbf{c} & \mathbf{d}\\
\hline \mathbf{a} & 0 & 0.1 & 0.2 & 0.4\\
\hline \mathbf{b} & 0.2 & 0 & 0.1 & 0.3\\
\hline \mathbf{c} & 0.4 & 0.2 & 0 & 0.1\\
\hline \mathbf{d} & (0.6) & 0.5 & 0.2 & 0
\\\hline \end{array}.
\]
We deal analogously with the third violation of the triangle inequality\index{inequality!triangle},
in the triangle $\mathbf{dcb}$, indexed $\left(1,22\right)$: 
\[
0.5=M^{\left(1,21\right)}\mathbf{db}>M^{\left(1,21\right)}\mathbf{dc}+M^{\left(1,21\right)}\mathbf{cb}=0.2+0.2.
\]
So $M^{\left(1,21\right)}\equiv M^{\left(1,20\right)}\equiv M^{\left(1,19\right)}$,
and 
\[
\begin{array}{c|c|c|c|c|}
M^{\left(1,21\right)} & \mathbf{a} & \mathbf{b} & \mathbf{c} & \mathbf{d}\\
\hline \mathbf{a} & 0 & 0.1 & 0.2 & 0.4\\
\hline \mathbf{b} & 0.2 & 0 & 0.1 & 0.3\\
\hline \mathbf{c} & 0.4 & 20. & 0 & 0.1\\
\hline \mathbf{d} & 0.6 & 0.5 & 0.2 & 0
\\\hline \end{array}\Rightarrow\begin{array}{c|c|c|c|c|}
M^{\left(1,22\right)} & \mathbf{a} & \mathbf{b} & \mathbf{c} & \mathbf{d}\\
\hline \mathbf{a} & 0 & 0.1 & 0.2 & 0.4\\
\hline \mathbf{b} & 0.2 & 0 & 0.1 & 0.3\\
\hline \mathbf{c} & 0.4 & 20. & 0 & 0.1\\
\hline \mathbf{d} & 0.6 & (0.4) & 0.2 & 0
\\\hline \end{array}.
\]
With the remaining two triangles before the sequence $\mathbf{T}^{\left(1\right)}$
has been exhausted no violations occur, so $M^{\left(1,24\right)}\equiv M^{\left(1,23\right)}\equiv M^{\left(1,22\right)}$
is the matrix with which the second sequence, $\mathbf{T}^{\left(2\right)}$,
begins. The first and only violation here occurs at the triangle indexed
$\left(2,5\right)$, $\mathbf{abd}$: 
\[
0.4=M^{(2,4)}\mathbf{ad}>M^{(2,4)}\mathbf{ab}+M^{(2,4)}\mathbf{bd}=0.1+0.2,
\]
So $M^{(2,4)}\equiv\ldots\equiv M^{(2,1)}\equiv M^{(1,24)}$, and
\[
\begin{array}{c|c|c|c|c|}
M^{\left(1,24\right)} & \mathbf{a} & \mathbf{b} & \mathbf{c} & \mathbf{d}\\
\hline \mathbf{a} & 0 & 0.1 & 0.2 & 0.4\\
\hline \mathbf{b} & 0.2 & 0 & 0.1 & 0.3\\
\hline \mathbf{c} & 0.4 & 0.2 & 0 & 0.1\\
\hline \mathbf{d} & 0.6 & 0.4) & 0.2 & 0
\\\hline \end{array}\Rightarrow\begin{array}{c|c|c|c|c|}
M^{\left(2,5\right)} & \mathbf{a} & \mathbf{b} & \mathbf{c} & \mathbf{d}\\
\hline \mathbf{a} & 0 & 0.1 & 0.2 & (0.3)\\
\hline \mathbf{b} & 0.2 & 0 & 0.1 & 0.3\\
\hline \mathbf{c} & 0.4 & 0.2 & 0 & 0.1\\
\hline \mathbf{d} & 0.6 & 0.4 & 0.2 & 0
\\\hline \end{array}.
\]
One can verify that $M^{\left(2,5\right)}$ is a quasimetric dissimilarity
on $\mathfrak{S}=\left\{ \mathbf{a},\mathbf{b},\mathbf{c},\mathbf{d}\right\} $,
so that $M^{\left(2,6\right)}$ and all higher-indexed matrices remain
equal to $M^{\left(2,5\right)}$. The latter therefore is the terminal
corrected dissimilarity, and its comparison with (\ref{eq:matrix G1 toy})
shows that it coincides with $G=G^{\left(1\right)}$, the quasimetric
induced by the initial dissimilarity function\index{dissimilarity function}
$D=\Psi^{\left(1\right)}$.

\section{Dissimilarity cumulation in path-connected spaces}

\index{dissimilarity cumulation!in path-connected spaces} 

\subsection{Chains-on-nets and paths}

We now turn to dissimilarity cumulation in stimulus spaces $\left(\mathfrak{S},D\right)$
in which points can be connected by \emph{paths}. A path is a continuous
function $\mathbf{f}:\left[a,b\right]\rightarrow\mathfrak{S}$. \index{path}Because
$\left[a,b\right]$ is a closed interval of reals, this function is
also uniformly continuous. The latter means that $\mathbf{f}\left(x\right)\leftrightarrow\mathbf{f}\left(y\right)$
if $x-y\rightarrow0$ ($x,y\in\left[a,b\right]$. We will present
this path more compactly as $\mathbf{f}|\left[a,b\right]$, and say
that it connects $\mathbf{f}\left(a\right)=\mathbf{a}$ to $\mathbf{f}\left(b\right)=\mathbf{b}$,
where $\mathbf{a}$ and $\mathbf{b}$ are allowed to coincide.

To introduce the notion of the length of the path $\mathbf{f}|\left[a,b\right]$,
we need the following auxiliary notions. A \emph{net\index{net}}
on $\left[a,b\right]$ is defined as a sequence of numbers 
\[
\mu=\left(a=x_{0}\leq x_{1}\leq\ldots\leq x_{k}\leq x_{k+1}=b\right),
\]
not necessarily pairwise distinct. The quantity 
\[
\delta\mu=\max_{i=0,1\ldots,k}\left(x_{i+1}-x_{i}\right)
\]
is called the net's\emph{ mesh\index{net!mesh of}}. A net $\mu=\left(a,x_{1},\ldots,x_{k},b\right)$
can be elementwise paired with a chain $\mathbf{X}=\mathbf{x}_{0}\mathbf{x}_{1}\ldots\mathbf{x}_{k}\mathbf{x}_{k+1}$
to form a \emph{chain-on-net\index{chain-on-net}} 
\[
\mathbf{X}^{\mu}=\left(\left(a,\mathbf{\mathbf{x}_{0}}\right),\left(x_{1},\mathbf{x}_{1}\right),\ldots,\left(x_{k},\mathbf{x}_{k}\right),\left(b,\mathbf{x}_{k+1}\right)\right).
\]
Note that the elements of the chain $\mathbf{X}$ need not be pairwise
distinct. The \emph{separation} of\emph{\ }the chain-on-net\emph{\ }$\mathbf{X}^{\mu}$
from the path $\mathbf{f}|\left[a,b\right]$ is defined as 
\[
\sigma\left(\mathbf{f},\mathbf{X}^{\mu}\right)=\max_{x_{i}\in\mu}D\mathbf{f}\left(x_{i}\right)\mathbf{x}_{i}.
\]

\begin{figure}
\begin{centering}
\includegraphics[scale=0.3]{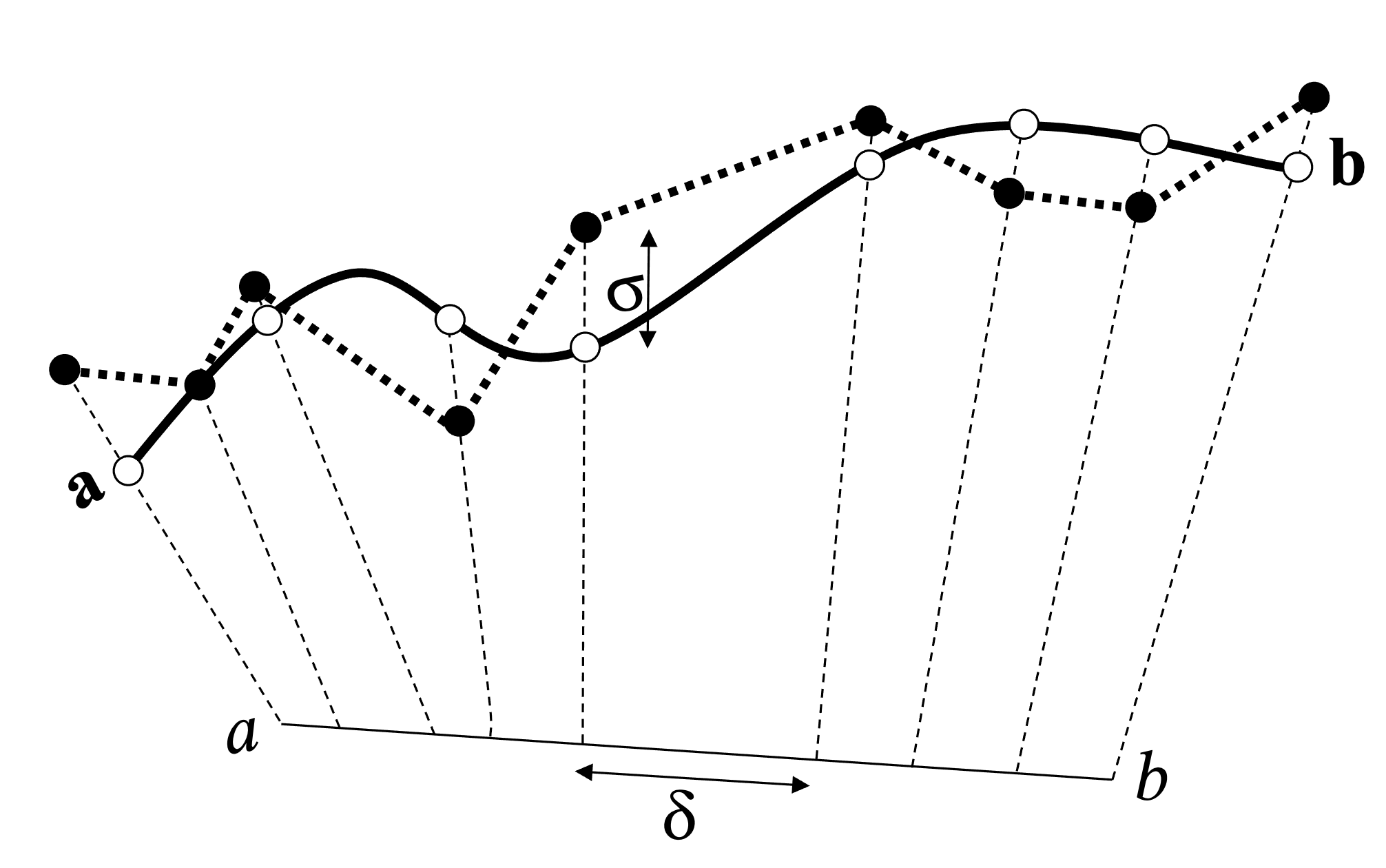} 
\par\end{centering}
\caption{\label{fig:Convergence}Chains-on-nets $\mathbf{X}^{\mu}$ are converging
to a path $\mathbf{f}|\left[a,b\right]$ as $\delta=\delta\mu\rightarrow0$
and $\sigma=\sigma\left(\mathbf{f},\mathbf{X}^{\mu}\right)\rightarrow0$.
The length $D\mathbf{X}$ of the chains has then the limit inferior
that is taken for the length of the path $\mathbf{f}$.}
\end{figure}

\begin{defn}
\label{def:D-length}The $D$\emph{-length\index{path!D-length of}}
of path $\mathbf{f}|\left[a,b\right]$ is defined as 
\[
D\mathbf{f}=\liminf_{\delta\mu\rightarrow0,\sigma\left(\mathbf{f},\mathbf{X}^{\mu}\right)\rightarrow0}D\mathbf{X}.
\]
\end{defn}

The limit inferior stands here for 
\[
\sup_{\varepsilon_{1}>0,\varepsilon_{2}>0}\inf\left\{ D\mathbf{X}:\delta\mu<\varepsilon_{1},\sigma\left(\mathbf{f},\mathbf{X}^{\mu}\right)<\varepsilon_{2}\right\} .
\]
Let us agree to say that $\mathbf{X}^{\mu}$ converges\index{convergence!to a path}
\emph{to} $\mathbf{f}$ (and write $\mathbf{X}^{\mu}\rightarrow\mathbf{f}$)
if $\delta\mu\rightarrow0$ and $\sigma\left(\mathbf{f},\mathbf{X}^{\mu}\right)\rightarrow0$.
We can then rewrite the definition above as 
\begin{equation}
D\mathbf{f}=\liminf_{\mathbf{X}^{\mu}\rightarrow\mathbf{f}}D\mathbf{X}.
\end{equation}
Using the properties of $\liminf$, for any path $\mathbf{f}$, there
exists a sequence $\left\{ \mathbf{X}_{n}^{\mu_{n}}\right\} $ of
chains-on-nets such that $\delta\mu_{n}\rightarrow0$ and $\sigma\left(\mathbf{f},\mathbf{X}_{n}^{\mu_{n}}\right)\rightarrow0$,
and $D\mathbf{X}_{n}\rightarrow D\mathbf{f}$.

Let us list some of the most basic properties of the $D$-length of
a path. 
\begin{thm}
The length $D\mathbf{f}$ of any path $\mathbf{f}|\left[a,b\right]$
has the following properties:

$\mathcal{L}1$ (nonnegativity) $D\mathbf{f}\geq0$;

$\mathcal{L}2$ (zero property) $D\mathbf{f}=0$ if and only if $\mathbf{f}\left(\left[a,b\right]\right)$
is a single point;

$\mathcal{L}3$ (additivity) for any $c\in\left[a,b\right]$, $D\mathbf{f}|\left[a,b\right]=D\mathbf{f}|\left[a,c\right]+D\mathbf{f}|\left[c,b\right]$. 
\end{thm}

Proofs of these statements are simple. Thus, to show the additivity
of $D\mathbf{f}$, add the point $c$ twice to all nets, 
\[
\widetilde{\mu}=\left\{ \overset{\text{\ensuremath{\alpha}}}{\overbrace{a=x_{0}\leq\ldots x_{i}\leq c}}=\underset{\beta}{\underbrace{c\leq x_{i+1}\leq\ldots\leq x_{k+1}=b}}\right\} ,
\]
and two corresponding point $\mathbf{c^{1}},\mathbf{c}^{2}$ to all
chains, 
\[
\widetilde{\mathbf{X}}=\overset{\mathbf{Y}}{\overbrace{\mathbf{x}_{0}\ldots\mathbf{x}_{i}\mathbf{c}^{1}}}\underset{\mathbf{Z}}{\underbrace{\mathbf{c}^{2}\mathbf{x}_{i+1}\ldots\mathbf{x}_{k+1}}}.
\]
Clearly, 
\[
\liminf_{\mathbf{\widetilde{X}}^{\widetilde{\mu}}\rightarrow\mathbf{f}|\left[a,b\right]}\widetilde{\mathbf{X}}=\liminf_{\mathbf{Y}^{\alpha}\rightarrow\mathbf{f}|\left[a,c\right]}\mathbf{Y}+\liminf_{\mathbf{Z}^{\beta}\rightarrow\mathbf{f}|\left[c,b\right]}\mathbf{Z}=D\mathbf{f}|\left[a,c\right]+D\mathbf{f}|\left[c,b\right].
\]
For any sequence $\left\{ \mathbf{X}_{n}^{\mu_{n}}\right\} $ of chains-on-nets
such that $\mathbf{X}_{n}^{\mu_{n}}\rightarrow\mathbf{\mathbf{f}}|\left[a,c\right]$,
and $D\mathbf{X}_{n}\rightarrow D\mathbf{f}$, we have $\mathbf{\widetilde{X}}_{n}^{\widetilde{\mu}_{n}}\rightarrow\mathbf{\mathbf{f}}|\left[a,c\right]$
for the corresponding sequence $\left\{ \mathbf{\widetilde{X}}_{n}^{\widetilde{\mu}_{n}}\right\} $,
assuming $\mathbf{c}_{n}^{1}\rightarrow\mathbf{f}\left(c\right)$
and $\mathbf{c}_{n}^{2}\rightarrow\mathbf{f}\left(c\right)$. We also
have 
\[
D\mathbf{\widetilde{X}}_{n}=D\mathbf{X}_{n}+\left(D\mathbf{x}_{i_{n}}\mathbf{c}_{n}^{1}+D\mathbf{c}_{n}^{1}\mathbf{c}_{n}^{2}+D\mathbf{c}_{n}^{2}\mathbf{x}_{i_{n}+1}-D\mathbf{x}_{i_{n}}\mathbf{x}_{i_{n}+1}\right),
\]
where each summand in the parentheses tends to zero by the uniform
continuity of $\mathbf{f}$ and $D$.

Note that $D\mathbf{f}$ is well-defined for any path $\mathbf{f}$,
but only on the extended set of nonnegative reals: the value of $D\mathbf{f}$
may very well be equal to $\infty$. This does not invalidate or complicate
any of the results presented in this chapter, but, for brevity sake,
we will tacitly assume that $D\mathbf{f}$ is finite.

\begin{figure}[ptbh]
\begin{centering}
\includegraphics[scale=0.25]{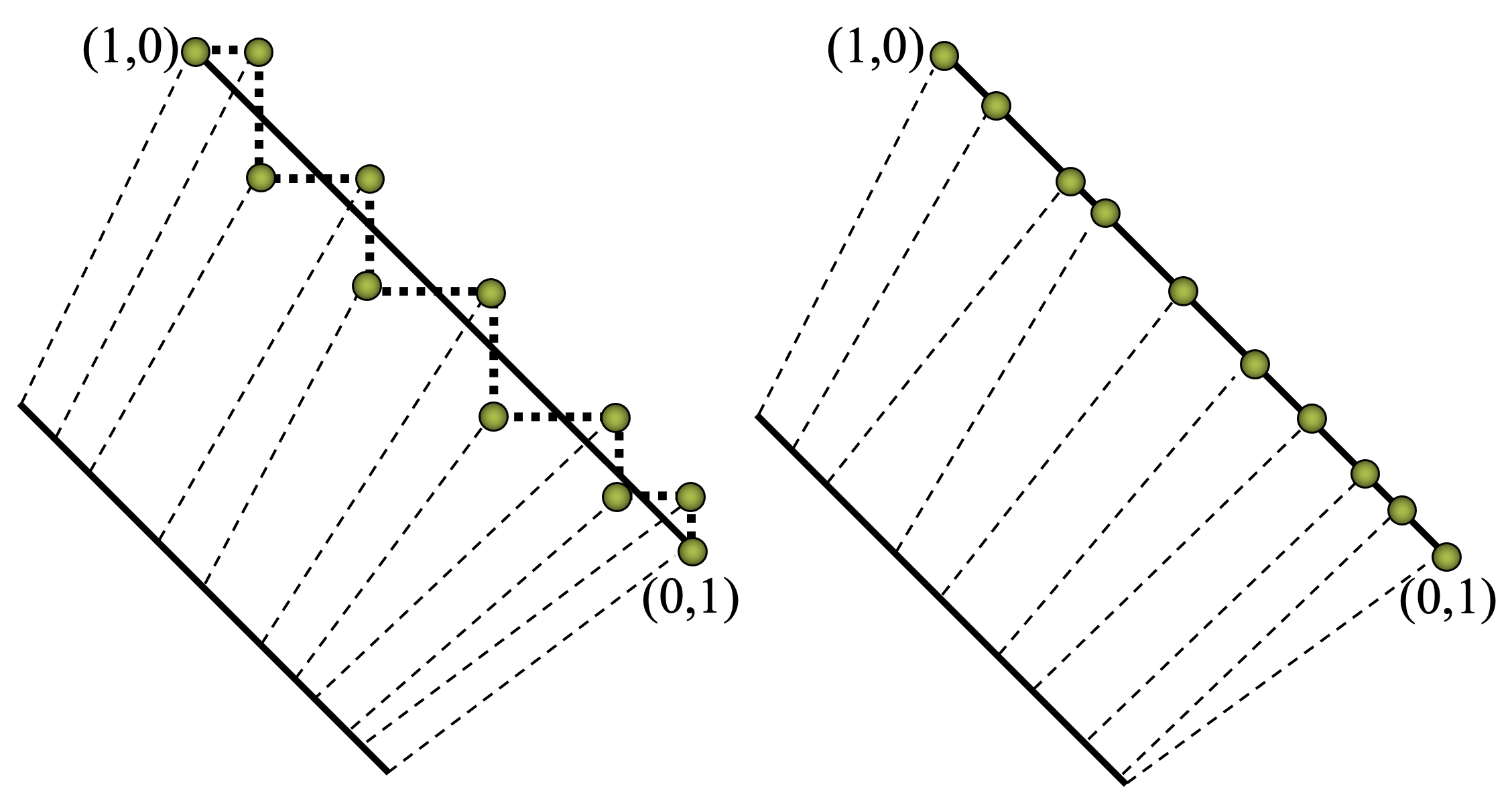} 
\par\end{centering}
\caption{\label{fig:hypothenuse}A demonstration of why for $D$-length computations
we need the ``meandering'' chains like in Figure \ref{fig:Convergence}
rather than just inscribed chains. Here, $D\mathbf{ab}$ for $\mathbf{a}=\left(a_{1},a_{2}\right)$
and $\mathbf{b}=\left(b_{1},b_{2}\right)$ is defined as $\left\vert a_{1}-b_{1}\right\vert +\left\vert a_{2}-b_{2}\right\vert +\min\left(\left\vert a_{1}-b_{1}\right\vert ,\left\vert a_{2}-b_{2}\right\vert \right)$.
All staircase chains $\mathbf{X}$, irrespective of the spacing of
their elements, have the cumulated dissimilarity $D\mathbf{X}=2$,
and $2$ is the true $D$-length of the path between $\left(1,0\right)$
and (0,1). All inscribed chains, irrespective of the spacing of their
elements, have the cumulated dissimilarity $3$. Explanations are
given in the text.}

\centering{}\label{FigExample} 
\end{figure}

The reader may wonder why, in the definition of $D\mathbf{f}$, it
is not sufficient to deal with the inscribed chains-on-nets, with
all elements of the chains belonging to the path $\mathbf{f}$. We
will see later that this is indeed sufficient if $D$ is a quasimetric
dissimilarity. However, in general, the inscribed chains-on-nets do
not reach the infimum of the $D$-lengths of the ``meandering''
chains-on-nets. Figure \ref{fig:hypothenuse} provides an illustration.
In this example, the stimuli are points in $\mathbb{R}^{2}$, and,
for $\mathbf{a}=\left(a_{1},a_{2}\right)$ and $\mathbf{b}=\left(b_{1},b_{2}\right)$,
\[
D\mathbf{ab}=\left\vert a_{1}-b_{1}\right\vert +\left\vert a_{2}-b_{2}\right\vert +\min\left(\left\vert a_{1}-b_{1}\right\vert ,\left\vert a_{2}-b_{2}\right\vert \right).
\]
It is easy to check that $D$ is a dissimilarity function\index{dissimilarity function}.
Thus, $\mathcal{D}3$ follows from the fact 
\[
D\mathbf{a}_{n}\mathbf{b}_{n}\rightarrow0\Longleftrightarrow\left|\mathbf{a}-\mathbf{b}\right|\rightarrow0,
\]
where $\left|\mathbf{a}-\mathbf{b}\right|$ is the usual Euclidean
norm. Also, for any chain $\mathbf{aXb},$ 
\[
D\mathbf{aXb}\geq\left\vert a_{1}-b_{1}\right\vert +\left\vert a_{2}-b_{2}\right\vert ,
\]
whence $D\mathbf{a}_{n}\mathbf{X}_{n}\mathbf{b}_{n}\rightarrow0$
implies $D\mathbf{a}_{n}\mathbf{b}_{n}\rightarrow0$. That is, $D$
satisfies $\mathcal{D}4$. By the same inequality, the length of the
line segment $\mathbf{f}$ shown in Figure \ref{fig:hypothenuse},
connecting $\mathbf{a}=\left(1,0\right)$ to $\mathbf{b}=\left(0,1\right)$,
cannot be less than 2. (The domain interval for $\mathbf{f}$ can
be chosen arbitrarily, e.g., $\left[0,1\right]$) Consider now chains-on-nets
$\mathbf{X}^{\mu}$ with the staircase chains, as in the left panel.
By decreasing the mesh of $\mu$ and the spacing of the elements of
$\mathbf{X}$, it can be made to converge to $\mathbf{f}$, and since
$D\mathbf{X}$ for all these chains equals 2, $D\mathbf{f}=2$. At
the same time, the inscribed chains, as in the right panel of the
figure, are easily checked to have the length 3.

\subsection{Path length through quasimetric dissimilarity}

Different dissimilarity functions $D$ lead to different quantifications
of path length. We know that the quasimetric dissimilarity $G$ defined
by (\ref{eq:Gdefined}) is a dissimilarity function\index{dissimilarity function}.
However, in this case, since $G$ is defined through $D$ by (\ref{eq:Gdefined}),
one should expect, for consistency, that the the path-length will
remain unchanged on replacing $D$ and with $G$. This will indeed
be established in Section \ref{sec:Df=00003D00003DGf}. We need several
preliminary results first, however.

Using $G$ in place of $D$ to define the $G$-length of paths, we
have 
\[
G\mathbf{f}=\liminf_{\mathbf{X}^{\mu}\overset{G}{\rightarrow}\mathbf{f}}G\mathbf{X}.
\]
The condition $\mathbf{X}^{\mu}\overset{G}{\rightarrow}\mathbf{f}$
here means $\delta\mu\rightarrow0$ and 
\[
\sigma_{G}\left(\mathbf{f},\mathbf{X}^{\mu}\right)=\max_{x_{i}\in\mu}G\mathbf{f}\left(x_{i}\right)\mathbf{x}_{i}\rightarrow0.
\]
But by Theorem \ref{thm:GandDinsmall}, the latter condition is equivalent
to 
\[
\sigma\left(\mathbf{f},\mathbf{X}^{\mu}\right)=\max_{x_{i}\in\mu}D\mathbf{f}\left(x_{i}\right)\mathbf{x}_{i}\rightarrow0.
\]
Therefore $\mathbf{X}^{\mu}\overset{G}{\rightarrow}\mathbf{f}$ and
$\mathbf{X}^{\mu}\rightarrow\mathbf{f}$ are equivalent, and we can
formulate 
\begin{defn}
\label{def:G-length}The $G$\emph{-length\index{path!G-length of}}
of path $\mathbf{f}|\left[a,b\right]$ is 
\[
G\mathbf{f}=\liminf_{\mathbf{X}^{\mu}\rightarrow\mathbf{f}}G\mathbf{X}.
\]
\end{defn}

Consider now chains-on-net{\small{}{}s} \emph{$\mathbf{Z}^{\nu}$
inscribed} in $\mathbf{f}|\left[a,b\right]$, that is, those with
\[
\nu=\left\{ a=z_{0},z_{1},\ldots,z_{k},z_{k+1}=b\right\} 
\]
and 
\[
\mathbf{Z}=\mathbf{f}\left(z_{0}\right)\ldots\mathbf{f}\left(z_{k+1}\right)=\mathbf{z}_{0}\ldots\mathbf{z}_{k+1}.
\]
Since $\sigma\left(\mathbf{f},\mathbf{Z}^{\nu}\right)=0$, the condition
$\mathbf{Z}^{\nu}\rightarrow\mathbf{f}$ here reduces to $\delta\nu\rightarrow0$.
Clearly, 
\begin{equation}
\liminf_{\delta\nu\rightarrow0}G\mathbf{Z}\geq\liminf_{\mathbf{X}^{\mu}\rightarrow\mathbf{f}}G\mathbf{X}=G\mathbf{f},\label{eq:GZ<Gf}
\end{equation}
because inscribed chains-on-net{\small{}{}s converging to $\mathbf{f}$
form a subset of all chains-on-nets converging to $\mathbf{f}$. We
will see now that in fact the two quantities in (\ref{eq:GZ<Gf})
are equal.} By the additivity property, 
\[
G\mathbf{f}|\left[a,b\right]=\sum_{i=0}^{k}G\mathbf{f}|\left[z_{i},z_{i+1}\right].
\]
Let $\mathbf{X}_{i}^{\mu_{i}}$ be an arbitrary chain-in-net with
$\mu_{i}\subset\left[z_{i},z_{i+1}\right]$. By the same reasoning
as in the proof of the additivity property, if $\mu_{i}$ is changed
into 
\[
\widetilde{\mu}_{i}=\left\{ z_{i},\overset{\mu_{i}}{\overbrace{\ldots}},z_{i+1}\right\} ,
\]
and $\mathbf{X}_{i}$ into 
\[
\widetilde{\mathbf{X}}_{i}=\mathbf{z}_{i}\mathbf{X}_{i}\mathbf{z}_{i+1},
\]
the conditions $\mathbf{X}_{i}^{\mu_{i}}\rightarrow\mathbf{f}|\left[z_{i},z_{i+1}\right]$
and $\mathbf{\widetilde{X}}_{i}^{\widetilde{\mu}_{i}}\rightarrow\mathbf{f}|\left[z_{i},z_{i+1}\right]$
are equivalent. Denoting by $\mathbf{X}^{\mu}$ the concatenation
of $\mathbf{X}_{i}^{\mu_{i}}$ for $i=0,\ldots,k$, and defining $\mathbf{\widetilde{X}}^{\widetilde{\mu}}$
analogously, we have 
\[
G\mathbf{f}=\liminf_{\mathbf{X}^{\mu}\rightarrow\mathbf{f}}G\mathbf{X}=\liminf_{\mathbf{\widetilde{X}}^{\widetilde{\mu}}\rightarrow\mathbf{f}}G\mathbf{\widetilde{X}}.
\]
At the same time, by the triangle inequality\index{inequality!triangle},
\[
G\mathbf{z}_{i}\mathbf{z}_{i+1}\leq G\mathbf{z}_{i}\mathbf{X}_{i}\mathbf{z}_{i+1},
\]
whence 
\[
G\mathbf{Z}\leq G\mathbf{\widetilde{X}}
\]
and 
\begin{equation}
\liminf_{\delta\nu\rightarrow0}G\mathbf{Z}\geq\liminf_{\mathbf{\widetilde{X}}^{\widetilde{\mu}}\rightarrow\mathbf{f}}G\mathbf{\widetilde{X}}=G\mathbf{f}.\label{eq:GZ>Gf}
\end{equation}
Together with (\ref{eq:GZ<Gf}), this establishes 
\begin{thm}
\label{thm:Gf=00003D00003DliminfGZ}For any path $\mathbf{f}$, 
\[
G\mathbf{f}=\liminf_{\delta\nu\rightarrow0}G\mathbf{Z},
\]
where $\mathbf{Z}^{\nu}$ are chains-on-nets inscribed in $\mathbf{f}$. 
\end{thm}

In other words, to approximate $G\mathbf{f}$ by $G$-lengths of chains-on-nets,
one does not need all possible chains converging to $\mathbf{f}$,
the inscribed ones only are sufficient. Recall that the analogous
statement is not correct for $D\mathbf{f}$. The equality in Theorem
\ref{thm:Gf=00003D00003DliminfGZ} critically owes to the fact that
$G$ satisfies the triangle inequality.

We can further clarify Theorem \ref{thm:Gf=00003D00003DliminfGZ}
as follows. 
\begin{thm}
For any path $\mathbf{f}$,

\begin{equation}
G\mathbf{f}=\sup G\mathbf{Z}=\lim_{\delta\nu\rightarrow0}G\mathbf{Z},
\end{equation}
where $\mathbf{Z}^{\nu}$ are chains-on-nets inscribed in $\mathbf{f}$. 
\end{thm}

In other words, $G\mathbf{f}$ is the lowest upper bound for the lengths
of all inscribed chains-on-nets; and any sequence of the inscribed
chains-on-nets converges to $G\mathbf{f}$ as their mesh decreases.

To prove the first equality, $G\mathbf{f}=\sup G\mathbf{Z}$, consider
a chain-on-net $\mathbf{Z}^{\nu}$ with $\sup G\mathbf{Z}-G\mathbf{Z}$
arbitrarily small. For every pair of successive $z_{i},z_{i+1}$ in
$\nu$, one can find an inscribed chain-on-net $\mathbf{V}_{i}^{\mu_{i}}$
such that $\mu_{i}=\left\{ z_{i_{n}},\ldots,z_{i_{n}+1}\right\} $
and $\left|G\mathbf{V}_{i}-G\mathbf{f}|\left[z_{i},z_{i+1}\right]\right|$
is arbitrarily small. By the additivity of $G$-length, denoting by
$\mathbf{V}^{\mu}$ the concatenation of all $\mathbf{V}_{i}^{\mu_{i}}$,
we can make $\left|G\mathbf{V}-G\mathbf{f}|\left[a,b\right]\right|$
arbitrarily small. From the triangle inequality\index{inequality!triangle}
it follows that $G\mathbf{V}\geq G\mathbf{Z}$, whence $G\mathbf{f}\geq\sup G\mathbf{Z}$.
But $G\mathbf{V}\leq\sup G\mathbf{Z}$, whence we also have $G\mathbf{f}\leq\sup G\mathbf{Z}$.

To prove that $G\mathbf{f}=\lim_{\delta\nu\rightarrow0}G\mathbf{Z}$,
deny it, and assume that there is a sequence of inscribed chains-on-nets
$\mathbf{V}_{n}^{\mu_{n}}$ such that $\delta\mu_{n}\rightarrow0$
but $G\mathbf{V}_{n}\not\rightarrow D\mathbf{f}$. Since $D\mathbf{f=}\sup D\mathbf{Z}$
across all possible inscribed chains-on-nets, $D\mathbf{V}_{n}\leq D\mathbf{f}$
for all $n$. Then one can find a $\Delta>0$ and a subsequence of
$\mathbf{V}_{n}^{\mu_{n}}$ (which, with no loss of generality, we
can assume to be $\mathbf{V}_{n}^{\mu_{n}}$ itself) such that 
\[
D\mathbf{V}_{n}\rightarrow D\mathbf{f}-\Delta.
\]
Let $\mathbf{Z}^{\nu}$ be an inscribed chain-on-net with 
\[
D\mathbf{Z}>D\mathbf{f}-\Delta/2.
\]
For every $z_{i}$ in $\nu$ and every $n$, let $v_{k_{i,n}}^{n},v_{k_{i,n}+1}^{n}$
be two successive elements of $\mu_{n}$ such that $v_{k_{i,n}}^{n}\leq z_{i}\leq v_{k_{i,n}+1}^{n}$.
For a sufficiently large $n$, $\delta\mu_{n}$ is sufficiently small
to ensure that $z_{i}$ is the only member of $\nu$ falling between
$v_{k_{i,n}}^{n}$ and $v_{k_{i,n}+1}^{n}$ (without loss of generality,
we can assume that $\nu$ contains no identical elements). Denote
by $\nu\uplus\mu_{n}$ the nets formed by the elements of $\nu$ inserted
into $\mu_{n}$. Consider the inscribed chains-on-nets $\mathbf{U}^{\nu\uplus\mu_{n}}$.
We have (denoting by $l$ the cardinality of $\nu$), 
\begin{multline*}
G\mathbf{U}=G\mathbf{V}_{n}\\
+\sum_{i=0}^{l}\left\{ G\mathbf{f}\left(v_{k_{i,n}}^{n}\right)\mathbf{f}\left(z_{i}\right)+G\mathbf{f}\left(z_{i}\right)\mathbf{f}\left(v_{k_{i,n}+1}^{n}\right)-G\mathbf{f}\left(v_{k_{i,n}}^{n}\right)\mathbf{f}\left(v_{k_{i,n}+1}^{n}\right)\right\} .
\end{multline*}
By the uniform continuity of $\mathbf{f}$, the expression under the
summation operator tends to zero, whence 
\[
G\mathbf{U}-G\mathbf{V}_{n}\rightarrow0,
\]
and then 
\[
G\mathbf{U}\rightarrow G\mathbf{f}-\Delta.
\]
But by the triangle inequality\index{inequality!triangle}, for all
$n$, 
\[
D\mathbf{U}\geq D\mathbf{Z}>D\mathbf{f}-\Delta/2.
\]
This contradiction completes the proof.

\subsection{\label{sec:Df=00003D00003DGf}The equality of the $D$-length and
$G$-length of paths}

As mentioned previously, one can expect that path length should not
depend on whether one chooses dissimilarity $D$ or the quasimetric
dissimilarity $G$ induced by $D$. 
\begin{thm}
\label{thm:Gf=00003D00003DDf}For any path $\mathbf{f}$\textbf{,}
\[
D\mathbf{f}=G\mathbf{f.}
\]
\end{thm}

Comparing Definitions \ref{def:D-length} and \ref{def:G-length},
since $D\mathbf{X}\geq G\mathbf{X}$ for any chain, we have $D\mathbf{f}\geq G\mathbf{f}$.
To see that $D\mathbf{f}\leq G\mathbf{f}$, we form a sequence of
inscribed chains-on-nets $\mathbf{Z}_{n}^{\nu_{n}}$ such that $\delta\nu_{n}\rightarrow0$,
and $G\mathbf{Z}_{n}\rightarrow G\mathbf{f}$. By the definition of
$G$, one can insert chains $\mathbf{X}_{i}^{n}$ between pairs of
successive elements $\mathbf{z}_{i}^{n},\mathbf{z}_{i+1}^{n}$ of
$\mathbf{Z}_{n}$, so that 
\[
D\mathbf{U}_{n}-G\mathbf{Z}_{n}\leq\frac{1}{n},
\]
where 
\[
\mathbf{U}_{n}=\mathbf{z}_{0}^{n}\mathbf{X}_{0}^{n}\mathbf{z}_{1}^{n}\ldots\mathbf{z}_{k_{n}}^{n}\mathbf{X}_{k_{n}}^{n}\mathbf{z}_{k_{n}+1}^{n}.
\]
In other words, $D\mathbf{U}_{n}\rightarrow G\mathbf{f}$. Let us
now create a net $\mu_{n}$ for every $\mathbf{U}_{n}$ as follows:
if $z_{i}^{n}\in\nu_{n}$ is associated with $\mathbf{z}_{i}^{n}\in\mathbf{Z}_{n}$,
we associate $z_{i}^{n}$ with every element of $\mathbf{X}_{i}^{n}$.
The resulting chain-on-net is 
\[
\mathbf{U}_{n}^{\mu_{n}}=\left(\ldots,\left(z_{i}^{n},\mathbf{z}_{i}^{n}\right),\left(z_{i}^{n},\mathbf{x}_{1}^{i,n}\right),\ldots,\left(z_{i}^{n},\mathbf{x}_{l_{i,n}}^{i,n}\right)\left(z_{i+1}^{n},\mathbf{z}_{i+1}^{n}\right),\ldots\right).
\]
We will show now that $\mathbf{U}_{n}^{\mu_{n}}\rightarrow\mathbf{f}$.
Since $\delta\mu_{n}=\delta\nu_{n}\rightarrow0$, we have to show
that $\sigma\left(\mathbf{f},\mathbf{U}_{n}^{\mu_{n}}\right)\rightarrow0$.
Let $\left(z_{i_{n}}^{n},\mathbf{m}_{i_{n}}^{n}\right)$ be an element
of $\mathbf{U}_{n}^{\mu_{n}}$ such that 
\[
\sigma\left(\mathbf{f},\mathbf{U}_{n}^{\mu_{n}}\right)=D\mathbf{f}\left(z_{i_{n}}^{n}\right)\mathbf{m}_{i_{n}}^{n}=D\mathbf{z}_{i_{n}}^{n}\mathbf{m}_{i_{n}}^{n}.
\]
By the uniform continuity of $\mathbf{f}$ and $G$, 
\[
G\mathbf{z}_{i_{n}}^{n}\mathbf{z}_{i_{n}+1}^{n}=G\mathbf{f}\left(z_{i_{n}}^{n}\right)\mathbf{f}\left(z_{i_{n}+1}^{n}\right)\rightarrow0
\]
as $\delta\mu_{n}=\delta\nu_{n}\rightarrow0$. By the construction
of $\mathbf{U}_{n}$, 
\[
D\mathbf{z}_{i_{n}}^{n}=D\mathbf{z}_{i_{n}}^{n}\overset{\mathbf{X}_{i_{n}}^{n}}{\overbrace{\mathbf{x}_{1}^{i_{n},n}\ldots\mathbf{m}_{i_{n}}^{n}\ldots\mathbf{x}_{l_{i_{n},n}}^{i_{n},n}}}\mathbf{z}_{i_{n}+1}^{n}\rightarrow0,
\]
implying 
\[
D\mathbf{z}_{i_{n}}^{n}\mathbf{x}_{1}^{i_{n},n}\ldots\mathbf{m}_{i_{n}}^{n}\rightarrow0.
\]
By the chain property of dissimilarity functions, 
\[
\sigma\left(\mathbf{f},\mathbf{U}_{n}^{\mu_{n}}\right)=D\mathbf{z}_{i_{n}}^{n}\mathbf{m}_{i_{n}}^{n}\rightarrow0.
\]
We have therefore a sequence of chains-on-nets $\mathbf{U}_{n}^{\mu_{n}}\rightarrow\mathbf{f}$
with $G\mathbf{f}$ as the limit point of $D\mathbf{U}_{n}$, and
then $G\mathbf{f}\geq D\mathbf{f}$ because $D\mathbf{f}$ is the
infimum of all such limit points. This completes the proof.

We see that although $D\mathbf{xy}$ and $G\mathbf{xy}$ are generally
distinct for points $\mathbf{x},\mathbf{y}$, when it comes to paths
$\mathbf{f}$, the quantities $D\mathbf{f}$ and $G\mathbf{f}$ can
be used interchangeably. One consequence of this result is that the
properties of the $D$-length of paths can now be established by replacing
it with the $G$-length, the advantage of this being that we acquire
the powerful triangle inequality\index{inequality!triangle} to use,
and also restrict chains-on-nets to the inscribed ones, more familiar
than the ``meandering'' chains in Figure \ref{fig:Convergence}.
However, the general definition of $D\mathbf{f}$ remains convenient
in many situations. We illustrate this on the important property of
\emph{lower semicontinuity} of the $D$-length. 
\begin{defn}
A sequence of paths $\mathbf{f}_{n}|\left[a,b\right]$ \emph{converges}
\index{convergence!to a path} to a path $\mathbf{f}|\left[a,b\right]$
(in symbols, $\mathbf{f}_{n}\rightarrow\mathbf{f}$) if 
\[
\sigma\left(\mathbf{f},\mathbf{f}_{n}\right)=\max_{x\in\left[a,b\right]}D\mathbf{f}\left(x\right)\mathbf{f}_{n}\left(x\right)\rightarrow0.
\]
\end{defn}

Consider any sequence of chains-on-nets $\mathbf{X}_{n}^{\mu_{n}}\rightarrow\mathbf{f}_{n}$
such that $\left\vert D\mathbf{X}_{n}-D\mathbf{f}_{n}\right\vert \rightarrow0$.
By the uniform continuity of $D$, 
\[
\left[\sigma\left(\mathbf{f}_{n},\mathbf{X}_{n}^{\mu_{n}}\right)\rightarrow0\right]\textnormal{ and }\left[\sigma\left(\mathbf{f},\mathbf{f}_{n}\right)\rightarrow0\right]\Longrightarrow\sigma\left(\mathbf{f},\mathbf{X}_{n}^{\mu_{n}}\right)\rightarrow0.
\]
Then $\mathbf{X}_{n}^{\mu_{n}}\rightarrow\mathbf{f}$, whence $\liminf_{n\rightarrow\infty}D\mathbf{X}_{n}\geq D\mathbf{f}$.
But $\liminf_{n\rightarrow\infty}D\mathbf{X}_{n}=\liminf_{n\rightarrow\infty}D\mathbf{f}_{n}$.
This proves 
\begin{thm}[Lower semicontinuity]
\index{semicontinuity!lower} \label{thm:LowerSemicontinuity}For
any sequence of paths $\mathbf{f}_{n}|\left[a,b\right]\rightarrow\mathbf{f}|\left[a,b\right]$,
\[
\liminf_{n\rightarrow\infty}D\mathbf{f}_{n}\geq D\mathbf{f.}
\]
\end{thm}

\subsection{Intrinsic metrics and spaces with intermediate points}

In a path-connected space, a metric\index{metric!intrinsic} is traditionally
called \emph{intrinsic} if the distance between two points is the
greatest lower bound for the length of all paths connecting the two
points. For instance, in $\mathbb{R}^{n}$ endowed with the Euclidean
geometry, the Euclidean distance 
\[
D\mathbf{ab}=\left|\mathbf{a}-\mathbf{b}\right|
\]
between points $\mathbf{a}$ and $\mathbf{b}$ is intrinsic, because
it is also the length of the shortest path connecting these points,
a straight line segment. By contrast, 
\[
D\mathbf{ab}=\sqrt{\left|\mathbf{a}-\mathbf{b}\right|}
\]
is also a metric, but it is not intrinsic: the path length $D\mathbf{f}$
induced by this metric is infinitely large for every path $\mathbf{f}$.
As an example of a non-intrinsic metric with a finite path length
function, consider 
\[
D\mathbf{ab}=\tan\left|a-b\right|
\]
on the interval $\left[0,\frac{\pi}{2}\right[$, where $a,b$ are
the values of $\mathbf{a},\mathbf{b}$, respectively. The length of
the (only) path connecting $\mathbf{a}$ to $\mathbf{b}$ here is
$\left|a-b\right|\not=\tan\left|a-b\right|$.

In this section we consider a generalization of the notion of intrinsic
metric to quasimetric dissimilarities. 
\begin{defn}
\label{def:intrinsic}The quasimetric dissimilarity $G$ defined in
a space $\left(\mathfrak{S},D\right)$ by (\ref{eq:Gdefined}) is
called intrinsic if, for any $\mathbf{a},\mathbf{b}\in$$\mathfrak{S}$,
\[
G\mathbf{ab}=\inf_{\mathbf{f}\in\mathcal{P}_{\mathbf{a}}^{\mathbf{b}}}D\mathbf{f},
\]
where $\mathcal{P}_{\mathbf{a}}^{\mathbf{b}}$ is the class of all
\emph{paths} connecting $\mathbf{a}$ to $\mathbf{b}$. 
\end{defn}

\begin{figure}[ptbh]
\begin{centering}
\includegraphics[scale=0.35]{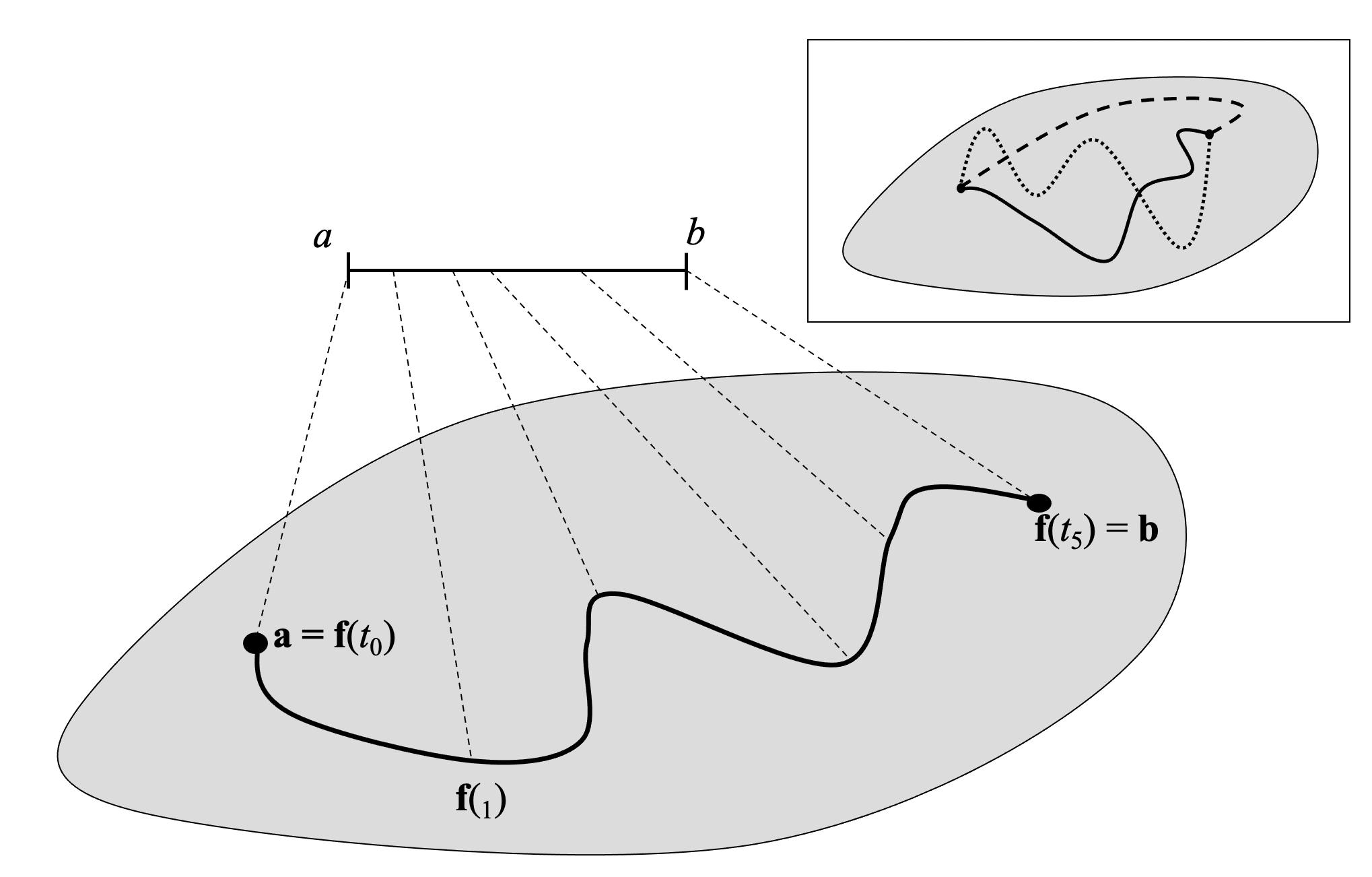} 
\par\end{centering}
\caption{\label{FigPathMetric}The metric $G$ induced by dissimilarity $D$
is intrinsic if the $G$-distance from $\mathbf{a}$ to $\mathbf{b}$
equal the infimum of $D$-lengths (equivalently, $G$-lengths) of
all paths connecting $\mathbf{a}$ to $\mathbf{b}$. }
\end{figure}

Figure \ref{FigPathMetric} provides an illustration.

We know that in Definition \ref{def:intrinsic} $D\mathbf{f}$ can
be replaced with $G\mathbf{f}$. We also know that $G\mathbf{f}$
for any $\mathbf{f}\in\mathcal{P}_{\mathbf{a}}^{\mathbf{b}}$ can
be arbitrarily closely approximated by $G\mathbf{aXb}$ for some inscribed
chain-on-net $\mathbf{X}^{\mu}$. By the triangle inequality\index{inequality!triangle},
$G\mathbf{ab}\leq G\mathbf{aXb}$. Therefore, in any space $\left(\mathfrak{S},D\right)$,
\begin{equation}
G\mathbf{ab}\leq\inf_{\mathbf{f}\in\mathcal{P}_{\mathbf{a}}^{\mathbf{b}}}D\mathbf{f}.\label{eq:G>inf}
\end{equation}
We need now to consider a special class of spaces in which this inequality
can be reversed.

\begin{figure}[ptbh]
\begin{centering}
\includegraphics[scale=0.5]{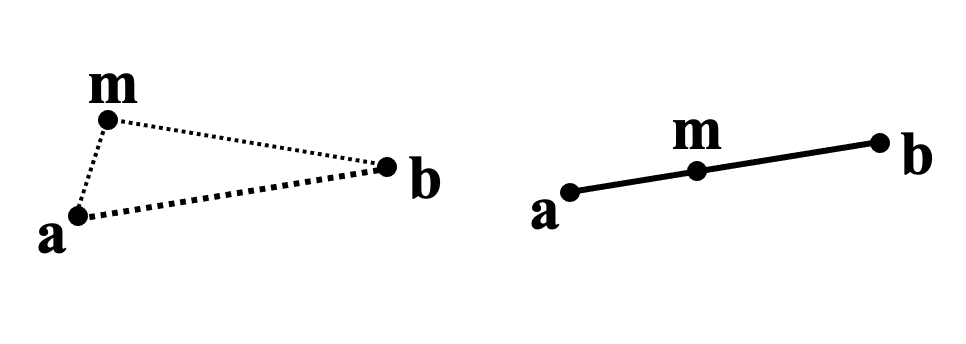} 
\par\end{centering}
\caption{\label{FigTriangle}If $D\mathbf{amb}\protect\leq D\mathbf{ab}$,
the point $\mathbf{m}$ is said to be intermediate to $\mathbf{a}$
and $\mathbf{b}$. As a special case, if $D$ is Euclidean distance
(right picture), any $\mathbf{m}$ on the straight line segment connecting
$\mathbf{a}$ and $\mathbf{b}$ is intermediate to $\mathbf{a}$ and
$\mathbf{b.}$}
\end{figure}

\begin{defn}
\label{def:Intermediate}A stimulus space\index{stimulus space!with intermediate points}
$\left(\mathfrak{S},D\right)$ is said to be a space \emph{with intermediate
points} if, for any distinct $\mathbf{a},\mathbf{b}$, one can find
an $\mathbf{m}$ such that $\mathbf{m\notin}\left\{ \mathbf{a,b}\right\} $
and $D\mathbf{amb}\leq D\mathbf{ab}$. 
\end{defn}

Fig. \ref{FigTriangle} provides an illustration. If $D$ is a metric\index{metric}
(or quasimetric dissimilarity), the inequality $D\mathbf{amb}\leq D\mathbf{ab}$
can only have the form 
\[
D\mathbf{amb}=D\mathbf{ab}.
\]
In this form the notion is know as \emph{Menger convexity}\index{Menger convexity}.

A sequence $\mathbf{x}_{\mathbf{1}},\mathbf{x}_{2},\ldots$ in $\left(\mathfrak{S},D\right)$
is called a \emph{Cauchy sequence} if\index{Cauchy sequence} 
\[
\lim_{\substack{k\rightarrow\infty\\
l\rightarrow\infty
}
}D\mathbf{x}_{k}\mathbf{x}_{l}=0,
\]
that is, if for any $\varepsilon>0$ one can find an $n$ such that
$D\mathbf{x}_{k}\mathbf{x}_{l}<\varepsilon$ whenever $k,l>n$. 
\begin{defn}
A space $\left(\mathfrak{S},D\right)$ is called \emph{$D$-complete}
(or simply, complete) if every Cauchy sequence in it converges to
a point. \index{stimulus space!D-complete} 
\end{defn}

That is, in a complete space, for any Cauchy sequence $\mathbf{x}_{\mathbf{1}},\mathbf{x}_{2},\ldots$,
there is a point $\mathbf{x}\in\mathfrak{S}$ such that $\mathbf{x}_{n}\rightarrow\mathbf{x}$.
For example, if stimuli are represented by points in a closed region
of $\mathbb{R}^{n}$, and the convergence $\mathbf{x}_{n}\rightarrow\mathbf{x}$
coincides with the usual convergence of $n$-element vectors, then
the space is complete.

The main mathematical fact we are interested in is as follows.

\begin{figure}[ptbh]
\begin{centering}
\includegraphics[scale=0.4]{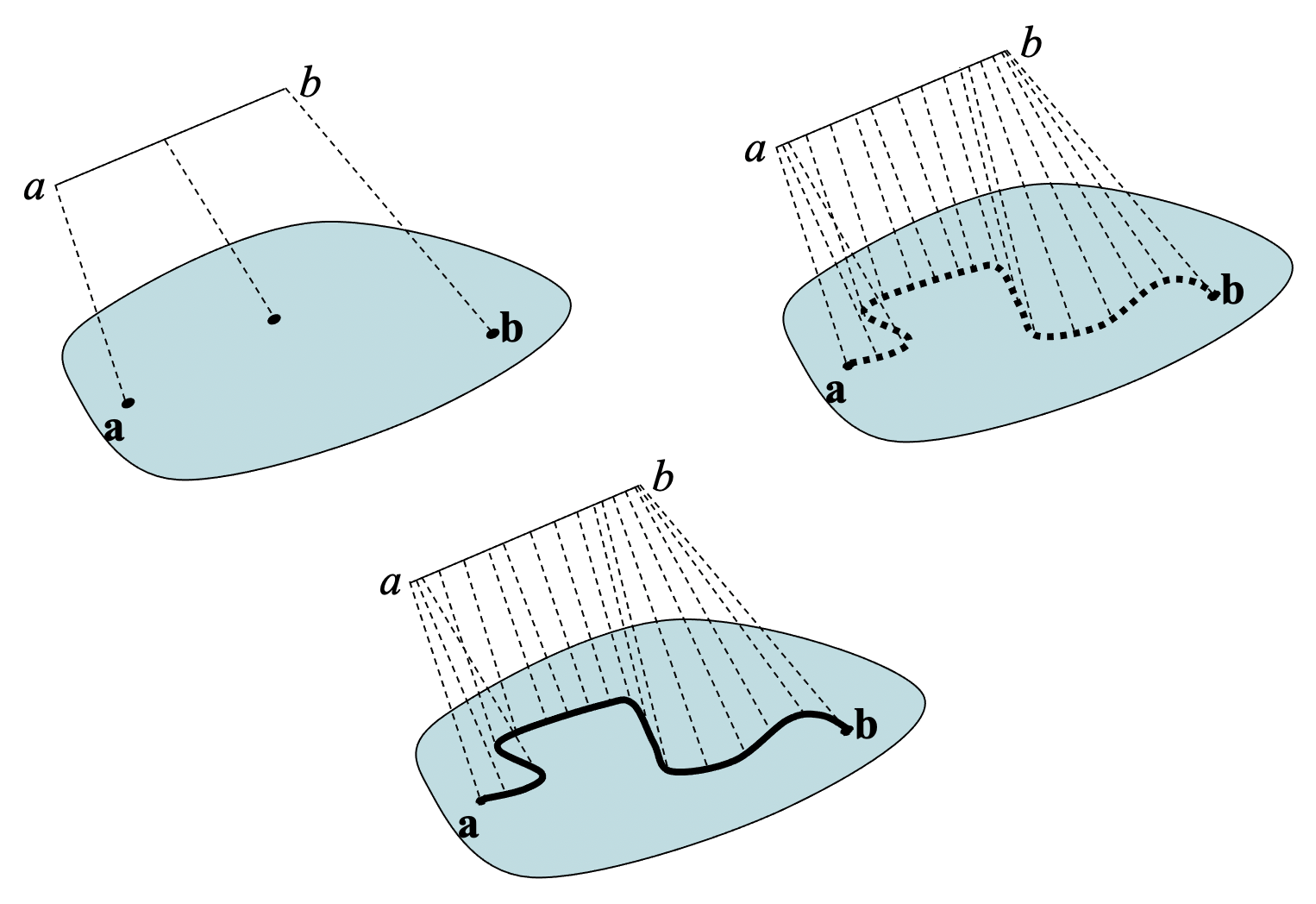} 
\par\end{centering}
\caption{\label{FigTheorem Intermediate}An informal illustration of Theorems
\ref{thm:Intermediate points} and \ref{thm:Gab is min of arcs}:
by adding intermediate points for every pair of successive points
one can create at the limit a path connecting $\mathbf{a}$ to $\mathbf{b}$,
with its $D$-length not exceeding $D\mathbf{ab}$. The infimum of
then D-lengths of all such paths equals $G\mathbf{a}\mathbf{b}$.}
\end{figure}

\begin{thm}
\label{thm:Intermediate points}In a complete space $\left(\mathfrak{S},D\right)$
with intermediate points, any point $\mathbf{a}$ can be connected
to any point $\mathbf{b}$ by a path $\mathbf{f}$ with 
\[
D\mathbf{f}\leq D\mathbf{ab}.
\]
\end{thm}

A proof of this statement known to us is rather involved (see Section
\ref{sec:Related-Literature} for a reference), and we will omit it
here. Figure \ref{FigTheorem Intermediate} provides an intuitive
illustration.

A consequence of this theorem that is of special importance for us
is as follows. In any sequence of chains-on-nets $\mathbf{X}_{n}$
connecting $\mathbf{a}$ to $\mathbf{b}$, with $D\mathbf{X}_{n}\rightarrow G\mathbf{ab}$,
each link $\mathbf{x}_{i_{n}}\mathbf{x}_{i_{n}+1}$ in each chain
$\mathbf{X}_{n}$ can be replaced with a path $\mathbf{f}_{i_{n}}$
connecting $\mathbf{x}_{i_{n}}$ to $\mathbf{x}_{i_{n}+1}$, such
that $D\mathbf{f}_{i_{n}}\leq D\mathbf{x}_{i_{n}}\mathbf{x}_{i_{n}+1}$.
This would create a path $\mathbf{f}_{n}$ connecting $\mathbf{a}$
to $\mathbf{b}$, with $D\mathbf{f}_{n}\leq D\mathbf{X}_{n}$. Hence
\begin{equation}
\inf_{\mathbf{f}\in\mathcal{P}_{\mathbf{a}}^{\mathbf{b}}}D\mathbf{f}\leq\liminf_{n\rightarrow\infty}D\mathbf{f}\leq\lim_{n\rightarrow\infty}D\mathbf{X}_{n}=G\mathbf{ab}.
\end{equation}
Combining this with (\ref{eq:G>inf}), we establish 
\begin{thm}
\label{thm:Gab is min of arcs}In a complete space $\left(\mathfrak{S},D\right)$
with intermediate points, the quasimetric dissimilarity $G$ is intrinsic:
\[
G\mathbf{ab}=\inf_{\mathbf{f}\in\mathcal{P}_{\mathbf{a}}^{\mathbf{b}}}D\mathbf{f}.
\]
\end{thm}

\section{\label{sec:Finsler}Dissimilarity Cumulation in Euclidean spaces }

\index{Euclidean space} 

\subsection{Introduction}

\index{dissimilarity cumulation!in Euclidean spaces} We are now prepared
to see how the general theory of path length can be specialized to
a variant of (Finsler) differential geometry\index{Finsler geometry}.
We assume that in the canonical space of stimuli $\left(\mathfrak{S},D\right)$,
the set $\mathfrak{S}$ is an an \emph{open connected} region of the
Euclidean $n$-space $\mathbb{R}^{n}$. The Euclidean $n$-space is
endowed with the global coordinate system, 
\[
\mathbf{x}=\left(x^{1},\ldots,x^{n}\right),
\]
and the conventional metric\index{metric!Euclidean} 
\begin{equation}
E\mathbf{ab}=\left\vert \mathbf{a-b}\right\vert .
\end{equation}
Recall that the connectedness of $\mathfrak{S}$ means that it cannot
be presented as a union of two open nonempty sets. In the Euclidean
space this notion is equivalent to \emph{path-connectedness}: any
two points can be connected by a path. \index{path connectedness}

\begin{figure}
\begin{centering}
\includegraphics[scale=0.4]{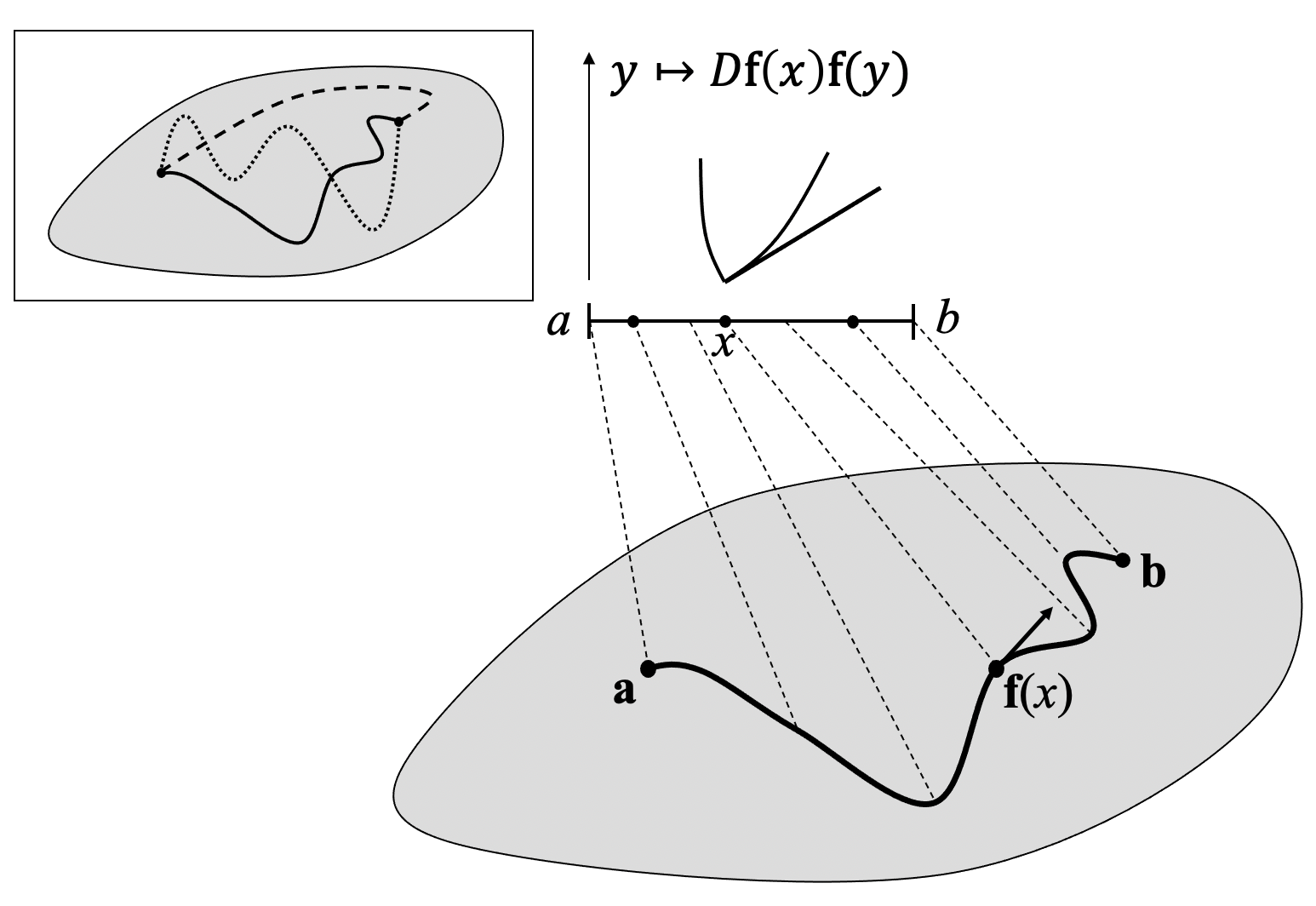} 
\par\end{centering}
\caption{As the point on the path moves away from a position $\mathbf{f}\left(x\right)$,
the dissimilarity $D\mathbf{f}\left(x\right)\mathbf{f}\left(y\right)$
increases from zero, and the rate of this increase, $\left.\mathrm{d}D\mathbf{f}\left(x\right)\mathbf{f}\left(x+s\right)/\mathrm{d}y\right|_{s=0}$
is shown by the slope of the tangent line in the graph of $y\protect\mapsto D\mathbf{f}\left(x\right)\mathbf{f}\left(y\right)$.
This derivative then is integrated with respect to $x$ from $a$
to $b$ to obtain the length of the path $\mathbf{f}$. If this derivative
only depends on $\mathbf{f}\left(x\right)$ and $\mathrm{d}\mathbf{f}\left(x\right)/\mathrm{d}x$
(assuming the path is continuously differentiable), then it can be
viewed as a way of measuring the tangent vector to the path as a point
moves along it, $F\left(\mathbf{f}\left(x\right),\mathrm{d}\mathbf{f}\left(x\right)/\mathrm{d}x\right)$.
The infimum of the lengths of all such smooth paths connecting $\mathbf{a}$
to $\mathbf{b}$ is then taken for the value of $G\mathbf{ab}$.}

\centering{}\label{fig:FigMFS} 
\end{figure}

Among all paths we focus on continuously differentiable ones. We develop
a way of measuring the value $F\left(\mathbf{f}\left(x\right),\dot{\mathbf{f}}\left(x\right)\right)$
of the tangent vector $\dot{\mathbf{f}}\left(x\right)$ to the path
$\mathbf{f}|\left[a,b\right]$ at point $x$, by showing (under certain
assumptions) that 
\[
\widehat{F}\left(\mathbf{f}\left(x\right),\dot{\mathbf{f}}\left(x\right)\right)=\lim_{s\rightarrow0+}\frac{G\mathbf{f}\left(x\right)\mathbf{f}\left(x+s\right)}{s}.
\]
The $D$-length of the path is then computed as 
\[
\intop_{a}^{b}\widehat{F}\left(\mathbf{f}\left(x\right),\dot{\mathbf{f}}\left(x\right)\right)\DD x.
\]
The idea is illustrated in Figure \ref{fig:FigMFS}.

We begin now a systematic development. 
\begin{defn}
\index{tangent!space} The \emph{tangent space} $\mathbb{T}_{\mathbf{p}}$
at a point $\mathbf{p}$ of $\mathfrak{S}$ is the set $\left\{ \mathbf{p}\right\} \times\mathbb{U}^{n}$,
where $\mathbb{U}^{n}$ is the \emph{vector space} 
\[
\left\{ \mathbf{u}=\mathbf{x}-\mathbf{p}:\mathbf{x}\in\mathbb{R}^{n},\mathbf{x}\neq\mathbf{p}\right\} 
\]
endowed with the Euclidean vector norm $\left\vert \mathbf{u}\right\vert $
and the standard topology. The $n$-vectors $\mathbf{u}\in\mathbb{U}^{n}$
are referred to as \emph{directions}, and the elements $\left(\mathbf{p},\mathbf{u}\right)$
of $\mathbb{T}_{\mathbf{p}}$ as \emph{line elements}. The set of
all line elements 
\[
\mathbb{T}=\mathfrak{S}\times\mathbb{U}^{n}=\bigcup_{\mathbf{p}\in\mathfrak{S}}\mathbb{T}_{\mathbf{p}}
\]
is called the \emph{tangent bundle} of the space $\mathfrak{S}$. 
\end{defn}

\index{tangent!bundle} This definition deviates from the traditional
one, which does not include the point $\mathbf{p}$ explicitly, but
it is more convenient for our purposes. In the more general case of
a \emph{differentiable manifold} the vector space $\mathbb{U}^{n}$
should be redefined. Note that the vectors in $\mathbb{U}^{n}$ do
not represent stimuli, but we still use boldface letters to denote
them. In the context of Euclidean spaces the boldface notation for
both stimuli and directions can simply be taken as indicating vectors.

For any $\mathbf{u}\in\mathbb{U}^{n}$ the notation $\overline{\mathbf{u}}$
will be used for the unit vector codirectional with $\mathbf{u}$:
\begin{equation}
\overline{\mathbf{u}}=\frac{\mathbf{u}}{\left\vert \mathbf{u}\right\vert },\qquad\left\vert \overline{\mathbf{u}}\right\vert =1.\label{eq:unit}
\end{equation}

\subsection{Submetric Function}

\index{submetric function} We make the following two assumptions
about the space $\left(\mathfrak{S},D\right)$ and its relation to
$\left(\mathfrak{S},E\right)$. 
\begin{lyxlist}{00.00.0000}
\item [{($\mathcal{E}1$)}] The topologies of $\left(\mathfrak{S},D\right)$
and $\left(\mathfrak{S},E\right)$ coincide. 
\end{lyxlist}
The coincidence of the $D$-topology and the Euclidean topology means
that the notion of convergence, 
\begin{equation}
\mathbf{a}_{n}\rightarrow\mathbf{a},
\end{equation}
means simultaneously $D\mathbf{a}_{n}\mathbf{a}\rightarrow0$ and
$\left|\mathbf{a}_{n}-\mathbf{a}\right|\rightarrow0$. As a result,
all topological concepts (openness, continuity, compactness, etc.)
can be used without the prefixes $D$, $G,$ or $E$. In particular,
dissimilarity $D\mathbf{xy}$ and metric $G\mathbf{xy}$ are continuous
in $\left(\mathbf{x,y}\right)$ with respect to the usual Euclidean
topology.

\begin{figure}[ptbh]
\begin{centering}
\includegraphics[scale=0.4]{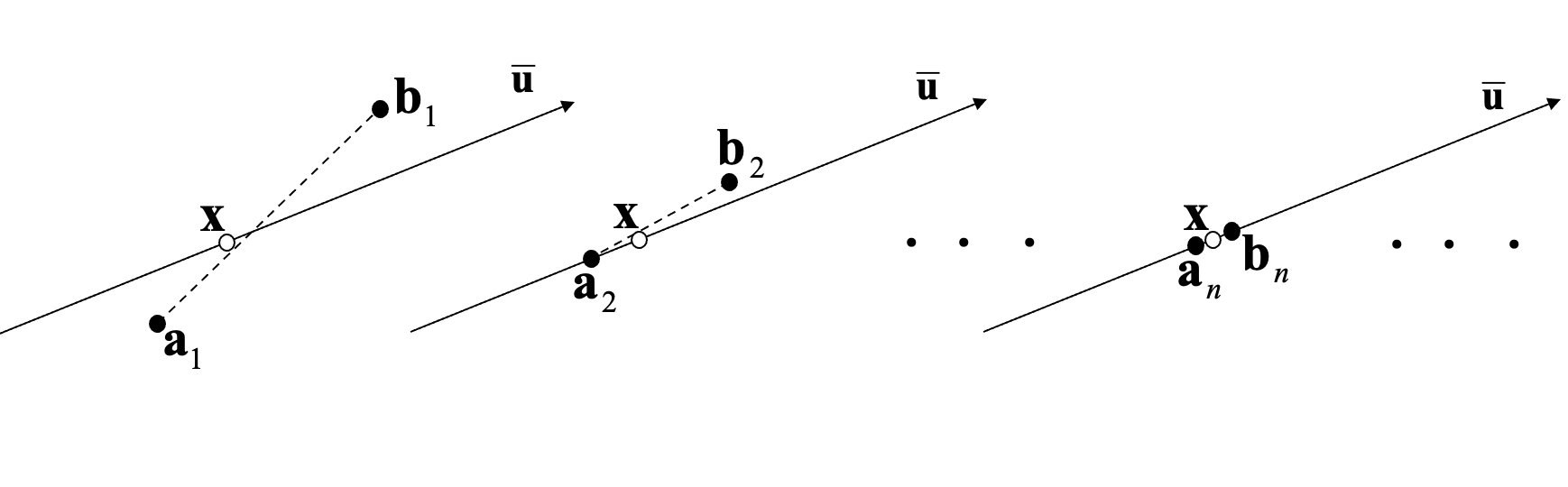} 
\par\end{centering}
\caption{An illustration for Assumption $\mathcal{E}2$. Shown are a point
$\mathbf{x}$ (open circle), a direction $\overline{\mathbf{u}}$
attached to it, and (in successive panels from left to right) pairs
of points $\left(\mathbf{a}_{1},\mathbf{b}_{1}\right)$, $\left(\mathbf{a}_{2},\mathbf{b}_{2}\right)$,
..., $\left(\mathbf{a}_{n},\mathbf{b}_{n}\right)$, ... gradually
converging to $\mathbf{x}$ so that the dashed line connecting them
(and directed from $\mathbf{a}_{n}$ to $\mathbf{b}_{n}$) gradually
aligns with the the direction $\overline{\mathbf{u}}$. The assumption
says that in this situation the dissimilarity $D\mathbf{a}_{n}\mathbf{b}_{n}$
and the Euclidean distance $\left\vert \mathbf{b}_{n}-\mathbf{a}_{n}\right\vert $
are comeasurable in the small: neither of them tends to zero infinitely
faster than the other.}

\centering{}\label{FigAxiom F properties} 
\end{figure}

Note, however, that the notions of uniform convergence in $\left(\mathfrak{S},D\right)$
and $\left(\mathfrak{S},E\right)$ are not assumed to coincide. Thus,
it is possible that $D\mathbf{a}_{n}\mathbf{b}_{n}\rightarrow0$ but
$\left\vert \mathbf{a}_{n}-\mathbf{b}_{n}\right\vert \not\rightarrow0,$
or vice versa. In particular, dissimilarity $D\mathbf{xy}$ and metric
$G\mathbf{xy}$ are not generally uniformly continuous in the Euclidean
sense. 
\begin{lyxlist}{00.00.0000}
\item [{($\mathcal{E}2$)}] For any $\mathbf{x},\mathbf{a}_{n},\mathbf{b}_{n}\in\mathfrak{S}$
($\mathbf{a}_{n}\neq\mathbf{b}_{n}$) and any unit vector $\overline{\mathbf{u}},$
if $\mathbf{a}_{n}\rightarrow\mathbf{x}$, $\mathbf{b}_{n}\rightarrow\mathbf{x},$
and $\overline{\mathbf{b}_{n}-\mathbf{a}_{n}}\rightarrow\overline{\mathbf{u}}$
(see Figure \ref{FigAxiom F properties}), then 
\[
\frac{D\mathbf{a}_{n}\mathbf{b}_{n}}{\left\vert \mathbf{b}_{n}-\mathbf{a}_{n}\right\vert }
\]
tends to a positive limit, denoted $F\left(\mathbf{x,}\overline{\mathbf{u}}\right)$. 
\end{lyxlist}

Putting $\mathbf{a}_{n}=\mathbf{x}$ and $\overline{\mathbf{b}_{n}-\mathbf{a}_{n}}=\overline{\mathbf{u}}$
in Assumption $\mathcal{E}2$, and denoting $\mathbf{b}_{n}=\mathbf{x}+\overline{\mathbf{u}}s,$
the function $F\left(\mathbf{x,}\overline{\mathbf{u}}\right)$ can
be presented as 
\begin{equation}
F\left(\mathbf{x,}\overline{\mathbf{u}}\right)=\lim_{s\rightarrow0+}\frac{D\mathbf{x}\left[\mathbf{x+}\overline{\mathbf{u}}s\right]}{s}.
\end{equation}
We now generalize this function to apply to any vector $\mathbf{u}$,
not just the unit one. 
\begin{defn}
The function 
\[
F:\mathbb{T}\cup\left\{ \left(\mathbf{x},\mathbf{0}\right):\mathbf{x}\in\mathfrak{S}\right\} \rightarrow\mathbb{R}
\]
defined as

\begin{equation}
F\left(\mathbf{x,u}\right)=\left\{ \begin{array}{lc}
\lim_{s\rightarrow0+}\frac{D\mathbf{x}\left[\mathbf{x+u}s\right]}{s} & \text{if }\mathbf{u\neq0}\\
0 & \text{if }\mathbf{u=0}
\end{array}\right.,\label{eq:F in general}
\end{equation}
is called a \emph{submetric function}. \index{submetric function} 
\end{defn}

The standard term for $F\left(\mathbf{x,u}\right)$ in differential
geometry is ``metric function.'' It can, however, be easily confused
with a metric on the space of stimuli, such as $G\mathbf{ab}$. To
prevent this confusion, we use the non-standard term ``submetric
function.''\index{submetric function} 
\begin{thm}
\label{thm:Submetric function}$F\left(\mathbf{x,u}\right)$ is well-defined
for any $\left(\mathbf{x,u}\right)\in\mathbb{T}\cup\left\{ \left(\mathbf{x},\mathbf{0}\right):\mathbf{x}\in\mathfrak{S}\right\} $.
It is positive for $\mathbf{u\neq0,}$ continuous in $\left(\mathbf{x,u}\right)$,
and Euler homogeneous in $\mathbf{u}$. \index{Euler homogeneity} 
\end{thm}

Euler homogeneity in $\mathbf{u}$ means that for any $k>0$, $F\left(\mathbf{x},k\mathbf{u}\right)=kF\left(\mathbf{x},\mathbf{u}\right).$
See Appendix for a proof. 

Assumption $\mathcal{E}2$ can now be strengthened as follows. 
\begin{thm}
\label{thm:DtoF and GtoF ratios}For any $\mathbf{a}_{n},\mathbf{b}_{n}\in\mathfrak{s}\subset\mathfrak{S}$,
if $\mathfrak{s}$ is compact and $\mathbf{a}_{n}\leftrightarrow\mathbf{b}_{n}$
($\mathbf{a}_{n}\neq\mathbf{b}_{n}$) then 
\[
\frac{D\mathbf{a}_{n}\mathbf{b}_{n}}{F\left(\mathbf{a}_{n},\mathbf{b}_{n}\mathbf{-a}_{n}\right)}\rightarrow1.
\]
\end{thm}

Indeed, rewrite 
\[
\frac{D\mathbf{a}_{n}\mathbf{b}_{n}}{F\left(\mathbf{a}_{n},\mathbf{b}_{n}\mathbf{-a}_{n}\right)}=\frac{D\mathbf{a}_{n}\mathbf{b}_{n}}{F\left(\mathbf{a}_{n}\mathbf{,}\overline{\mathbf{b}_{n}-\mathbf{a}_{n}}\right)\left\vert \mathbf{b}_{n}-\mathbf{a}_{n}\right\vert },
\]
and denote either $\lim\inf$ or $\lim\sup$ of this ratio by $l$.
There is an infinite subsequence of $\left(\mathbf{a}_{n},\mathbf{b}_{n}\right)$
(without loss of generality, the sequence itself) for which 
\[
\frac{D\mathbf{a}_{n}\mathbf{b}_{n}}{F\left(\mathbf{a}_{n}\mathbf{,}\overline{\mathbf{b}_{n}-\mathbf{a}_{n}}\right)\left\vert \mathbf{b}_{n}-\mathbf{a}_{n}\right\vert }\rightarrow l.
\]
But within a compact set $\mathfrak{s}$ one can always select from
this sequence $\left(\mathbf{a}_{n},\mathbf{b}_{n}\right)$ a subsequence
with $\mathbf{a}_{n}\leftrightarrow\mathbf{x}$, $\mathbf{b}_{n}\leftrightarrow\mathbf{x}$,
for some $\mathbf{x}$; and due to the compactness of the set $\overline{\mathfrak{u}}$
of all unit directions, one can always select a subsequence of this
subsequence with $\overline{\mathbf{b}_{n}-\mathbf{a}_{n}}\rightarrow\overline{\mathbf{u}},$
for some $\overline{\mathbf{u}}$. In this resulting subsequence (again,
without changing the indexation for convenience), 
\[
F\left(\mathbf{a}_{n}\mathbf{,}\overline{\mathbf{b}_{n}-\mathbf{a}_{n}}\right)\rightarrow F\left(\mathbf{a,\overline{u}}\right),
\]
whence 
\[
\frac{D\mathbf{a}_{n}\mathbf{b}_{n}}{\left\vert \mathbf{b}_{n}-\mathbf{a}_{n}\right\vert }\rightarrow lF\left(\mathbf{a,\overline{u}}\right).
\]
By Assumption $\mathcal{E}2$ then, $l=1$. Since this result holds
for both $\lim\inf$ and $\lim\sup$ of the original ratio, the statement
of the theorem follows.

\subsection{Indicatrices}

\index{indicatrix} 
\begin{defn}
The function 
\[
\mathbf{1}:\mathbb{T}\rightarrow\mathbb{U}^{n}
\]
defined by 
\[
\mathbf{\mathbf{1}}\left(\mathbf{a},\mathbf{u}\right)=\frac{\mathbf{u}}{F\left(\mathbf{a},\mathbf{u}\right)}
\]
is called the \emph{radius-vector function} associated with (or corresponding\index{function!radius-vector}
to) the submetric function\index{submetric function} $F\left(\mathbf{a},\mathbf{u}\right)$.
The values of this function are referred to as \emph{radius-vectors}.
For a fixed $\mathbf{a}\in\mathfrak{S}$, the function $\mathbf{u}\mapsto\mathbf{\mathbf{1}}\left(\mathbf{a},\mathbf{u}\right)$
is called the \emph{indicatrix} \emph{centered at} (or \emph{attached
to}) the point $\mathbf{a}$. The set

\[
\mathbb{I}_{\mathbf{a}}=\left\{ \mathbf{u}\in\mathbb{U}^{n}:F\left(\mathbf{a},\mathbf{u}\right)\leq1\right\} 
\]
is called the \emph{body} of this indicatrix, and the set 
\[
\delta\mathbb{I}_{\mathbf{a}}=\left\{ \mathbf{u}\in\mathbb{U}^{n}:F\left(\mathbf{a},\mathbf{u}\right)=1\right\} 
\]
is called its \emph{boundary}. 
\end{defn}

\begin{figure}
\begin{centering}
\includegraphics[scale=0.25]{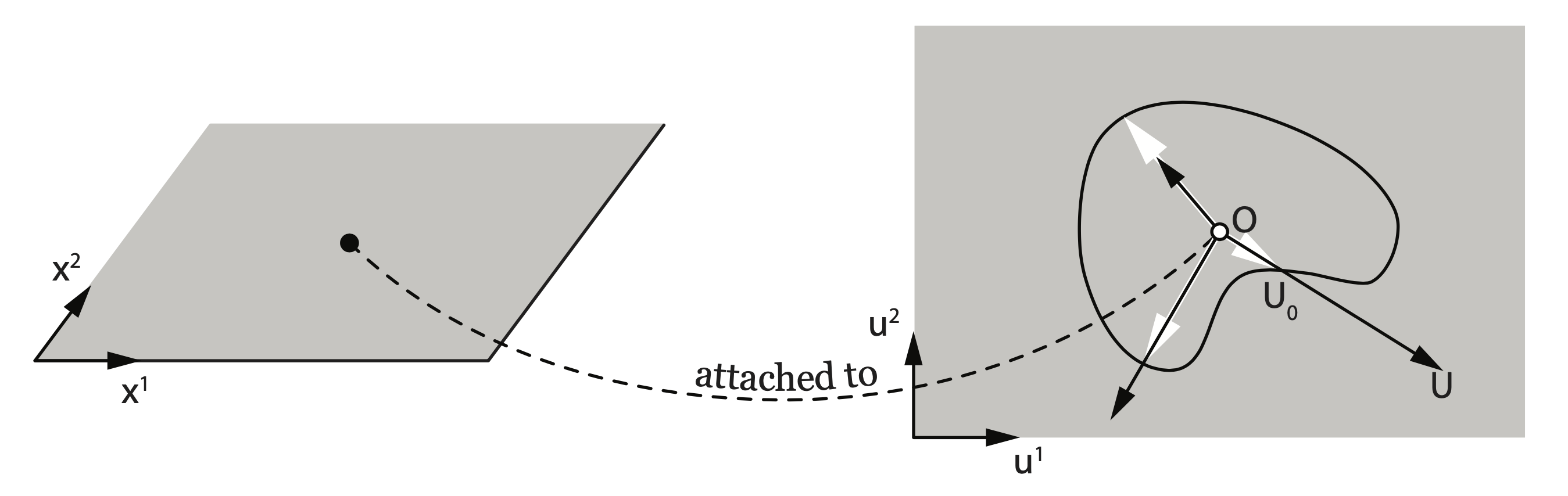} 
\par\end{centering}
\caption{An indicatrix (right) attached to a point in plane (left). The value
of the submetric function\index{submetric function} $F$ at this
point and any vector $\protect\overrightarrow{OU}$ is computed as
the ratio of $\protect\overrightarrow{OU}$ to the codirectional radius-vector
of the indicatrix, $\protect\overrightarrow{OU_{0}}$ (shown in white).}

\centering{}\label{fig:attached} 
\end{figure}

Figure \ref{fig:attached} provides an illustration for the relationship
between $F\left(\mathbf{a},\mathbf{u}\right)$ and $\mathbf{1}\left(\mathbf{a},\mathbf{u}\right)$.

\begin{figure}
\noindent \begin{centering}
\includegraphics[scale=0.2]{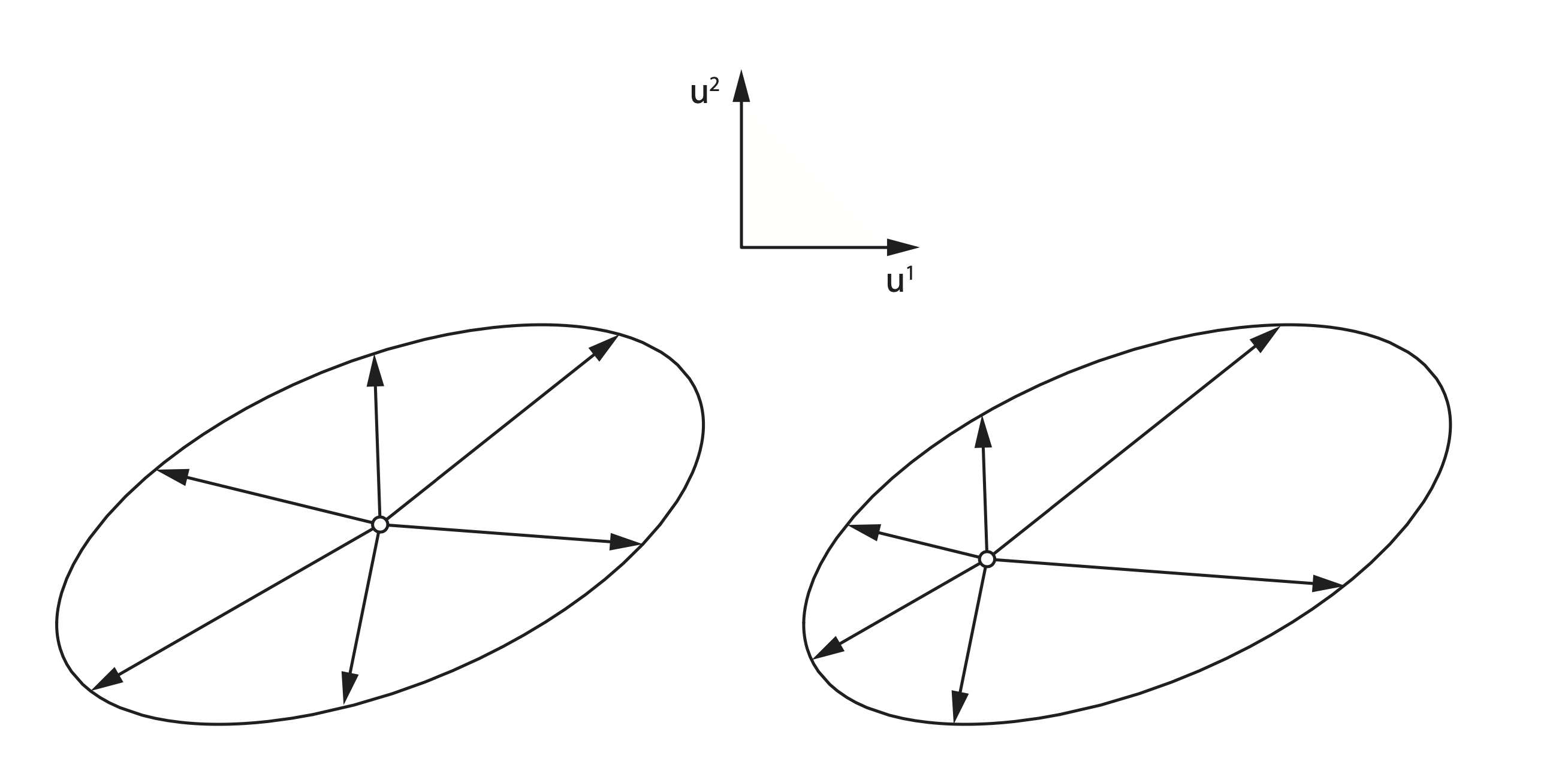} 
\par\end{centering}
\caption{The two indicatrices are different (consist of different vectors)
although they have identical sets of endpoints.}

\noindent \centering{}\label{fig:2indicatrices} 
\end{figure}

\begin{figure}
\noindent \begin{centering}
\includegraphics[scale=0.15]{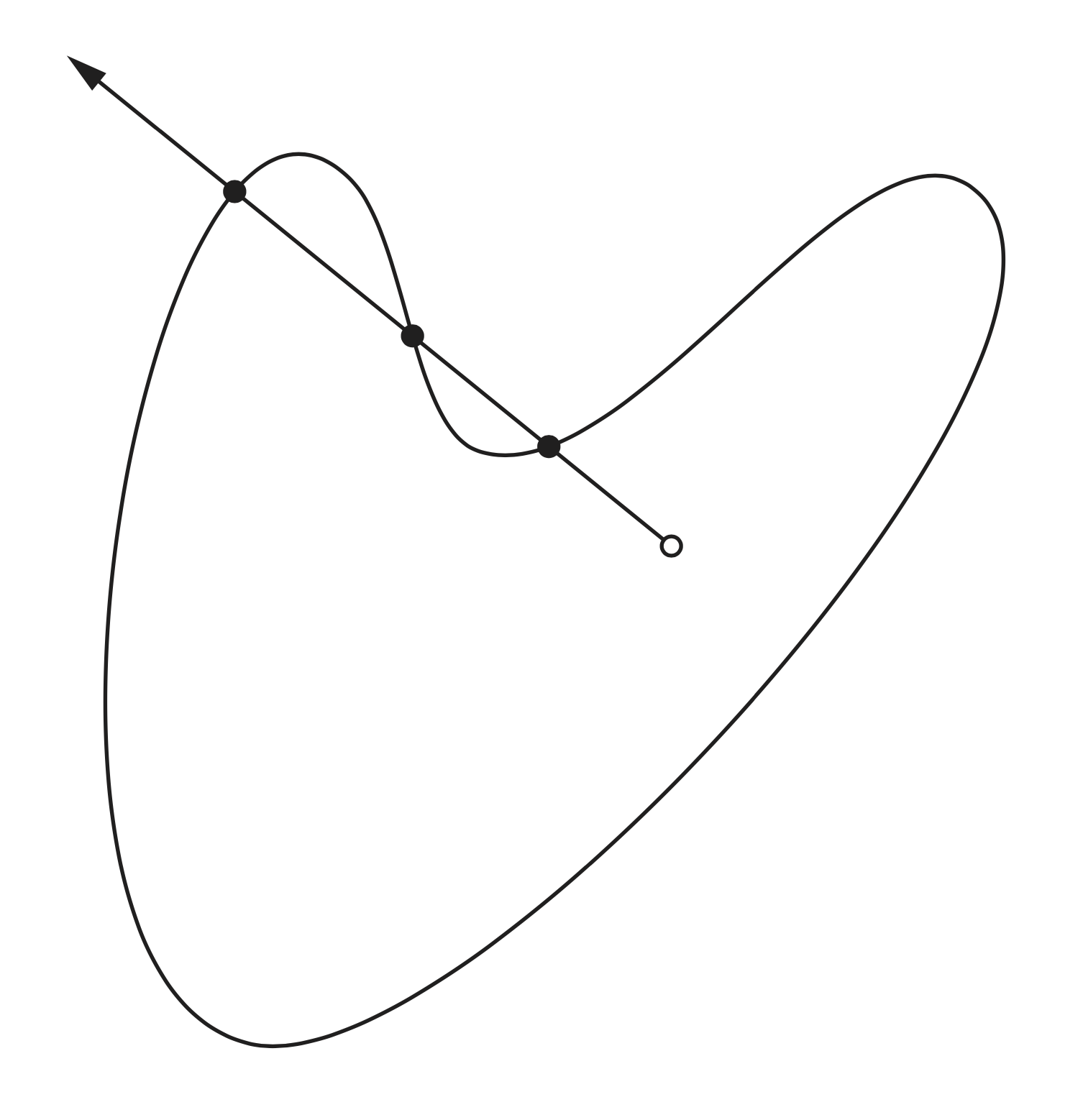} 
\par\end{centering}
\caption{This combination of a set of endpoints with a position of the origin
does not form an indicatrix, because a radius-vector from the origin
(shown by the open circle) intersects the boundary at more than one
point.}

\noindent \centering{}\label{fig:NotIndicatrix} 
\end{figure}

Note that $\left\{ \mathbf{a}\right\} \times\mathbb{I}_{\mathbf{a}}$
is a subset of the tangent space $\mathbb{T}_{\mathbf{a}}$. Note
also that the body (or the boundary) of an indicatrix is a set of
vectors in $\mathbb{U}^{n}$ emanating from a common origin. The boundary
should not be thought of as the set of the endpoints of the radius-vectors:
the latter set does not determine the indicatrix uniquely, as one
should also know the position of the origin within the boundary (see
Figure \ref{fig:2indicatrices}). Not all points within a given set
of endpoints may serve as points of origin: by definition, there can
be no endpoint $A$ on the boundary which is not connected to the
origin $O$ by a vector $\overrightarrow{OA}\in\delta\mathbb{I}_{\mathbf{a}}$,
and the boundary cannot have two codirectional but non-identical vectors
$\overrightarrow{OA}$ and $\overrightarrow{OB}$ (see Figure \ref{fig:NotIndicatrix}):
indeed, if 
\[
\frac{\overrightarrow{OA}}{\overrightarrow{OB}}=k\neq1,
\]
then 
\[
\frac{F\left(\mathbf{a},\overrightarrow{OA}\right)}{F\left(\mathbf{a},\overrightarrow{OB}\right)}=k,
\]
so one of the vectors $\overrightarrow{OA}$ and $\overrightarrow{OB}$
does not belong to $\delta\mathbb{I}_{\mathbf{a}}$.

\begin{figure}
\noindent \begin{centering}
\includegraphics[scale=0.3]{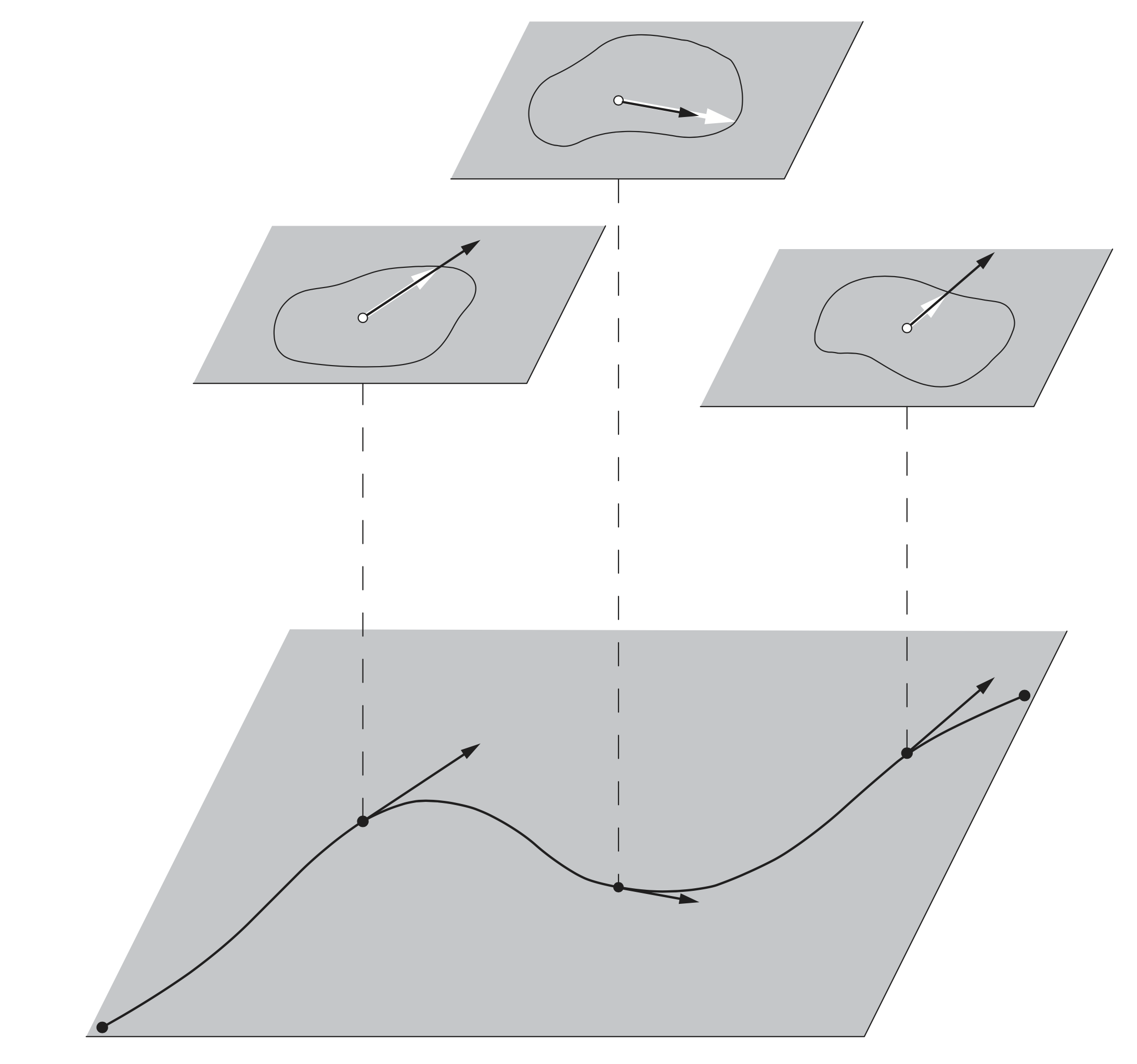} 
\par\end{centering}
\caption{Geometric interpretation of how indicatrices measure tangents to a
smooth path: by centering the indicatrix $\mathbb{I}_{\mathbf{f}\left(x\right)}$
at each point $\mathbf{f}\left(x\right)$, one measures the magnitude
of the tangent at this point by relating it to the codirectional radius-vector
of the indicatrix, as explained in Figure \ref{fig:attached}. The
length of the path $\mathbf{f}$|$\left[a,b\right]$ then is obtained
by integrating this magnitude from $a$ to $b$. For the conventional
Euclidean length all indicatrices should be unit-radius circles.}

\noindent \centering{}\label{fig:IndicatrixMeasure} 
\end{figure}

Figure \ref{fig:IndicatrixMeasure} offers a geometric interpretation
for measuring the length of a smooth path, to be rigorously justified
later.

We now list basic, almost obvious, properties of the unit vector function\index{function!unit vector}
and the corresponding indicatrices. 
\begin{thm}
\label{thm:properties of indicatrix}The following statements hold
true:

(i) $\mathbf{1}\left(\mathbf{a},\mathbf{u}\right)$ is continuous;

(ii) \textup{$\mathbf{1}\left(\mathbf{a},k\mathbf{u}\right)=\mathbf{1}\left(\mathbf{a},\mathbf{u}\right)$
for all $\left(\mathbf{a},\mathbf{u}\right)\in\mathbb{T}$ and all
$k>0$ (Euler homogeneity in }\textbf{\textup{u}}\textup{ of order
zero);}\index{Euler homogeneity}

(iii) for any $\mathbf{a}\in\mathfrak{S}$, the mapping $\mathbf{\overline{u}}\mapsto\mathbf{1}\left(\mathbf{a},\mathbf{\overline{u}}\right)$
is a homeomorphism;

(iv) $\mathbb{I}_{\mathbf{a}}$ is a compact set in $\mathbb{U}^{n}$;

(v) $\delta\mathbb{I}_{\mathbf{a}}$ is a compact set in $\mathbb{U}^{n}$;

(vi) for any $\mathbf{a}\in\mathfrak{S}$, there are two positive
reals $k_{\mathbf{a}},K_{\mathbf{a}}$ such that 
\[
k_{\mathbf{a}}\leq\left|\mathbf{1}\left(\mathbf{a},\mathbf{u}\right)\right|\leq K_{\mathbf{a}}
\]
for all $\mathbf{u}\in\mathbb{U}$\textup{, and the values} $k_{\mathbf{a}},K_{\mathbf{a}}$
are attained by $\mathbf{1}\left(\mathbf{a},\mathbf{u}\right)$ at
some $\mathbf{u}$. 
\end{thm}

The proof of Propositions (i) and (ii) follow from the continuity
and Euler homogeneity of $F\left(\mathbf{a},\mathbf{u}\right)$. Denoting
$\mathbf{1}\left(\mathbf{a},\mathbf{\overline{u}}\right)$ by $\mathbf{\widetilde{u}}$,
Proposition (iii) follows from the relations 
\[
\frac{\mathbf{\widetilde{u}}}{\left|\mathbf{\widetilde{u}}\right|}=\mathbf{\overline{u}}
\]
and 
\[
\widetilde{\mathbf{u}}=\frac{\mathbf{\overline{u}}}{F\left(\mathbf{a},\mathbf{\overline{u}}\right)},
\]
because both these functions are injective and continuous. The continuous
function $\mathbf{\overline{u}}\mapsto\widetilde{\mathbf{u}}$ induces
the continuous function $k\mathbf{\overline{u}}\mapsto k\widetilde{\mathbf{u}}$
for all $k\in\left[0,1\right]$, and (iv)-(v) then follow from the
compactness of the unit Euclidean ball $\left\{ k\mathbf{\overline{u}}:k\in\left[0,1\right]\right\} $
and the unit Euclidean sphere $\left\{ \mathbf{\overline{u}}\right\} $.
The continuous mapping $\mathbf{\overline{u}}\mapsto\mathbf{1}\left(\mathbf{a},\mathbf{\overline{u}}\right)$
of the compact unit Euclidean sphere should attain a maximum value
$K_{\mathbf{a}}$and a minimum value $k_{\mathbf{a}}$, and we get
(vi) due to (ii).

\begin{figure}[ptbh]
\begin{centering}
\includegraphics[scale=0.3]{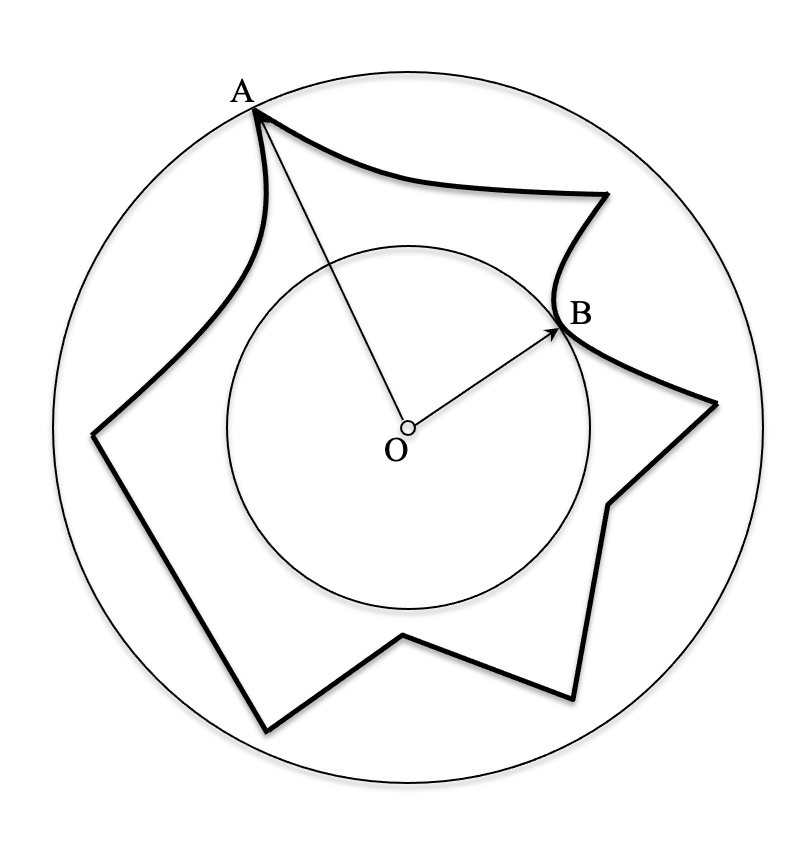} 
\par\end{centering}
\caption{A planar indicatrix (whose origin point $O$ is attached to a point
$\mathbf{a}$ in $\mathfrak{S}$) is sandwiched between two concentric
circles of radii $\left|\protect\overrightarrow{OA}\right|=K_{\mathbf{a}}$
and $\left|\protect\overrightarrow{OB}\right|=k_{\mathbf{a}}$.}

\centering{}\label{fig:2circles} 
\end{figure}

Based on Theorem \ref{thm:properties of indicatrix}, we can think
of an indicatrix boundary as a homeomorphically ``deformed'' Euclidean
$\left(n-1\right)$-sphere ``sandwiched'' between two concentric
Euclidean $\left(n-1\right)$-spheres of radii $k_{\mathbf{a}}>0$
and $K_{\mathbf{a}}\geq k_{\mathbf{a}}$. Figure \ref{fig:2circles}
illustrates this for $n=2$.

\subsection{Convex combinations and hulls}

\index{convex!combination} To further investigate the properties
of indicatrices, we need to recall certain notions from linear algebra.
In the vector space\index{vector!space} $\mathbb{U}^{n}$, a linear
combination 
\begin{equation}
\mathbf{u}=\lambda_{1}\mathbf{v}_{1}+\ldots+\lambda_{m}\mathbf{v}_{m},\;m\geq1,\label{eq:convexcombination}
\end{equation}
is called a \emph{convex!combination} of $\mathbf{v}_{1},\ldots,\mathbf{v}_{m}$
if $\lambda_{i}\geq0$ for $i=1,\ldots,m$, and 
\[
\lambda_{1}+\ldots+\lambda_{m}=1.
\]
From a geometric point of view, the set of convex combinations of
$\mathbf{v}_{1},\ldots,\mathbf{v}_{m}$ forms an $\left(m-1\right)$-dimensional
facet with vertices $\mathbf{v}_{1},\ldots,\mathbf{v}_{m}$. The following
therefore is obviously true. 
\begin{lem}
\label{lem:damned}If $\alpha\mathbf{u}$ is a convex combination
of $a_{1}\mathbf{v}_{1},\ldots,a_{m}\mathbf{v}_{m}$ and $\beta\mathbf{u}$
is a convex combination of $b_{1}\mathbf{v}_{1},\ldots,b_{m}\mathbf{v}_{m}$,
with $a_{i}\geq b_{i}$ for \textup{$i=1,\ldots,m$} and at least
one inequality being strict, then $\alpha>\beta$. 
\end{lem}

\begin{figure}[ptbh]
\begin{centering}
\includegraphics[scale=0.4]{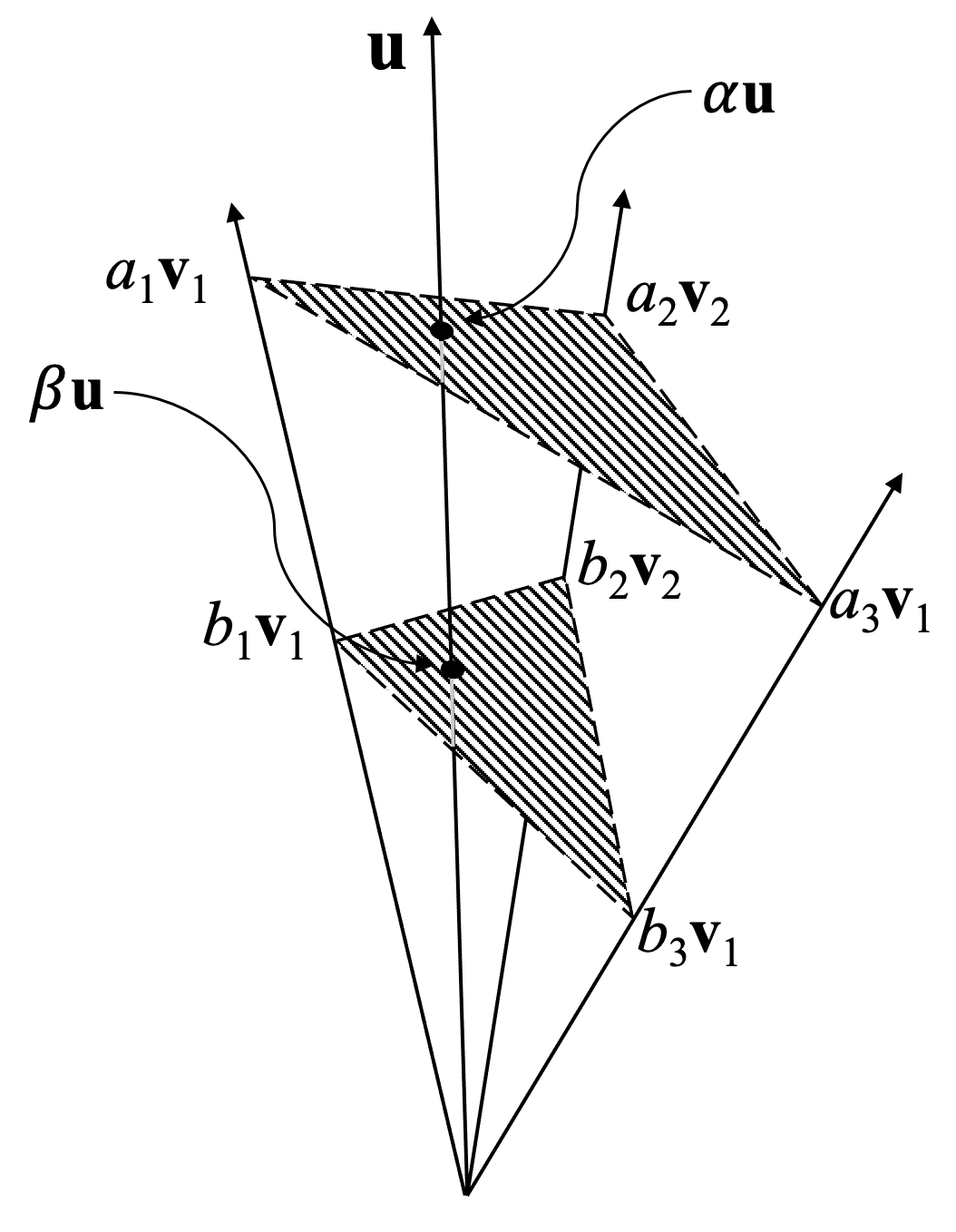} 
\par\end{centering}
\caption{Illustration for Lemma \ref{lem:damned}: a direction within the cone
formed by $\mathbf{v}_{1},\mathbf{v}_{2},\mathbf{v}_{3}$ first crosses
the lower facet and then the higher facet.}

\centering{}\label{fig:damned} 
\end{figure}

Figure \ref{fig:damned} provides an illustration.

Vectors $\mathbf{v}_{1},\ldots,\mathbf{v}_{m}$ are called \emph{affinely
dependent} if, for some $\gamma_{1},\ldots,\gamma_{m}$, not all zero,
\index{vector!affinely dependent} 
\begin{equation}
\begin{array}{c}
\gamma_{1}\mathbf{v}_{1}+\ldots+\gamma_{m}\mathbf{v}_{m}=\mathbf{0}\\
\gamma_{1}+\ldots+\gamma_{m}=0
\end{array}.
\end{equation}
If $\mathbf{u}$ is a convex combination of affinely dependent vectors,
we have simultaneously 
\[
\left\{ \begin{array}{c}
\lambda_{1}\mathbf{v}_{1}+\ldots+\lambda_{m}\mathbf{v}_{m}=\mathbf{u}\\
\gamma_{1}\mathbf{v}_{1}+\ldots+\gamma_{m}\mathbf{v}_{m}=\mathbf{0}
\end{array}\right.,
\]
where 
\[
\left\{ \begin{array}{c}
\lambda_{1}+\ldots,+\lambda_{m}=1\\
\gamma_{1}+\ldots,+\gamma_{m}=0
\end{array}\right.,
\]
all $\lambda'$s are nonnegative and some $\gamma'$s are nonzero
(which means that at least one of them is positive and at least one
negative). To exclude trivial cases, let $\mathbf{v}_{1},\ldots,\mathbf{v}_{m}$
be pairwise distinct and let $\lambda_{i}>0$ for $i=1,\ldots,m$.
Let $c$ be the minimum $\left|\frac{\lambda_{i}}{\gamma_{i}}\right|$
among all negative ratios $\frac{\lambda_{i}}{\gamma_{i}}$. Then
at least one of the coefficients in the representation 
\[
\mathbf{u}=\left(\lambda_{1}+c\gamma_{1}\right)\mathbf{v}_{1}+\ldots,+\left(\lambda+c\gamma_{m}\right)\mathbf{v}_{m}
\]
is zero, while all other coefficients are nonnegative and sum to 1.
This means that $\mathbf{u}$ is a convex combination of at most $m-1$
elements of $\left\{ \mathbf{v}_{1},\ldots,\mathbf{v}_{m}\right\} $,
and we have 
\begin{lem}
\label{lem:almostCarath}If $\mathbf{u}\in\mathbb{U}^{n}$ is a convex
combination of affinely dependent $\mathbf{v}_{1},\ldots,\mathbf{v}_{m}\in\mathbb{U}^{n}$,
then $\mathbf{u}$ is a convex combination of some $m'<m$ elements
of $\mathbf{v}_{1},\ldots,\mathbf{v}_{m}$. 
\end{lem}

The following corollary of the lemma is known as a Carathéodory theorem.\index{Carathéodory theorem} 
\begin{cor}
\label{cor:Carath}If $\mathbf{u},\mathbf{v}_{1},\ldots,\mathbf{v}_{m}\in\mathbb{U}^{n}$,
$m>n+1$, and $\mathbf{u}$ is a convex combination of $\mathbf{v}_{1},\ldots,\mathbf{v}_{m}$,
then $\mathbf{u}$ is a convex combination of at most $n+1$ elements
of $\mathbf{v}_{1},\ldots,\mathbf{v}_{m}$. 
\end{cor}

This follows from the fact that if $m>n+1$, any $\mathbf{v}_{1},\ldots,\mathbf{v}_{m}$
in $\mathbb{U}^{n}$ are affinely dependent. Indeed, since $\mathrm{rank}\left(\mathbf{v}_{1},\ldots,\mathbf{v}_{m}\right)\leq n,$
there should exist reals $\alpha_{1},\ldots,\alpha_{m}$, not all
zero, such that the system of $n+1$ linear equations 
\[
\left\{ \begin{array}{c}
\alpha_{1}\mathbf{v}_{1}+\ldots+\alpha_{m}\mathbf{v}_{m}=\mathbf{0}\\
\alpha_{1}+\ldots+\alpha_{m}=0
\end{array}\right.
\]
is satisfied.

A subset $\mathbb{V}$ of $\mathbb{U}^{n}$ is said to be \emph{convex}\index{convex!subset}
if it contains any convex combination 
\[
\lambda\mathbf{x}+(1-\lambda)\mathbf{y},\quad\lambda\in\left[0,1\right],
\]
of any two of its elements $\mathbf{x,y}.$ By induction from 2 to
$(n+1)$-element subsets of $\mathbb{V}$ (which is sufficient by
Corollary \ref{cor:Carath}), we see that a convex set $\mathfrak{X\subset\mathbb{U}^{\mathfrak{\mathit{n}}}}$contains
all convex combinations of \emph{all} finite subsets of $\mathbb{V}$.

For any $\mathfrak{X\subset\mathbb{U}^{\mathfrak{\mathit{n}}}}$ the
set of all convex combinations of all $\left(n+1\right)$-tuples of
elements of $\mathbb{V}$ is called the \emph{convex hull} of $\mathbb{V}$
and is denoted $\mathrm{conv}\mathbb{V}$. Again, $\mathrm{conv}\mathbb{V}$
is, clearly, the set of all convex combinations of all finite subsets
of $\mathbb{V}$, and it is the smallest convex subset of $\mathbb{U}^{n}$
containing $\mathbb{V}$.\index{convex!hull}

Consider now an indicatrix $\mathbb{I}_{\mathbf{a}}$ and its convex
hull. The following is obvious. 
\begin{lem}
For any indicatrix $\mathbb{I}_{\mathbf{a}}$, $\mathrm{conv}\mathbb{I}_{\mathbf{a}}$
is compact in $\mathbb{U}^{n}$. 
\end{lem}

Let now $\mathbf{u}\in\mathrm{conv}\mathbb{I}_{\mathbf{a}}$. Then,
for some $\mathbf{v}_{1},\ldots,\mathbf{v}_{m}\in\mathbb{I}_{\mathbf{a}}$
and some nonnegative reals $\lambda_{1},\ldots,\lambda_{m}$ that
sum to 1, 
\[
\mathbf{u}=\lambda_{1}\mathbf{v}_{1}+\ldots+\lambda_{m}\mathbf{v}_{m}.
\]
But then 
\begin{multline*}
\left|\mathbf{u}\right|=\left|\lambda_{1}\mathbf{v}_{1}+\ldots+\lambda_{m}\mathbf{v}_{m}\right|\leq\lambda_{1}\left|\mathbf{v}_{1}\right|+\ldots+\lambda_{m}\left|\mathbf{v}_{m}\right|\\
\leq\left(\lambda_{1}+\ldots,+\lambda\right)K_{\mathbf{a}}=K_{\mathbf{a}},
\end{multline*}
where $K_{\mathbf{a}}$ denotes $\max_{\mathbf{u}\in\mathbb{I}_{\mathbf{a}}}\left|\mathbf{u}\right|$
(whose existence is stated in Theorem \ref{thm:properties of indicatrix},
v). We have therefore 
\begin{lem}
\label{lem:max-radius}For any $\mathbf{a}\in\mathfrak{S}$, 
\[
\max_{\mathbf{u}\in\mathrm{conv}\mathbb{I}_{\mathbf{a}}}\left|\mathbf{u}\right|=\max_{\mathbf{u}\in\mathbb{I}_{\mathbf{a}}}\left|\mathbf{u}\right|.
\]
\end{lem}

\begin{defn}
For any $\left(\mathbf{a},\mathbf{u}\right)\in\mathbb{T},$ the quantity
\[
\kappa\left(\mathbf{a},\mathbf{u}\right)=\max\left\{ \alpha>0:\alpha\mathbf{1}\mathbf{\left(\mathbf{a},\mathbf{u}\right)}\in\mathrm{conv}\mathbb{I}_{\mathbf{a}}\right\} 
\]
is called the \emph{maximal production factor} for $\mathbf{u}$ in
$\mathbb{I}_{\mathbf{a}}$, and the vector\index{vector!maximal production}
$\kappa\left(\mathbf{a},\mathbf{u}\right)\mathbf{1}\mathbf{\left(\mathbf{a},\mathbf{u}\right)}$
is called the \emph{maximal production} of (or \emph{maximally produced})
$\mathbf{u}$ in $\mathbb{I}_{\mathbf{a}}$. 
\end{defn}

This is clearly a well-defined function, because it follows from the
compactness of $\mathrm{conv}\mathbb{I}_{\mathbf{a}}$ that 
\begin{lem}
\label{lem:max production exists}For any $\mathbf{a}\in\mathfrak{S}$,
every $\mathbf{u}\in\mathbb{U}^{n}$ has its maximal production in
$\mathbb{I}_{\mathbf{a}}$. 
\end{lem}

The following statement holds because $\mathbf{\alpha u}$ and $\mathbf{u}$
have one and the same maximal production in $\mathbb{I}_{\mathbf{a}}$. 
\begin{lem}
The function \textup{$\kappa\left(\mathbf{a},\mathbf{u}\right)$}
is Euler homogeneous of zero order, 
\[
\kappa\left(\mathbf{a},\mathbf{\alpha u}\right)=\kappa\left(\mathbf{a},\mathbf{u}\right).
\]
\end{lem}

Finally, we need to observe the following. 
\begin{lem}
\label{lem:max prod n}For any $\mathbf{\left(\mathbf{a},\mathbf{u}\right)}\in\mathbb{T}$,
the maximal production \textup{of $\mathbf{u}$ in $\mathbb{I}_{\mathbf{a}}$}
can be presented as a convex combination of $n$ (not necessarily
distinct) radius-vectors $\mathbf{v}_{1},\ldots,\mathbf{v}_{n}\in\delta\mathbb{I}_{\mathbf{a}}$. 
\end{lem}

See Appendix for a proof.

Figure \ref{fig:3Dindicatrices} provides an illustration for this
lemma on three-dimensional indicatrices. (It also illustrates the
useful notion of the degree of flatness for a radius vector within
the body of the indicatrix.)

\begin{figure}[ptbh]
\begin{centering}
\includegraphics[scale=0.35]{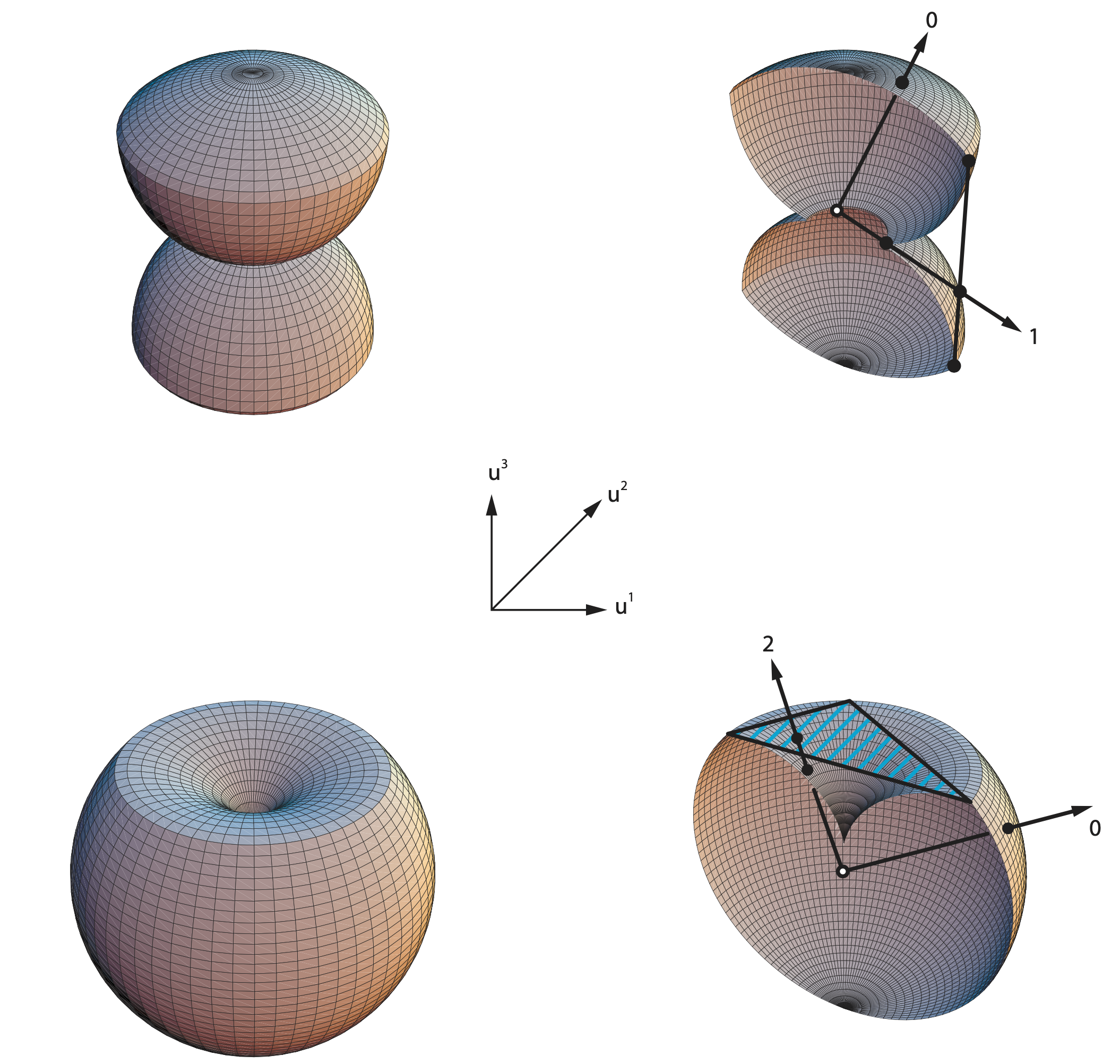} 
\par\end{centering}
\caption{Two indicatrices in $\mathbb{U}^{3}$ (left) and their cross-sections
(right) showing the position of the origin (white dots). The maximal
productions of two vectors are shown in each of the indicatrices,
as parts of the vectors between the origin to the farthest black dot.
The number attached to a vector $\mathbf{v}$ shows the degree of
flatness $r-1$ of the indicatrix in the direction $\mathbf{v}$,
where $r$ is the maximum number of linearly independent radius-vectors
whose convex combination equals the maximum production of $\mathbf{v}$
in the body of the indicatrix.}

\centering{}\label{fig:3Dindicatrices} 
\end{figure}

\subsection{Minimal submetric function\index{submetric function!minimal} and
convex hulls of indicatrices}

In this section we consider the problem of finding a \emph{geodesic
in the small}, a shortest path connecting stimuli $\mathbf{a}$ and
$\mathbf{a}+\mathbf{u}s$ as $s\rightarrow0$. It will be established
later (Section \ref{sec:Length-and-Metric}) that $G\mathbf{a}\left(\mathbf{a}+\mathbf{u}s\right)$
in $\mathfrak{S}\subseteq\mathbb{R}^{n}$ can be approximated by concatenation
of $m\leq n$ straight line segments with lengths $F\left(\mathbf{a},\mathbf{u}_{i}\right)s$
for some vectors $\mathbf{u}_{1},\ldots,\mathbf{u}_{m}$ summing to
$\mathbf{u}$. So we begin with investigating the minimal value for
certain sums of $F\left(\mathbf{a},\mathbf{u}_{i}\right)$. 
\begin{defn}
A sequence of vectors $\left(\mathbf{u}_{1},\ldots,\mathbf{u}_{m}\right)$
in $\mathbb{U}^{n}$, $m\geq1$, is said to form a \emph{minimizing
vector chain} for a line element $\left(\mathbf{a},\mathbf{u}\right)\in\mathbb{T}$,
\index{chain!minimizing vector} 
\[
\mathbf{u}=\mathbf{u}_{1}+\ldots+\mathbf{u}_{m}
\]
and 
\[
F\left(\mathbf{a},\mathbf{u}_{1}\right)+\ldots+F\left(\mathbf{a},\mathbf{u}_{m}\right)=\min\left\{ F\left(\mathbf{a},\mathbf{v}_{1}\right)+\ldots+F\left(\mathbf{a},\mathbf{v}_{k}\right)\right\} ,
\]
where the minimum is taken over all $k\geq1$ and all finite sequences
$\left(\mathbf{v}_{1},\ldots,\mathbf{v}_{k}\right)$ in $\mathbb{U}^{n}$
such that 
\[
\mathbf{u}=\mathbf{v}_{1}+\ldots+\mathbf{v}_{k}.
\]
\end{defn}

Note that this definition does not require that $\mathbf{u}_{1},\ldots,\mathbf{u}_{m}$
be pairwise distinct, so a minimizing chain for $\mathbf{\left(\mathbf{a},\mathbf{u}\right)}$
may, e.g., be $\left\{ \frac{1}{n}\mathbf{u},\ldots,\frac{1}{n}\mathbf{u}\right\} $(which
is equivalent to $\mathbf{u}$ alone being a minimizing vector chain
for $\mathbf{\left(\mathbf{a},\mathbf{u}\right)}$ too). Note also,
that if $\left(\mathbf{u}_{1},\ldots,\mathbf{u}_{m}\right)$ is a
minimizing chain, then so is any permutation thereof. 
\begin{thm}
\label{thm:minimizing set}A minimizing chain for any $\mathbf{\left(a,u\right)\in\mathbb{T}}$
exists and consists of $n$ (not necessarily distinct) nonzero vectors
$\mathbf{u}_{1},\ldots,\mathbf{u}_{m}$, such that 
\[
F\left(\mathbf{a},\mathbf{u}_{1}\right)+\ldots+F\left(\mathbf{a},\mathbf{u}_{n}\right)=\frac{F\left(\mathbf{a,u}\right)}{\kappa\left(\mathbf{a,u}\right)},
\]
where $\kappa\left(\mathbf{a,u}\right)$ is the maximal production
factor for $\mathbf{u}$ in \textup{$\mathbb{I}_{\mathbf{a}}$}. 
\end{thm}

To prove this, we fix $\kappa\left(\mathbf{a,u}\right)=\kappa$ as
we deal with a fixed $\mathbf{\left(a,u\right)}.$ Consider the maximal
production $\kappa\mathbf{1}\left(\mathbf{a,u}\right)$ of $\mathbf{u}$.
By Lemma \ref{lem:max prod n}, it can be presented as a convex combination
of some $n$ radius-vectors $\mathbf{\widetilde{v}}_{1},\ldots,\mathbf{\widetilde{v}}_{n}$
in $\delta\mathbb{I}_{\mathbf{a}}$, 
\[
\kappa\mathbf{1}\left(\mathbf{a,u}\right)=\lambda_{1}\mathbf{\widetilde{v}}_{1}+\ldots+\lambda_{n}\mathbf{\widetilde{v}}_{n},
\]
where all coefficients are nonnegative and sum to 1. Then, denoting
\[
\mathbf{v}_{i}=\frac{\lambda_{i}}{\kappa}\mathbf{\widetilde{v}}_{i},\qquad i=1,\ldots,n,
\]
we have 
\[
\mathbf{1}\left(\mathbf{a,u}\right)=\mathbf{v}_{1}+\ldots+\mathbf{v}_{n}
\]
and 
\[
F\left(\mathbf{a},\mathbf{v}_{1}\right)+\ldots+F\left(\mathbf{a},\mathbf{v}_{n}\right)=\frac{1}{\kappa}.
\]
We prove now that for any $\mathbf{w}_{1},\ldots,\mathbf{w}_{m}$
in $\mathbb{U}^{n}$, if 
\[
\mathbf{1}\left(\mathbf{a,u}\right)=\mathbf{w}_{1}+\ldots+\mathbf{w}_{m},
\]
then 
\[
F\left(\mathbf{a},\mathbf{w}_{1}\right)+\ldots+F\left(\mathbf{a},\mathbf{w}_{m}\right)=\delta\geq\frac{1}{\kappa}.
\]
Indeed, we have 
\[
\mathbf{1}\left(\mathbf{a,u}\right)=F\left(\mathbf{a},\mathbf{w}_{1}\right)\mathbf{1}\left(\mathbf{a,}\mathbf{w}_{1}\right)+\ldots+F\left(\mathbf{a},\mathbf{w}_{m}\right)\mathbf{1}\left(\mathbf{a,}\mathbf{w}_{m}\right)
\]
and 
\[
\frac{1}{\delta}\mathbf{1}\left(\mathbf{a,u}\right)=\frac{F\left(\mathbf{a},\mathbf{w}_{1}\right)}{\delta}\mathbf{1}\left(\mathbf{a,}\mathbf{w}_{1}\right)+\ldots+\frac{F\left(\mathbf{a},\mathbf{w}_{n+1}\right)}{\delta}\mathbf{1}\left(\mathbf{a,}\mathbf{w}_{m}\right).
\]
That is, $\frac{1}{\delta}\mathbf{1}\left(\mathbf{a,u}\right)$ is
a convex combination of $m$ radius-vectors of $\delta\mathbb{I}_{\mathbf{a}}$.
But then 
\[
\frac{1}{\delta}\leq\kappa.
\]
It follows that $\left(\mathbf{v}_{1},\ldots,\mathbf{v}_{n}\right)$
is a minimizing vector chain for $\left(\mathbf{a},\mathbf{1}\left(\mathbf{a,u}\right)\right)$,
with 
\[
F\left(\mathbf{a},\mathbf{v}_{1}\right)+\ldots+F\left(\mathbf{a},\mathbf{v}_{m}\right)=\frac{1}{\kappa}.
\]
The statement of the theorem obtains by putting $\mathbf{u}_{i}=F\left(\mathbf{a,u}\right)\mathbf{v}_{i}$,
$i=1,\ldots,n$.

We introduce now one of the central notions of the theory. 
\begin{defn}
For any $\left(\mathbf{a},\mathbf{u}\right)\in\mathbb{T}$$\cup\left\{ \left(\mathbf{x},\mathbf{0}\right):\mathbf{x}\in\mathfrak{S}\right\} $,
the function 
\[
\widehat{F}\left(\mathbf{a,u}\right)=\left\{ \begin{array}{cc}
\frac{F\left(\mathbf{a,u}\right)}{\kappa\left(\mathbf{a,u}\right)} & \mathrm{if\:\mathbf{u}\neq\mathbf{0}}\\
0 & \mathrm{if\:\mathbf{u}=\mathbf{0}}
\end{array}\right.
\]
is called the \emph{minimal submetric function}. \index{submetric function!minmal} 
\end{defn}

Clearly, 
\[
\widehat{F}\left(\mathbf{a,u}\right)\leq F\left(\mathbf{a,u}\right).
\]

\begin{thm}
\label{thm:Fhat is submetric}The minimal submetric function\index{submetric function!minimal}
$\widehat{F}\left(\mathbf{a,u}\right)$ has all the properties of
a submetric function\index{submetric function}: it is positive for
$\mathbf{u}\neq\mathbf{0}$, Euler homogeneous, and continuous. 
\end{thm}

See Appendix for a proof.
\begin{thm}
\label{thm:minFconvI}The indicatrix at $\mathbf{a}\in\mathfrak{S}$
associated with $\widehat{F}\left(\mathbf{a},\mathbf{u}\right)$\textup{,
} 
\[
\mathbf{u}\mapsto\mathbf{\mathbf{\widehat{1}}}\left(\mathbf{a},\mathbf{u}\right)=\frac{\mathbf{u}}{\widehat{F}\left(\mathbf{a},\mathbf{u}\right)},
\]
has the body 
\[
\widehat{\mathbb{I}}_{\mathbf{a}}=\left\{ \mathbf{u}\in\mathbb{U}^{n}:\widehat{F}\left(\mathbf{a},\mathbf{u}\right)\leq1\right\} =\mathrm{conv\mathbb{I}_{\mathbf{a}}},
\]
where $\mathbb{I}_{\mathbf{a}}$ is the body of the indicatrix $\mathbf{\mathbf{u}\mapsto\mathbf{1}\left(\mathbf{a},\mathbf{u}\right)}$
associated with \textup{$F\left(\mathbf{a},\mathbf{u}\right)$. }\textup{\emph{The
boundary 
\[
\widehat{\delta\mathbb{I}}_{\mathbf{a}}=\left\{ \mathbf{u}\in\mathbb{U}^{n}:\widehat{F}\left(\mathbf{a},\mathbf{u}\right)=1\right\} 
\]
of the indicatrix $\mathbf{u}\mapsto\mathbf{\mathbf{\widehat{1}}}\left(\mathbf{a},\mathbf{u}\right)$
is the set of all maximally produced radius-vectors of the indicatrix
}}\emph{$\mathbf{\mathbf{u}\mapsto\mathbf{1}\left(\mathbf{a},\mathbf{u}\right)}$}.\textup{
} 
\end{thm}

This is essentially a summary of the results established so far. To
prove the second statement of the theorem, by Lemma \ref{lem:max production exists}
and Theorem \ref{thm:minimizing set}, the maximal production $\kappa\left(\mathbf{a},\mathbf{u}\right)\mathbf{\mathbf{1}\left(\mathbf{a},\mathbf{u}\right)}$
of $\mathbf{u}$ in $\mathbb{I}_{\mathbf{a}}$ exists for every $\mathbf{u}$,
and 
\[
\widehat{F}\left(\mathbf{a},\mathbf{\mathbf{1}\left(\mathbf{a},\mathbf{u}\right)}\right)=\frac{1}{\kappa\left(\mathbf{a},\mathbf{u}\right)}.
\]
It follows that $\widehat{F}\left(\mathbf{a},\mathbf{u}\right)=1$
if and only if 
\[
\mathbf{u=\kappa\left(\mathbf{a},\mathbf{u}\right)\mathbf{\mathbf{1}\left(\mathbf{a},\mathbf{u}\right)}.}
\]
To prove the first statement of the theorem, by Lemma \ref{lem:max prod n},
$\kappa\left(\mathbf{a},\mathbf{u}\right)\mathbf{\mathbf{1}\left(\mathbf{a},\mathbf{u}\right)}$
is a convex combination of some vectors $\mathbf{v}_{1},...,\mathbf{v}_{n}$
in $\mathbb{I}_{\mathbf{a}}$. But then $c\kappa\left(\mathbf{a},\mathbf{u}\right)\mathbf{\mathbf{1}\left(\mathbf{a},\mathbf{u}\right)}$
is a convex combination of $c\mathbf{v}_{1},...,c\mathbf{v}_{n}\in\mathbb{I}_{\mathbf{a}}$
for any $c\in\left[0,1\right]$. It is clear then that $\mathrm{conv}\mathbb{I}_{\mathbf{a}}$
consists of all vectors 
\[
\mathbf{u}=c\mathbf{\kappa\left(\mathbf{a},\mathbf{u}\right)\mathbf{\mathbf{1}\left(\mathbf{a},\mathbf{u}\right)}},\qquad c\in\left[0,1\right].
\]
But these are precisely the vectors satisfying $\widehat{F}\left(\mathbf{a},\mathbf{u}\right)\leq1$.
This completes the proof.

\begin{figure}[ptbh]
\begin{centering}
\includegraphics[scale=0.3]{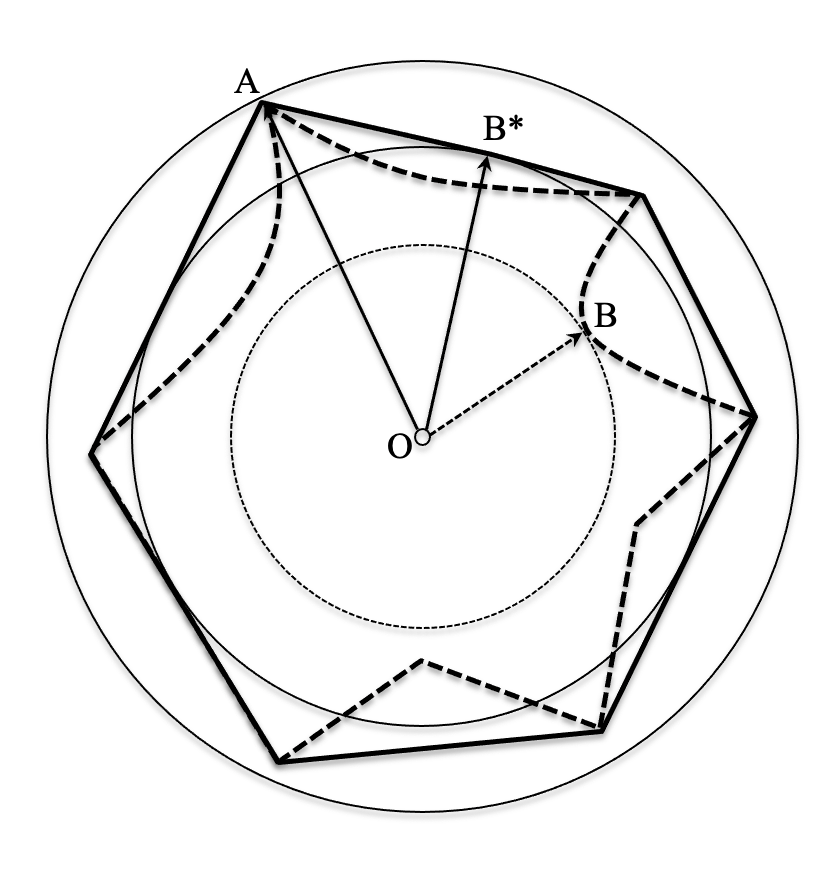} 
\par\end{centering}
\caption{The convex hull of the indicatrix body shown in Figure \ref{fig:2circles}
is sandwiched between $\left|\protect\overrightarrow{OA}\right|=K_{\mathbf{a}}$(the
same as for the indicatrix itself) and $\left|\protect\overrightarrow{OB^{*}}\right|=k_{\mathbf{a}}^{*}$
which is greater than $\left|\protect\overrightarrow{OB}\right|=k_{\mathbf{a}}$.}

\centering{}\label{fig:2circlesConvex} 
\end{figure}

\begin{figure}[ptbh]
\begin{centering}
\includegraphics[scale=0.35]{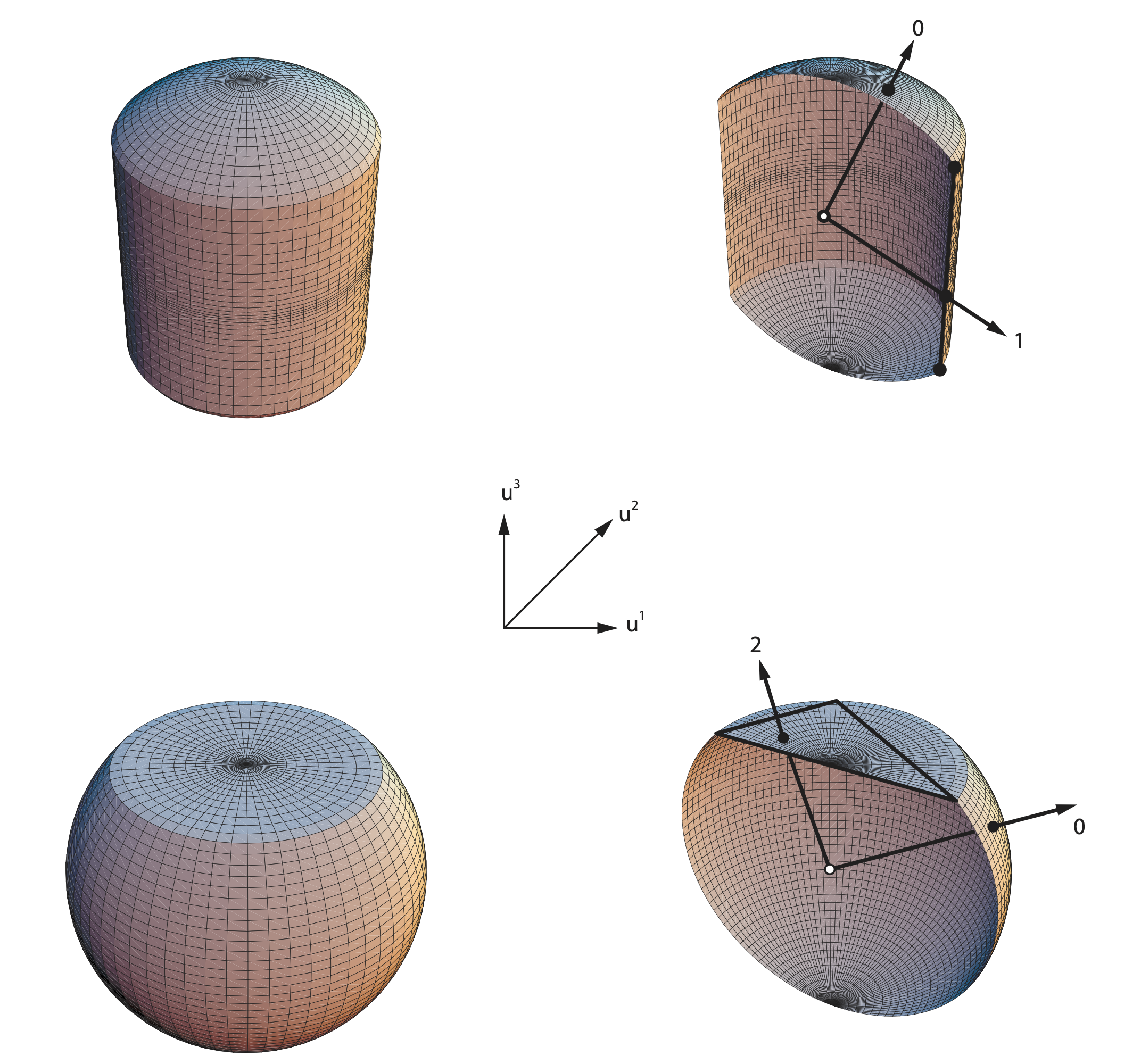} 
\par\end{centering}
\caption{The convex hulls of the indicatrices shown in Figure \ref{fig:3Dindicatrices}.
The degree of flatness of codirectional radius-vectors remains unchanged.}

\centering{}\label{fig:3Dconvex} 
\end{figure}

It follows from this theorem that $\mathbf{\mathbf{\widehat{1}}}\left(\mathbf{a},\mathbf{u}\right)$,
$\widehat{\mathbb{I}}_{\mathbf{a}}$, and $\widehat{\delta\mathbb{I}}_{\mathbf{a}}$
have all the properties listed in Theorem \ref{thm:properties of indicatrix}.
If $\mathrm{\delta\mathbb{I}_{\mathbf{a}}}$ is a homeomorphically
deformed Euclidean sphere sandwiched between two Euclidean spheres
of radii $k_{\mathbf{a}}$ and $K_{\mathbf{a}}$, then $\mathrm{\widehat{\delta\mathbb{I}}_{\mathbf{a}}}$
is a a homeomorphically deformed (but convex) Euclidean sphere sandwiched
between two Euclidean spheres of radii $k_{\mathbf{a}}^{*}$ and $K_{\mathbf{a}}$
(where $k_{\mathbf{a}}^{*}\geq k_{\mathbf{a}}$ and $K_{\mathbf{a}}$
is the same for $\mathrm{\delta\mathbb{I}_{\mathbf{a}}}$ and $\mathrm{\widehat{\delta\mathbb{I}}_{\mathbf{a}}}$,
as stated in Lemma \ref{lem:max-radius}). Figure \ref{fig:2circlesConvex}
illustrates this using the indicatrix shown in Figure \ref{fig:2circles}.
Figure \ref{fig:3Dconvex} shows the convex hulls of the indicatrices
shown in Figure \ref{fig:3Dindicatrices}.

\subsection{\label{sec:Length-and-Metric}Length and Metric in Euclidean spaces}
\begin{defn}
\label{def:convexF}A submetric function $F\left(\mathbf{a,u}\right)$
is called \emph{convex}\index{submetric function!convex} if for any
$\mathbf{a}\in\mathfrak{S}$ and $\mathbf{u}_{1},\mathbf{u}_{2}\in\mathbb{U}^{n}$,
\[
F\left(\mathbf{a,u}_{1}+\mathbf{u}_{2}\right)\leq F\left(\mathbf{a,u}_{1}\right)+F\left(\mathbf{a,u}_{2}\right).
\]
\end{defn}

Assume, excluding the trivial case, that $\mathbf{u}_{1},\mathbf{u}_{2}$
are not both zero. If $\mathbb{I}_{\mathbf{a}}$ is convex, then the
vector 
\begin{equation}
\frac{F\left(\mathbf{a,u}_{1}\right)}{F\left(\mathbf{a,u}_{1}\right)+F\left(\mathbf{a,u}_{2}\right)}\mathbf{1}\left(\mathbf{a,u}_{1}\right)+\frac{F\left(\mathbf{a,u}_{2}\right)}{F\left(\mathbf{a,u}_{1}\right)+F\left(\mathbf{a,u}_{2}\right)}\mathbf{1}\left(\mathbf{a,u}_{2}\right)\in\mathbb{I}_{\mathbf{a}}.\label{eq:belongs to}
\end{equation}
This is equivalent to 
\[
F\left(\mathbf{a},\frac{F\left(\mathbf{a,u}_{1}\right)}{F\left(\mathbf{a,u}_{1}\right)+F\left(\mathbf{a,u}_{2}\right)}\mathbf{1}\left(\mathbf{a,u}_{1}\right)+\frac{F\left(\mathbf{a,u}_{2}\right)}{F\left(\mathbf{a,u}_{1}\right)+F\left(\mathbf{a,u}_{2}\right)}\mathbf{1}\left(\mathbf{a,u}_{2}\right)\right)\leq1.
\]
But the lefthand side expression equals 
\[
\frac{F\left(\mathbf{a,u}_{1}+\mathbf{u}_{2}\right)}{F\left(\mathbf{a,u}_{1}\right)+F\left(\mathbf{a,u}_{2}\right)},
\]
whence we see that $F\left(\mathbf{a,u}\right)$ is convex. Conversely,
if the expression above is $\leq1$, then (\ref{eq:belongs to}) holds.
Since it holds for any $\mathbf{u}_{1},\mathbf{u}_{2}$, it also holds
for $\lambda\mathbf{u}_{1},\left(1-\lambda\right)\mathbf{u}_{2}$
for $0\leq\lambda\leq1$. But, as $\lambda$ changes from 0 to 1,
the expression 
\[
\frac{F\left(\mathbf{a,\lambda u}_{1}\right)}{F\left(\mathbf{a,\lambda u}_{1}\right)+F\left(\mathbf{a},\mathbf{u}_{2}\right)}=\frac{\lambda F\left(\mathbf{a,u}_{1}\right)}{\lambda F\left(\mathbf{a,\lambda u}_{1}\right)+\left(1-\lambda\right)F\left(\mathbf{a,\left(1-\lambda\right)u}_{2}\right)}
\]
runs through all values from 0 to 1 too. Since 
\[
\mathbf{1}\left(\mathbf{a,\lambda u}_{1}\right)=\mathbf{1}\left(\mathbf{a,u}_{1}\right),\mathbf{1}\left(\mathbf{a},\left(1-\lambda\right)\mathbf{u}_{2}\right)=\mathbf{1}\left(\mathbf{a,u}_{2}\right),
\]
we have 
\[
\theta\mathbf{1}\left(\mathbf{a,u}_{1}\right)+\left(1-\theta\right)\mathbf{1}\left(\mathbf{a,u}_{2}\right)\in\mathbb{I}_{\mathbf{a}},
\]
for any $0\leq\theta\leq1$. This means that $\mathbb{I}_{\mathbf{a}}$
is convex, and we have proved 
\begin{thm}
$F\left(\mathbf{a,u}\right)$ is convex if and only if the body of
the associated indicatrix $\mathbb{I}_{\mathbf{a}}$ at any point
$\mathbf{a}$ is convex. 
\end{thm}

From this and Theorem \ref{thm:minFconvI} we immediately have 
\begin{cor}
For every submetric function $F$,

(i) the corresponding minimal submetric function\index{submetric function!minimal}
$\widehat{F}$ is convex,

(ii) $F\equiv\widehat{F}$ if and only if $F$ is convex. 
\end{cor}

We also have 
\begin{cor}
If a submetric function $F$ is convex, then $\left\{ \mathbf{u}\right\} $
is a minimizing vector chain for any line element $\left(\mathbf{a,u}\right)\in\mathbb{T}$. 
\end{cor}

This follows from $F\left(\mathbf{a,u}\right)=\widehat{F}\left(\mathbf{a,u}\right)$.

Of course, if $F$ is convex, the following are also minimizing vector
chains for $\mathbf{\mathbf{u}}\in\mathbb{U}^{n}$ : $\left\{ \frac{1}{2}\mathbf{u},\frac{1}{2}\mathbf{u}\right\} $,
$\left\{ \frac{1}{3}\mathbf{u},\frac{2}{3}\mathbf{u}\right\} $,$\left\{ \frac{1}{n}\mathbf{u},\ldots,\frac{1}{n}\mathbf{u}\right\} $,
etc. Moreover, if $F$ is not strictly convex (i.e., the inequality
in Definition \ref{def:convexF} may be equality for some $\mathbf{u}_{1},\mathbf{u}_{2}$),
there may very well be minimizing chains involving vectors that are
not collinear with $\mathbf{u}$.

We have now arrived at one of the central theorems in the theory. 
\begin{thm}
\label{thm:G(x,x+us)/s}The distance $G\left(\mathbf{x},\mathbf{x}+\mathbf{u}s\right)$
is differentiable at $s=0+$ for any $\left(\mathbf{x},\mathbf{u}\right)\in\mathbb{T}$,
and 
\[
\left.\frac{\mathrm{d}G\left(\mathbf{x},\mathbf{x}+\mathbf{u}s\right)}{\mathrm{d}s+}\right|_{s=0}=\lim_{s\rightarrow0+}\frac{G\left(\mathbf{x},\mathbf{x}+\mathbf{u}s\right)}{s}=\widehat{F}\left(\mathbf{x},\mathbf{u}\right).
\]
\end{thm}

See Appendix for a proof.

An important corollary to this theorem is as follows. Let $\mathbf{f}|\left[a,b\right]$
be a continuously differentiable path. Consider 
\[
\frac{G\mathbf{\mathbf{f}}\left(t\right)\mathbf{\mathbf{f}}\left(\tau\right)}{\widehat{F}\left(\mathbf{\mathbf{f}}\left(t\right),\mathbf{\mathbf{f}}\left(\tau\right)\mathbf{-\mathbf{f}}\left(t\right)\right)},t<\tau.
\]
By presenting it 
\[
\frac{G\mathbf{x}\left(t\right)\left(\mathbf{\mathbf{f}}\left(t\right)+\frac{\mathbf{\mathbf{f}}\left(\tau\right)-\mathbf{\mathbf{f}}\left(t\right)}{\tau-t}\left(\tau-t\right)\right)}{\widehat{F}\left(\mathbf{\mathbf{f}}\left(t\right),\frac{\mathbf{\mathbf{f}}\left(\tau\right)-\mathbf{x}\left(t\right)}{\tau-t}\mathbf{f}\left(\tau-t\right)\right)}=\frac{G\mathbf{x}\left(t\right)\left(\mathbf{\mathbf{f}}\left(t\right)+\mathbf{\dot{\mathbf{f}}}\left(\theta\right)\left(\tau-t\right)\right)}{\widehat{F}\left(\mathbf{\mathbf{f}}\left(t\right),\mathbf{\dot{\mathbf{f}}}\left(\theta\right)\mathbf{f}\left(\tau-t\right)\right)},
\]
with $t\leq\theta\leq\tau$, we see that if $\tau-t\rightarrow0+$
on $\left[a,b\right]$, the ratio tends to 1 (by Theorem \ref{thm:G(x,x+us)/s}
and because all functions involved are uniformly continuous on $\left[a,b\right]$).
This establishes 
\begin{cor}
\label{cor:arc to distance}For any smooth path $\mathbf{f}|\left[a,b\right]$
and $\left[t,\tau\right]\subset\left[a,b\right]$, 
\[
\lim_{\tau-t\rightarrow0+}\frac{G\mathbf{\mathbf{f}}\left(t\right)\mathbf{\mathbf{f}}\left(\tau\right)}{\widehat{F}\left(\mathbf{\mathbf{f}}\left(t\right),\mathbf{\mathbf{f}}\left(\tau\right)\mathbf{-\mathbf{f}}\left(t\right)\right)}=1.
\]
\end{cor}

We are ready now to formulate the standard differential-geometric
computation of the length of a continuously differentiable path by
integration of the submetric function\index{submetric function} applied
to its points and tangents. 
\begin{thm}
\label{thm:StandardComputation}For any continuously differentiable
path $\mathbf{f}|\left[a,b\right]$, 
\[
D\mathbf{f}|\left[a,b\right]=\int_{a}^{b}\widehat{F}\left(\mathbf{f}\left(t\right),\mathbf{\dot{f}}\left(t\right)\right)\mathrm{d}t.
\]
\end{thm}

Indeed, by definition, 
\[
D\mathbf{f}|\left[a,b\right]=\lim_{\delta\mu\rightarrow0}\sum G\mathbf{f}\left(t_{i}\right)\mathbf{f}\left(t_{i+1}\right)
\]
across all nets $\mu=\left\{ ...,t_{i},t_{i+1}...\right\} $ partitioning
$\left[a,b\right]$. This limit can be presented as 
\[
\lim_{\delta\mu\rightarrow0}\sum\widehat{F}\left(\mathbf{f}\left(t_{i}\right),\mathbf{f}\left(t_{i+1}\right)-\mathbf{f}\left(t_{i}\right)\right)\frac{G\mathbf{f}\left(t_{i}\right)\mathbf{f}\left(t_{i+1}\right)}{\widehat{F}\left(\mathbf{f}\left(t_{i}\right),\mathbf{f}\left(t_{i+1}\right)-\mathbf{f}\left(t_{i}\right)\right)}.
\]
By Corollary \ref{cor:arc to distance}, 
\[
\lim_{\delta\mu\rightarrow0}\frac{G\mathbf{f}\left(t_{i}\right)\mathbf{f}\left(t_{i+1}\right)}{\widehat{F}\left(\mathbf{f}\left(t_{i}\right),\mathbf{f}\left(t_{i+1}\right)-\mathbf{f}\left(t_{i}\right)\right)}=1.
\]
Then 
\begin{multline*}
D\mathbf{f}|\left[a,b\right]=\lim_{\delta\mu\rightarrow0}\sum\widehat{F}\left(\mathbf{f}\left(t_{i}\right),\mathbf{f}\left(t_{i+1}\right)-\mathbf{f}\left(t_{i}\right)\right)\\
=\lim_{\delta\mu\rightarrow0}\sum\widehat{F}\left(\mathbf{f}\left(t_{i}\right),\frac{\mathbf{f}\left(t_{i+1}\right)-\mathbf{f}\left(t_{i}\right)}{t_{i+1}-t_{i}}\right)\left(t_{i+1}-t_{i}\right).
\end{multline*}
But 
\[
\lim_{\delta\mu\rightarrow0}\widehat{F}\left(\mathbf{f}\left(t_{i}\right),\frac{\mathbf{f}\left(t_{i+1}\right)-\mathbf{f}\left(t_{i}\right)}{t_{i+1}-t_{i}}\right)=\widehat{F}\left(\mathbf{f}\left(t\right),\mathbf{\dot{f}}\left(t\right)\right)
\]
and $\widehat{F}\left(\mathbf{f}\left(t\right),\mathbf{\dot{f}}\left(t\right)\right)$
is uniformly continuous on $\left[a,b\right]$. Hence 
\[
D\mathbf{f}|\left[a,b\right]=\lim_{\delta\mu\rightarrow0}\sum\widehat{F}\left(\mathbf{f}\left(t_{i}\right),\mathbf{\dot{f}}\left(t_{i}\right)\right)\left(t_{i+1}-t_{i}\right)=\int_{a}^{b}\widehat{F}\left(\mathbf{f}\left(t\right),\mathbf{\dot{f}}\left(t\right)\right)\mathrm{d}t,
\]
completing the proof.

Since 
\[
\lim_{\tau-t\rightarrow0+}\frac{\int_{t}^{\tau}\widehat{F}\left(\mathbf{f}\left(x\right),\mathbf{\dot{f}}\left(x\right)\right)\mathrm{d}x}{\widehat{F}\left(\mathbf{f}\left(t\right),\frac{\mathbf{f}\left(\tau\right)-\mathbf{f}\left(t\right)}{\tau-t}\right)\left(\tau-t\right)}=1,
\]
we also have 
\begin{cor}
For any continuously differentiable path $\mathbf{f}|\left[a,b\right]$,
and $\left[t,\tau\right]\subset\left[a,b\right]$, 
\[
\lim_{\tau-t\rightarrow0+}\frac{G\mathbf{\mathbf{f}}\left(t\right)\mathbf{\mathbf{f}}\left(\tau\right)}{D\mathbf{f}|\left[t,\tau\right]}=1.
\]
\end{cor}

\subsection{Continuously differentiable paths and intrinsic metric $G$}

\index{metric!intrinsic} Before proceeding, we need an auxiliary
observation. The space $\left(\mathfrak{S},E\right)$ being open,
each point $\mathbf{p}$ in $\mathfrak{S}$ can be enclosed in a compact
Euclidean ball 
\[
\mathfrak{B}\left(\mathbf{p},r\right)=\left\{ \mathbf{x}\in\mathbb{R}^{n}:\left|\mathbf{x}-\mathbf{p}\right|\leq r\right\} \subseteq\mathfrak{S},
\]
and we can associate with any $\mathbf{p}$ the ball $\mathfrak{B}\left(\mathbf{p},r\right)$
with the supremal value of $r_{\sup}\left(\mathbf{p}\right)$ (including
$\infty$). The observation is that, given any compact subset $\mathfrak{s}$
of $\mathfrak{S}$, 
\[
\inf_{\mathbf{p}\in\mathfrak{s}}r_{\sup}\left(\mathbf{p}\right)=\min_{\mathbf{p}\in\mathfrak{s}}r_{\sup}\left(\mathbf{p}\right)>0.
\]

A \emph{straight line segment} is defined as 
\[
\mathbf{s}\left(x\right)=\mathbf{a+u}x,\qquad x\in\left[a,b\right],\left(\mathbf{a},\mathbf{u}\right)\in\mathbb{T}.
\]
If $\mathbf{x}$ and $\mathbf{y}$ are within any ball $\mathfrak{B}\left(\mathbf{p},r\right)$
they can be connected by the straight line segment 
\[
\mathbf{s}\left(x\right)=\mathbf{x}+\frac{\mathbf{y}-\mathbf{x}}{b-a}\left(x-a\right),\qquad x\in\left[a,b\right].
\]
Concatenations of straight line segments forms piecewise linear paths,
about which we have the following result.
\begin{thm}
\label{thm:MDFS}For every path $\mathbf{h}|\left[a,b\right]$ connecting
$\mathbf{a}$ to $\mathbf{b}$ one can find a piecewise linear path
from $\mathbf{a}$ to $\mathbf{b}$ which is arbitrarily close to
$\mathbf{h}|\left[a,b\right]$ pointwise and in its length. 
\end{thm}

See Appendix for a proof.

\begin{figure}[ptbh]
\begin{centering}
\includegraphics[scale=0.2]{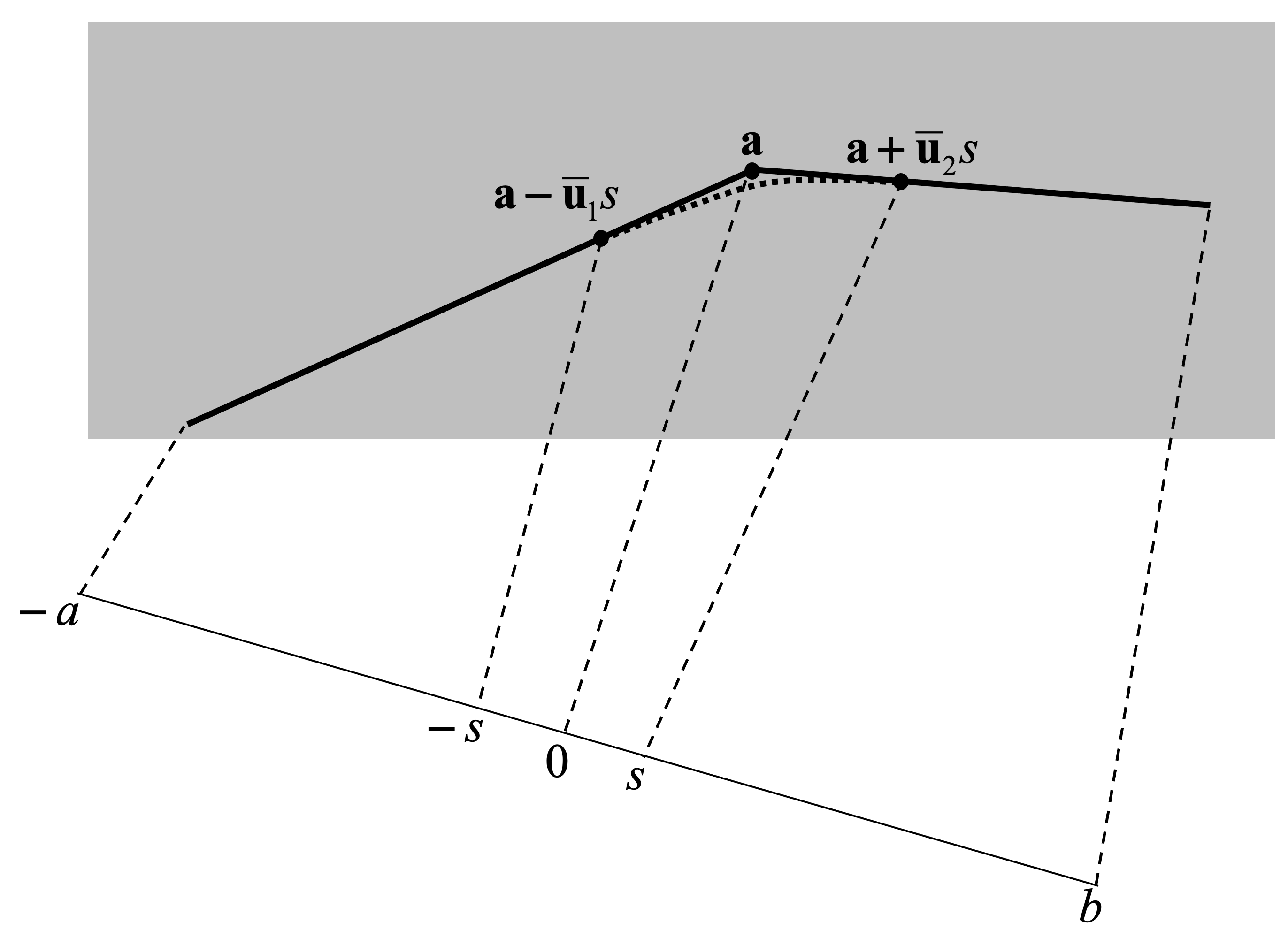} 
\par\end{centering}
\caption{An illustration for the corner-rounding procedure. The piecewise linear
path is shown as a mapping of the interval $\left[-a,b\right]$ into
Euclidean plane (gray area). At $0$ the two segments meet, and around
this point they are replaced by the path shown by the dotted line
of an arbitrarily close length.}

\centering{}\label{FigCornerRounding} 
\end{figure}

The straight-line segments are not indispensable in such an approximation.
In fact, we can use the following ``corner-rounding'' procedure
to replace any piecewise linear path with a continuously differentiable
path. It is illustrated in Figure \ref{FigCornerRounding}.

Let two adjacent straight line segments be presented as 
\[
\mathbf{p}\left(t\right)=\left\{ \begin{array}{cc}
\mathbf{a}+\overline{\mathbf{u}}_{1}t & \text{if }t\in\left[-a,0\right]\\
\mathbf{a}+\overline{\mathbf{u}}_{2}t & \text{if }t\in\left[0,b\right]
\end{array}\right.,
\]
with $a,b>0.$ On a small interval $\left[-s,s\right]$, 
\[
D\mathbf{p}|\left[-s,s\right]=\int_{-s}^{0}\widehat{F}\left(\mathbf{a}+\overline{\mathbf{u}}_{1}t,\overline{\mathbf{u}}_{1}\right)\mathrm{d}t+\int_{0}^{s}\widehat{F}\left(\mathbf{a}+\overline{\mathbf{u}}_{2}t,\overline{\mathbf{u}}_{2}\right)\mathrm{d}t.
\]
Corner-rounding consists in replacing $\mathbf{p}|\left[-s,s\right]$
with a continuously differentiable path 
\begin{equation}
\begin{array}{cc}
\mathbf{q}\left(t\right)=\mathbf{a+u}\left(t\right)t, & t\in\left[-s,s\right],\end{array}\label{eq:C1 curve}
\end{equation}
such that 
\begin{equation}
\begin{array}{l}
\mathbf{u}\left(-s\right)=\overline{\mathbf{u}}_{1},\mathbf{u}\left(s\right)=\overline{\mathbf{u}}_{2}\\
\mathbf{\dot{u}}\left(-s\right)=\mathbf{\dot{u}}\left(s\right)=\mathbf{0}
\end{array}\label{eq:Const1}
\end{equation}
and 
\begin{equation}
\lim_{s\rightarrow0+}D\mathbf{x}|\left[-s,s\right]=0.\label{eq:Const2}
\end{equation}
The requirements (\ref{eq:Const1}) ensure that the modified path
$\mathbf{r}|\left[-a,b\right]$ defined by 
\[
\mathbf{r}\left(t\right)=\left\{ \begin{array}{cl}
\mathbf{p}\left(t\right) & \text{if }t\not\in\left[-s,s\right]\\
\mathbf{q}\left(t\right) & \text{if }t\in\left[-s,s\right]
\end{array}\right.
\]
is continuously differentiable. The requirement (\ref{eq:Const2})
ensures that the difference 
\[
\left|D\mathbf{p}|\left[-a,b\right]-D\mathbf{r}|\left[-a,b\right]\right|
\]
can be made arbitrarily small by choosing $s$ sufficiently small.
One example of (\ref{eq:C1 curve}) is given by 
\[
\mathbf{u}\left(t\right)=\frac{\overline{\mathbf{u}}_{1}+\overline{\mathbf{u}}_{2}}{2}+\frac{\left(\frac{t}{s}\right)^{3}-3\left(\frac{t}{s}\right)}{4}\left(\overline{\mathbf{u}}_{1}-\overline{\mathbf{u}}_{2}\right).
\]
We can now reformulate Theorem \ref{thm:MDFS} as follows. 
\begin{thm}
\label{thm:MDFS-general}For every path $\mathbf{h}|\left[a,b\right]$
connecting $\mathbf{a}$ to $\mathbf{b}$ one can find a continuously
differentiable path from $\mathbf{a}$ to $\mathbf{b}$ which is arbitrarily
close to $\mathbf{h}|\left[a,b\right]$ pointwise and in its length. 
\end{thm}

As an immediate consequence, we have the following. 
\begin{thm}
If $G$ in $\left(\mathfrak{S},D\right)$ is an intrinsic metric\index{metric!intrinsic},
then, for any $\mathbf{a},\mathbf{b}$ in $\mathfrak{S}$, 
\[
G\mathbf{ab}=\inf\int_{a}^{b}\widehat{F}\left(\mathbf{f}\left(t\right),\mathbf{\dot{f}}\left(t\right)\right)\mathrm{d}t,
\]
where the infimum is taken across all continuously differentiable
paths (or piecewise continuously differentiable, if more convenient)
connecting $\mathbf{a}$ to $\mathbf{b}$. 
\end{thm}

Recall that $G$ is defined as intrinsic $G\mathbf{ab}$ is an infimum
of the length of all paths connecting $\mathbf{a}$ to $\mathbf{b}$.
This property is not derivable from the assumptions $\mathcal{E}1$
and $\mathcal{E}2$ we made about the relationship between $\left(\mathfrak{S},D\right)$
and $\left(\mathfrak{S},E\right)$. It should therefore be stipulated
as an additional assumption or derived from other additional assumptions,
e.g., that $\left(\mathfrak{S},D\right)$ is a complete space with
intermediate points.

\section{\label{sec:Data-analytic}Dissimilarity cumulation: Extensions and
applications}

In this section we give a few examples of extensions of the dissimilarity
cumulation theory aimed at broadening the scope of its applicability.

\subsection{\label{subsec: Sorit}Example 1: Observational sorites ``paradox''}

\index{sorites paradox} The issue of pairwise discrimination is the
main application of Fechnerian Scaling and the original motivation
for its development. As we know from Sections \ref{sec:Observation-areas}
and \ref{sec:Same-different-judgments}, it is a fundamental fact
that two stimuli being compared must belong to distinct observation
areas, say, one being on the left and the other on the right in visual
field, or one being first and the other second in time. Without this
one would not be able to speak, e.g., of a stimulus with value $\mathbf{x}$
being compared to a stimulus with the same value, because then we
would simply have a single stimulus. Similarly, without the distinct
observation areas there would be no operational meaning in distinguishing
$\left(\mathbf{x},\mathbf{y}\right)$ from $\left(\mathbf{y},\mathbf{x}\right)$.
Throughout this chapter the observation areas in our notation were
implicit: e.g., we assumed that the stimulus written first in $\left(\mathbf{x},\mathbf{y}\right)$
belongs to the first observation area\index{observation area}, or
that $\mathbf{x}$ always denotes a stimulus in the first observation
area. Here, however, we will need to indicate observation areas explicitly:
$\mathbf{v}^{\left(o\right)}$ means a stimulus with value $\mathbf{v}$
in observation area $o$. If we assume that the observation areas
are fixed, we can denote them $1$ and $2$, so that every value $\mathbf{v}$
may be part of the stimuli $\mathbf{v}^{\left(1\right)}$ and $\mathbf{v}^{\left(2\right)}$.
Note that with this notation any pair $\left\{ \mathbf{x}^{\left(1\right)},\mathbf{y}^{\left(2\right)}\right\} $
can be considered unordered, because $\left\{ \mathbf{y}^{\left(2\right)},\mathbf{x}^{\left(1\right)}\right\} $
represents the same pair.

There is an apparent ``paradox'' related to pairwise comparisons
that seems so compelling that many describe it as a well-known empirical
fact. Quoting from R. Duncan Luce (1956): 
\begin{quote}
It is certainly well known from psychophysics that if ``preference''
is taken to mean which of two weights a person believes to be heavier
after hefting them, and if ``adjacent'' weights are properly chosen,
say a gram difference in a total weight of many grams, then a subject
will be indifferent between any two ``adjacent'' weights. If indifference
were transitive, then he would be unable to detect any weight differences,
however great, which is patently false. 
\end{quote}
In other words, one can have a sequence of weights in which every
two successive weights subjectively match each other, but the first
and the last one do not. In philosophy, this seemingly paradoxical
situation is referred to as \emph{observational sorites}. The term
``sorites'' means ``heap'' in Greek, and the paradox is traced
back to the Greek philosopher Eubulides (4th century BCE). In fact,
Eubulides dealt with another form of the paradox, one in which stimuli
are mapped into one of two categories one at a time. This form of
sorites requires a different analysis. In our case, we have pairs
of stimuli mapped into categories ``match'' or ``do not match.''
The resolution of this paradox is based on two considerations: 
\begin{enumerate}
\item The relationship ``$\mathbf{x}^{\left(1\right)}$ matches $\mathbf{y}^{\left(2\right)}$''
(or vice versa) is computed from an ensemble of responses rather than
observed as an individual response. Individual responses to the same
pair $\left\{ \mathbf{x}^{\left(1\right)},\mathbf{y}^{\left(2\right)}\right\} $
vary, and the pair can only be associated to a probability of a response,
say, 
\begin{equation}
\psi^{*}\left(\mathbf{x}^{\left(1\right)},\mathbf{y}^{\left(2\right)}\right)=\Pr\left[\mathbf{x}^{\left(1\right)}\textnormal{ is judged to be different from }\mathbf{y}^{\left(2\right)}\right].\label{eq:psistar12}
\end{equation}
\item Stimuli $\mathbf{v}^{\left(1\right)}$ and $\mathbf{v}^{\left(2\right)}$
have the same value $\mathbf{v}$ but they are different. To repeat
the same stimulus, it should be presented in the same observation
area in addition to having the same value. 
\end{enumerate}
Applying these considerations to the above quotation from Luce, let
\[
w_{1},w_{2},w_{3},w_{4},\ldots,w_{n}
\]
be the sequence of weights about which Luce (and many others) think
as one in which $w_{k-1}$ and $w_{k}$ match (for $k=2,\ldots,n$)
but $w_{1}$ and $w_{n}$ do not. Such a sequence is called a \emph{(comparative)
soritical sequence}\index{soritical sequence}. Let us, however, assign
the weights to observation areas, as they should be. One can, e.g.,
place one weight in an observer's left hand and another weight in
her right hand to be hefted simultaneously, in which case $w^{\left(1\right)}=w^{\left(left\right)}$
and $w^{\left(2\right)}=w^{\left(right\right)}$. Or the observer
can heft one weight first and the other weight after a short interval,
in which case $w^{\left(1\right)}=w^{\left(first\right)}$ and $w^{\left(2\right)}=w^{\left(second\right)}$.
Whichever the case, since two adjacent weights in our sequence are
to be compared, they should belong to different observation areas,
\[
w_{1}^{\left(1\right)},w_{2}^{\left(2\right)},w_{3}^{\left(1\right)},w_{4}^{\left(2\right)},\ldots,w_{n}^{\left(2\right)}.
\]
The last and the first stimuli also should belong to different observation
areas if they are to be compared, so $n$ must be an even number.
Assuming that the discrimination here is of the ``greater-less''
variety, we have a function 
\[
\gamma\left(x^{\left(1\right)},y^{\left(2\right)}\right)=\Pr\left[x^{\left(1\right)}\textnormal{ is judged to be lighter than }y^{\left(2\right)}\right],
\]
and the match is determined by 
\[
\gamma\left(x^{\left(1\right)},y^{\left(2\right)}\right)=\frac{1}{2}.
\]
So we have 
\[
\gamma\left(w_{1}^{\left(1\right)},w_{2}^{\left(2\right)}\right)=\gamma\left(w_{2}^{\left(2\right)},w_{3}^{\left(1\right)}\right)=\gamma\left(w_{3}^{\left(1\right)},w_{4}^{\left(2\right)}\right)=\ldots=\gamma\left(w_{n-1}^{\left(1\right)},w_{n}^{\left(2\right)}\right)=\frac{1}{2}.
\]
It is not obvious now that we can have $w_{1}<w_{2}<w_{3}<w_{4}<\ldots<w_{n}$.
In fact, if we accept the usual model of a psychometric function $\gamma$,
as in Figure \ref{fig:ConstErrorGamma} and \ref{fig:RegMediality},
$w_{k}$ is uniquely determined as a match for $w_{k-1}$, and, moreover,
\[
\begin{array}{l}
w_{1}^{\left(1\right)}=w_{3}^{\left(1\right)}=\ldots=w_{n-1}^{\left(1\right)},\\
w_{2}^{\left(2\right)}=w_{4}^{\left(2\right)}=\ldots=w_{n}^{\left(2\right)}.
\end{array}
\]
The sequence clearly is not soritical, because $w_{1}^{\left(1\right)}$
and $w_{n}^{\left(2\right)}$ (for an even $n$) necessarily match.

Generalizing, if one explicitly considers observation areas as part
of stimuli's identity, the idea of soritical sequences becomes unfounded.
If one further accepts the principles stipulated in Section \ref{sec:Observation-areas},
enabling one to construct a canonical space $\left(\mathfrak{S},D\right)$,
then soritical sequences become impossible. Essentially we are dealing
with the problem of a reasonable definition of a match (PSE). We outline
below an axiomatic scheme that defines stimulus spaces in which soritical
sequences are impossible.

Not to be constrained to just two fixed observation areas, we consider
a union of stimulus spaces indexed by observation areas: 
\[
\mathcal{S}=\bigcup_{\alpha\in\Omega}\mathfrak{S}_{\omega}^{*}.
\]
We indicate the elements of $\mathfrak{S}_{\omega}^{*}$ by the corresponding
superscript, say $\mathbf{x}^{(\omega)}$. The set $\mathcal{S}$
is endowed with a binary relation $\mathbf{x}^{(\alpha)}\mathrm{M}\mathbf{y}^{(\beta)}$
(read as ``$\mathbf{x}$ in $\alpha$ is matched by $\mathbf{y}$
in $\beta$''). The most basic property of $\mathrm{M}$ is 
\begin{equation}
\mathbf{x}^{(\alpha)}\mathrm{M}\mathbf{y}^{(\beta)}\Longrightarrow\alpha\neq\beta.\label{EQ:basicM}
\end{equation}

\begin{defn}
Given a space $(\mathcal{S},\mathrm{M})$, we call a sequence $\mathbf{x}_{1}^{(\omega_{1})},\ldots,\mathbf{x}_{n}^{(\omega_{n})}$
\emph{well-matched} if 
\begin{equation}
\omega_{i}\neq\omega_{j}\Longrightarrow\mathbf{x}_{i}^{(\omega_{i})}\mathrm{M}\mathbf{x}_{j}^{(\omega_{j})}
\end{equation}
for all $i,j\in\{1,\ldots n\}$. The stimulus space $(\mathcal{S},\mathrm{M})$
is \emph{well-matched}\index{stimulus space!well-matched} if, for
any sequence $\alpha,\beta,\gamma\in\Omega$ and any $\mathbf{a}^{(\alpha)}\in\mathcal{S}$,
there is a well-matched sequence $\mathbf{a}^{(\alpha)},\mathbf{b}^{(\beta)},\mathbf{c}^{(\gamma)}$. 
\end{defn}

In particular, in a well-matched space, for any $\mathbf{a}^{(\alpha)}$
and any $\beta\in\Omega$, one can find a $\mathbf{b}^{(\beta)}\in\mathcal{S}$
such that $\mathbf{a}^{(\alpha)}\mathrm{M}\mathbf{b}^{(\beta)}$ and
$\mathbf{b}^{(\beta)}\mathrm{M}\mathbf{a}^{(\alpha)}$. 
\begin{defn}
Two stimuli $\mathbf{a}^{(\omega)},\mathbf{b}^{(\omega)}$ in $(\mathcal{S},\mathrm{M})$
are called \emph{equivalent}, in symbols $\mathbf{a}^{(\omega)}\mathrm{E}\mathbf{b}^{(\omega)}$,
if for any $\mathbf{c}^{(\iota)}\in\mathcal{S}$, 
\begin{equation}
\mathbf{c}^{(\iota)}\mathrm{M}\mathbf{a}^{(\omega)}\iff\mathbf{c}^{(\iota)}\mathrm{M}\mathbf{b}^{(\omega)}.
\end{equation}
$(\mathcal{S},\mathrm{M})$ is a \emph{regular space} if, for any
$\mathbf{a}^{(\omega)},\mathbf{b}^{(\omega)},\mathbf{c}^{(\omega^{\prime})}\in\mathcal{S}$
with $\omega\neq\omega^{\prime}$, 
\begin{equation}
\mathbf{a}^{(\omega)}\mathrm{M}\mathbf{c}^{(\omega^{\prime})}\wedge\mathbf{b}^{(\omega)}\mathrm{M}\mathbf{c}^{(\omega^{\prime})}\Longrightarrow\mathbf{a}^{(\omega)}\mathrm{E}\mathbf{b}^{(\omega)}.
\end{equation}
\end{defn}

This is a generalization of the notion of psychological equality introduced
in Section \ref{sec:Observation-areas}. 
\begin{defn}
Given a space $(\mathcal{S},\mathrm{M})$, a sequence $\mathbf{x}_{1}^{(\omega_{1})},\ldots,\mathbf{x}_{n}^{(\omega_{n})}$
with $\mathbf{x}_{i}^{(\omega_{i})}\in\mathcal{S}$ for $i=1,\ldots,n$,
is called \emph{soritical} if 
\end{defn}

\begin{enumerate}
\item $\mathbf{x}_{i}^{(\omega_{i})}\mathrm{M}\mathbf{x}_{i+1}^{(\omega_{i+1})}$
for $i=1,\ldots,n-1,$ 
\item $\omega_{1}\neq\omega_{n},$ 
\item but it is not true that $\mathbf{x}_{1}^{(\omega_{1})}\mathrm{M}\mathbf{x}_{n}^{(\omega_{n})}.$ 
\end{enumerate}
Well-matchedness and regularity can be shown to be independent properties.
Our interest is in the spaces that are both regular and well-matched.
It can be proved that 
\begin{thm}
\label{thm:DzhDzh}In a regular well-matched space it is impossible
to form a soritical sequence. 
\end{thm}

\subsection{\label{subsec:Thurstonian-type}Example 2: Thurstonian-type representations}

\index{Thurstonian model}\index{same-different judgments} Consider
now the special case of the regular well-matched spaces, when the
matching (PSE) relation is defined through minima of a same-different
discrimination probability function $\psi^{\star}:\mathfrak{S}_{1}^{\star}\times\mathfrak{S}_{2}^{\star}\rightarrow\left[0,1\right]$
in (\ref{eq:psistar12}). The issue discussed in this example is $\psi^{\star}$
can be ``explained'' by a \emph{random-utility} (or \emph{Thurstonian})
\emph{model}, according to which each stimulus is mapped into a random
variable in some perceptual space, and the decision ``same'' or
``different'' is determined by the values of these random variables
for the stimuli $\mathbf{x}^{\left(1\right)}$ and $\mathbf{y}^{\left(2\right)}$.

Let us assume that both $\mathfrak{S}_{1}^{\star}\times\mathfrak{S}_{2}^{\star}$
are open connected regions of $\mathbb{R}^{n}$, and we present the
property of Regular Minimality (\ref{eq:RegMinLaw}) in the following
special form: there is a homeomorphism $\mathbf{h}:\mathfrak{S}_{1}^{\star}\rightarrow\mathfrak{S}_{2}^{\star}$
(a continuous function with a continuous $\mathbf{h}^{-1}$) such
that 
\begin{equation}
\left\{ \begin{array}{l}
\arg\min_{\mathbf{y}}\psi^{\star}\left(\mathbf{x},\mathbf{y}\right)=\mathbf{h}\left(\mathbf{x}\right),\\
\arg\min_{\mathbf{x}}\psi^{\star}\left(\mathbf{x},\mathbf{y}\right)=\mathbf{h}^{-1}\left(\mathbf{y}\right).
\end{array}\right.\label{eq:RegMinAgain}
\end{equation}
Here we once again drop the superscripts in $\mathbf{x}^{\left(1\right)}$
and $\mathbf{y}^{\left(2\right)}$. The function $\arg\min_{a_{i}}f\left(a_{1},\ldots,a_{n}\right)$
indicate the value of the argument $a_{i}$ at which $f$ reaches
its minimum (at fixed values of the remaining arguments). Empirical
studies show that generally the \emph{minimum-level function} \emph{$\psi^{\star}\left(\mathbf{x},\mathbf{\mathbf{h}\left(\mathbf{x}\right)}\right)$
varies with $\mathbf{x}$,} 
\begin{equation}
\psi^{\star}\left(\mathbf{x},\mathbf{\mathbf{h}\left(\mathbf{x}\right)}\right)\not=const.\label{eq:NCSD}
\end{equation}
Equivalently written, 
\[
\psi^{\star}\left(\mathbf{\mathbf{h}^{-1}\left(\mathbf{y}\right)},\mathbf{y}\right)\not=const.
\]
We call this property \emph{nonconstant self-dissimilarity} \index{nonconstant self-dissimilarity}
of $\psi^{\star}$.

\begin{figure}
\begin{centering}
\includegraphics[scale=0.25]{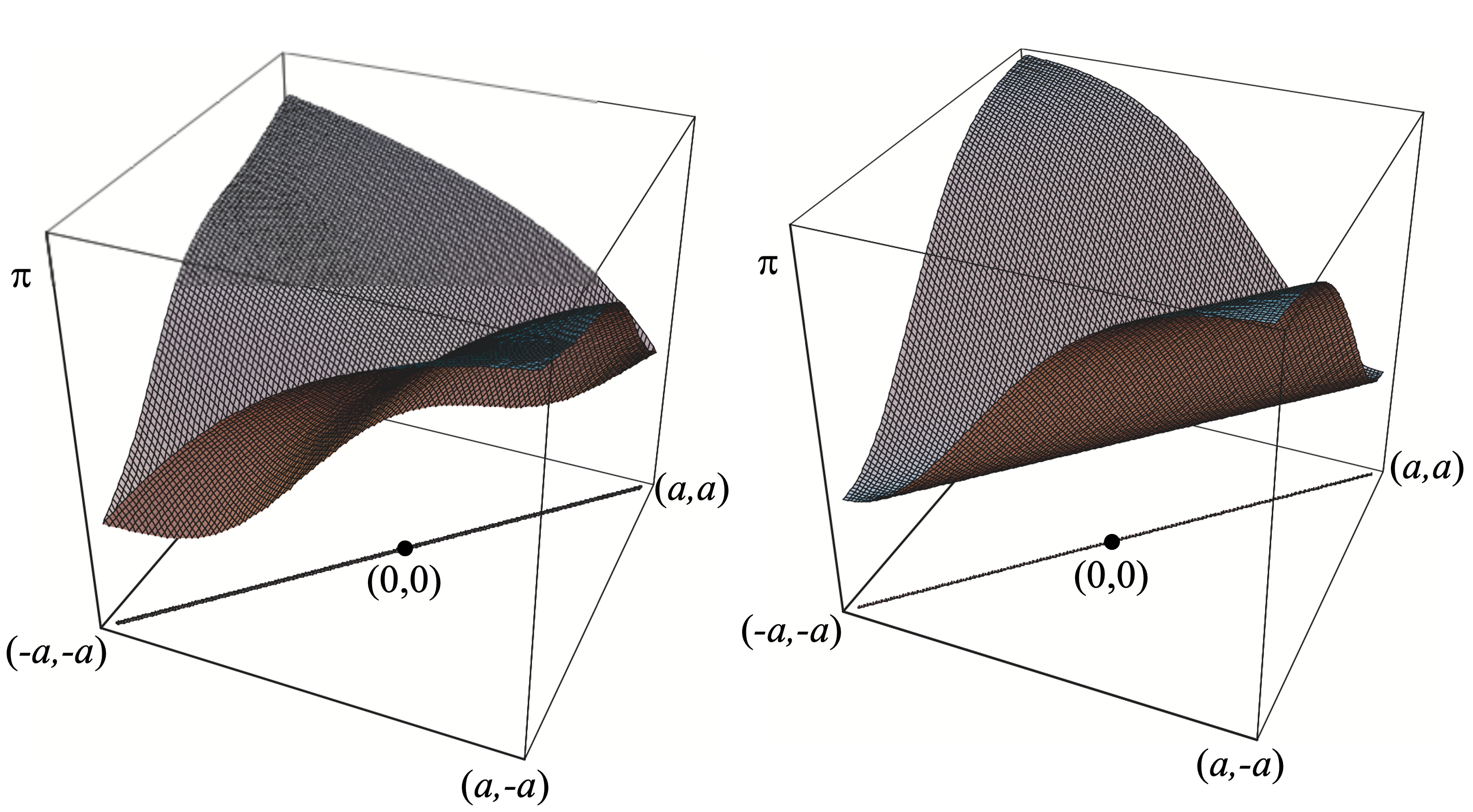} 
\par\end{centering}
\caption{\label{fig:patches} A typical patch (left) and an atypical patch
(right) on a small square $\left[-a,a\right]^{2}$.}
\end{figure}

Rather than using Regular Minimality (\ref{eq:RegMinAgain}) to bring
the stimulus space\index{stimulus space} to a canonical form, we
will consider the following construction. Consider a point $\left(\mathbf{p},\mathbf{\mathbf{h}\left(\mathbf{p}\right)}\right)$
in $\mathfrak{S}_{1}^{\star}\times\mathfrak{S}_{2}^{\star}$ and a
direction $\mathbf{u}$ in

\[
\mathbb{U}^{n}=\left\{ \mathbf{u}=\mathbf{x}-\mathbf{p}:\mathbf{x}\in\mathbb{R}^{n},\mathbf{x}\neq\mathbf{p}\right\} .
\]
For $\left(x,y\right)\in\left[-a,a\right]^{2}$, where $a$ is a small
positive number, the function 
\[
\lambda\left(x,y\right)=\psi^{\star}\left(\mathbf{p}+\mathbf{u}x,\mathbf{\mathbf{h}}\left(\mathbf{p}+\mathbf{u}y\right)\right)
\]
is called a \emph{patch} of the function $\psi^{\star}\left(\mathbf{x},\mathbf{\mathbf{y}}\right)$
at $\left(\mathbf{p},\mathbf{\mathbf{h}\left(\mathbf{p}\right)}\right)$.
Note that the $\left(\mathbf{p},\mathbf{\mathbf{h}\left(\mathbf{p}\right)}\right)$
itself corresponds to $\left(x=0,y=0\right)$, and the graph of the
PSE function $\left(\mathbf{x},\mathbf{\mathbf{h}\left(\mathbf{x}\right)}\right)$
in the vicinity of $\mathbf{x}=\mathbf{p}$ is mapped into the diagonal
$\left\{ \left(x,y\right):x=y\right\} $. We have therefore the following
``patch-wise'' version of the Regular Minimality\index{regular minimality}
and nonconstant self-dissimilarity: 
\[
\left\{ \begin{array}{l}
\arg\min_{y}\lambda\left(x,y\right)=x,\\
\arg\min_{x}\lambda\left(x,y\right)=y,
\end{array}\right.
\]
and 
\[
\lambda\left(x,x\right)\not=const
\]
for $\left(x,y\right)\in\left[-a,a\right]^{2}$. We will call a patch
\emph{typical}\index{patch!typical} if $\lambda\left(x,x\right)$
is nonconstant for all sufficiently small positive $a$. Figure \ref{fig:patches}
illustrates the notion.

In a Thurstonian-type model (called so in honor of Leon Thurstone
who introduced such models in psychology in the 1920s), there is some
internal space of images $\mathsf{P}$, and each stimulus $\mathbf{x}\in\mathfrak{S}_{1}^{\star}$
(hence also any $x$ representing $\mathbf{x}$ in a patch) is mapped
into a random variable $A$ with values in $\mathsf{P}$, and, similarly,
$\mathbf{y}\in\mathfrak{S}_{2}^{\star}$ (hence also any $y$ representing
$\mathbf{y}$ in a patch) is mapped into a random variables $B$ with
values in $\mathsf{P}$. We will denote these random variables $A\left(\mathbf{x}\right)$
and $B\left(\mathbf{y}\right)$, and their sets of possible values
$\mathsf{a}$ and $\mathsf{b}$, respectively. We will consider first
the case when $A\left(\mathbf{x}\right)$ and $B\left(\mathbf{y}\right)$
are stochastically independent. According to the model, there is a
function 
\[
d:\mathsf{a}\times\mathsf{b}\rightarrow\left\{ \textnormal{same, different}\right\} ,
\]
determining which response will be given in a given presentation of
the stimuli. In complete generality, with no constraints imposed,
such a model is not falsifiable.

\begin{figure}
\begin{centering}
\includegraphics[scale=0.2]{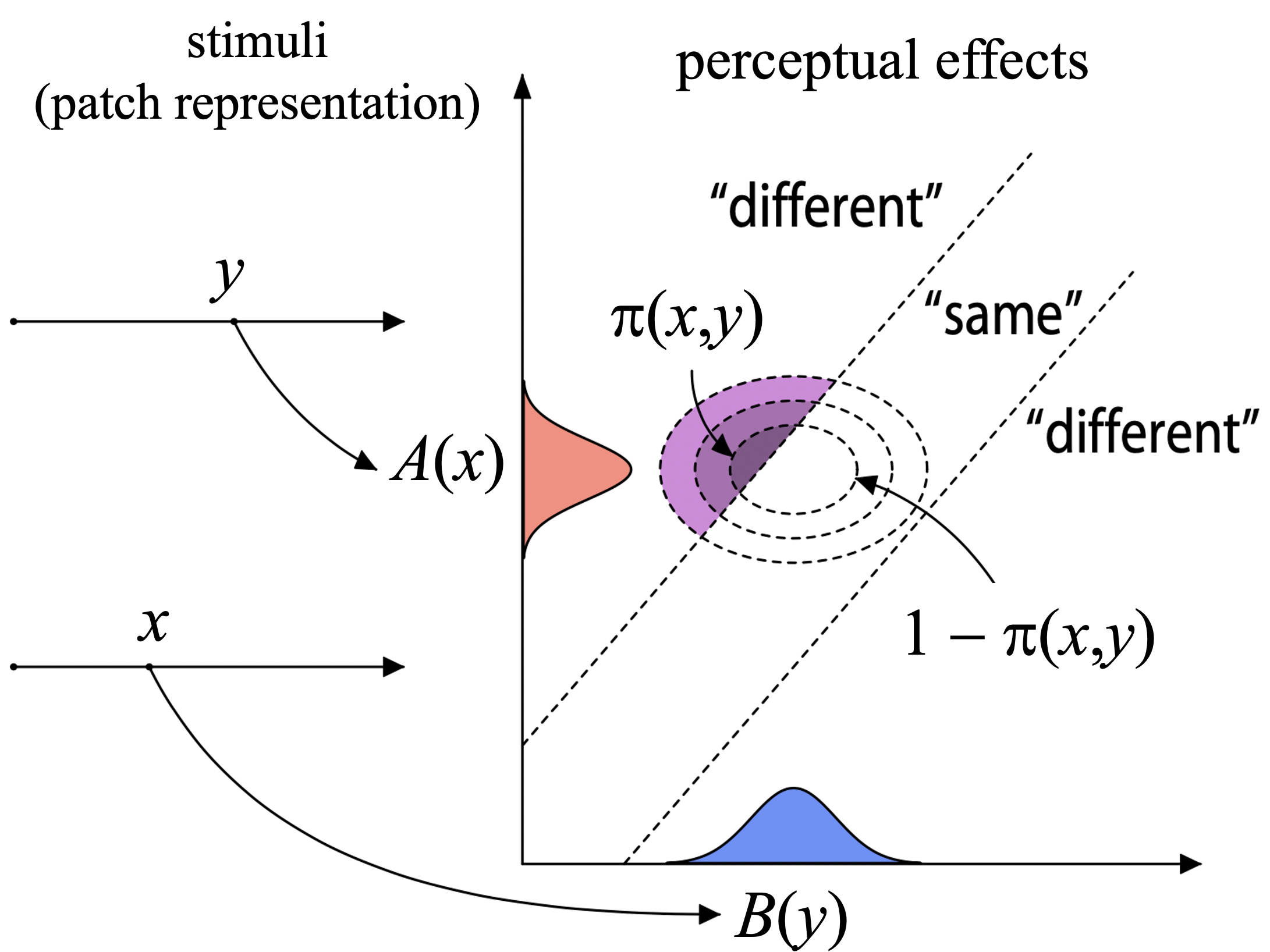} 
\par\end{centering}
\caption{\label{fig:Thurstonian} A schematic representation of a Thurstonian-type
model. The stimuli are represented by their patch variables $x$ and
$y$, and their perceptual effects are points in an interval of reals.
The response ``same'' is given if and only if both random variables
$A\left(x\right)$ and $B\left(y\right)$ fall within the area between
the two dashed lines.}
\end{figure}

\begin{thm}
Any psychometric function $\psi^{\star}:\mathfrak{S}_{1}^{\star}\times\mathfrak{S}_{2}^{\star}\rightarrow\left[0,1\right]$
can be generated by a Thurstonian-type model with stochastically independent
random variables $A\left(\mathbf{x}\right)$ and $B\left(\mathbf{y}\right)$. 
\end{thm}

This is not, however, very interesting, because one normally would
want to deal only with sufficiently ``well-behaved'' Thurstonian-type
models. The intuition here is that, as $\mathbf{x}$ and $\mathbf{y}$
continuously change, the random variables $A\left(\mathbf{x}\right)$
and $B\left(\mathbf{y}\right)$ change sufficiently smoothly. Consider,
e.g., Figure \ref{fig:Thurstonian}, depicting a common way of modeling
same-different comparisons. If the patch variables $x$ and $y$ change
by a small amount, on should expect that the shapes of the probability
density functions not change in an abrupt way. To formalize this intuition,
denote, for any $A$-measurable set $\mathsf{a}$ in the perceptual
space, 
\[
A_{x}\left(\mathsf{a}\right)=\Pr\left[A\left(x\right)\in\mathsf{a}\right],
\]
and analogously, for any $B$-measurable set $\mathsf{b}$ in the
perceptual space, 
\[
B_{y}\left(\mathsf{b}\right)=\Pr\left[B\left(y\right)\in\mathsf{b}\right].
\]

\begin{defn}
Given a patch $\lambda\left(x,y\right)$, a Thurstonian type model
generating it is said to be \emph{well-behaved}\index{Thurstonian model!well-behaved}
if, for every $A$-measurable set $\mathsf{a}$ and $B$-measurable
set $\mathsf{b}$, the left-hand and right-hand derivatives 
\[
\frac{\DD A_{x}\left(\mathsf{a}\right)}{\DD x\pm},\frac{\DD B_{y}\left(\mathsf{b}\right)}{\DD y\pm}
\]
exist, and are bounded across all measurable sets. 
\end{defn}

The latter means that there is a constant $c$ such that 
\[
\left|\frac{\DD A_{x}\left(\mathsf{a}\right)}{\DD x\pm}\right|<c,\left|\frac{\DD B_{y}\left(\mathsf{b}\right)}{\DD y\pm}\right|<c
\]
for all measurable $\mathsf{a}$ and $\mathsf{b}$. The ``textbook''
distributions (such as normal, Weibull, etc.) with parameters depending
on $x$ and $y$ in a piecewise differentiable way will always satisfy
this definition. 
\begin{defn}
A patch $\lambda\left(x,y\right)$ is called \emph{near-smooth}\index{patch!near-smooth}
if he left-hand and right-hand derivatives 
\[
\frac{\partial\lambda\left(x,y\right)}{\partial x\pm}
\]
exist and are continuous in $y$; and similarly, 
\[
\frac{\partial\lambda\left(x,y\right)}{\partial y\pm}
\]
exist and are continuous in $x$. 
\end{defn}

It turns out that, perhaps not surprisingly, 
\begin{thm}
A well-behaved Thurstonian representation can only generate near-smooth
patches. 
\end{thm}

A critical point in the development is created by the following fact. 
\begin{thm}
No near-smooth patch can be typical, i.e. satisfy simultaneously the
Regular Minimality and nonconstant self-dissimilarity properties. 
\end{thm}

This means that for Thurstonian-type modeling of discrimination probabilities
one cannot use well-behaved models, which in turn means the models
should be quite complex mathematically (or else one should reject
either Regular Minimality or nonconstant self-dissimilarity). With
appropriate modifications of the definitions, this conclusion has
been extended to Thurstonian models with \emph{stochastically interdependent}
(but \emph{selectively influenced}) random variables, and to Thurstonian
models in which the mapping of perceptual effects into responses is
probabilistic too.

\subsection{\label{subsec:DzhDzh}Example 3: Universality of corrections for
violations of the triangle inequality\index{inequality!triangle}. }

In Section \ref{sec:Discrete} we described the Floyd-Warshall algorithm\index{Floyd-Warshall algorithm}
for finite stimulus spaces. It turns out that it can be extended to
arbitrary sets, generally infinite and not necessarily discrete. This
is done by using the Axiom of Choice of the set theory to index all
triangles in a stimulus set by \emph{ordinals}. An \emph{ordinal}
is a set $\alpha$ such that each $\beta\in\alpha$ is a set, and
$\beta\subseteq\alpha$. Thus, 
\begin{equation}
\emptyset,\{\emptyset\},\{\emptyset,\{\emptyset\}\},\{\emptyset,\{\emptyset\},\{\emptyset,\{\emptyset\}\}\},\ldots\label{eq:ordinals}
\end{equation}
are (finite) ordinals. For any two ordinals $\alpha$ and $\beta$,
one and only one of the following is true: $\alpha=\beta$, $\alpha\in\beta$,
or $\beta\in\alpha$. The ordinals are ordered in the following way:
if $\alpha\in\beta$, we write $\alpha<\beta$; if either $\alpha\in\beta$
or $\alpha=\beta$, we write $\alpha\leq\beta$. For each ordinal
$\alpha$, $\alpha\cup\{\alpha\}$ is also an ordinal, called the
\emph{successor} of $\alpha$ and denoted $\alpha+1$. There are two
types of ordinals: 
\begin{enumerate}
\item \emph{successor} ordinals $\alpha$, such that $\alpha$ is the successor
of another ordinal, 
\item \emph{limit} ordinals, those that do not succeed other ordinals. 
\end{enumerate}
Thus, we can identify $\emptyset$ in (\ref{eq:ordinals}) with $0$,
and identify $n\cup\{n\}$ with $n+1$ for any ordinal identified
with $n$. We have then that $0$ is a limit ordinal, and each of
$1,2,3,\ldots$ is a successor ordinal. The ordinal 
\[
\omega=\{0,1,2,3,\ldots\}
\]
is the smallest limit ordinal after $0$, and the smallest infinite
ordinal. The ordinals $\omega+1,\omega+2,$ etc. are again successor
ordinals, $\omega+\omega$ is a limit ordinal, and so on. Theorems
involving ordinals are often proved by \emph{transfinite induction}\index{transfinite induction}:
if a certain property holds for $0$, and it holds for any ordinal
$\alpha$ whenever it holds for all ordinals $\beta<\alpha$, then
this property holds for all ordinals. Similarly, definitions of a
property of ordinals can be given by means of \emph{transfinite recursion}:
if it is defined for $0,$and if, having defined it for all $\beta<\alpha$,
we can use our definition to define it for $\alpha$, then we define
it for all ordinals. Thus, in Definition \ref{def:cor}, the procedure
of correcting dissimilarity functions for violations of the triangle
inequalities is described by means of the usual mathematical induction.
It can be replaced with transfinite recursion as follows. We index
the triangles $\mathbf{xyz}$ with pairwise distinct elements by ordinals,
so that for for every ordinal $\alpha$ there is an ordinal $\beta>\alpha$
indexing the same triangle. In other words, each triangle occurs an
infinite number of times. 
\begin{defn}
\label{DefM} Define for each ordinal $\alpha$ a function $M^{\left(\alpha\right)}:\mathfrak{S}\times\mathfrak{S}\rightarrow\mathbb{R}$
as follows:

(i) $M^{\left(0\right)}\equiv D$;

(ii) for any successor ordinal $\alpha=\beta+1$, and for all $\mathbf{a},\mathbf{b}\in\frak{\mathfrak{S}}$,
\[
M^{\left(\alpha\right)}\mathbf{ab}=\begin{cases}
\min\{M^{\left(\beta\right)}\mathbf{ab},M^{\left(\beta\right)}\mathbf{ax}+M^{\left(\beta\right)}\mathbf{xb}\} & \text{ if }\mathbf{axb}\textnormal{ is indexed by }\beta,\\
M^{(\beta)}\mathbf{ab} & \text{ otherwise};
\end{cases}
\]

(iii) if $\alpha$ is a limit ordinal, then, for all $\mathbf{a},\mathbf{b}\in\frak{\mathfrak{S}}$,
\[
M^{(\alpha)}\mathbf{ab}=\inf_{\beta<\alpha}M^{(\beta)}\mathbf{ab}.
\]
It turns out that all results presented in Section \ref{sec:Discrete}
have their transfinite analogous in this generalization. In particular,
``eventually'' (i.e., at some ordinal $\alpha$) the procedure is
terminated with $M^{\left(\alpha\right)}$ coinciding with the quasimetric
dissimilarity $G$, as defined in (\ref{eq:Gdefined}). 
\end{defn}

\subsection{\label{subsec:Data}Example 4: Data Analysis}

Multidimensional Scaling (MDS)\index{multidimensional scaling (MDS)}
and clustering are among the widely used tools of data analysis and
data visualization. The departure point of MDS is a matrix 
\[
\left\{ d_{ij}:i,j=1,2,\ldots,n\right\} 
\]
whose entries are values of a dissimilarity function\index{dissimilarity function}
on the set of objects $\mathfrak{S}=\left\{ 1,2,\ldots,n\right\} $.
This requires that, for all $i\not=j$, 
\[
d_{ii}=0,\textnormal{and }d_{ij}>0.
\]
If this is not the case, but Regular Minimality\index{regular minimality}
is satisfied, the matrix can be brought first to a canonical form,
so that $d_{ii}$ is the smallest value both in the $i$th raw and
in the $i$th columns. Then one can replace $d_{ij}$ with 
\[
\delta_{ij}^{\left(1\right)}=d_{ij}-d_{ii},
\]
or with 
\[
\delta_{ij}^{\left(2\right)}=d_{ji}-d_{ii}.
\]
The choice between the two corresponds to the choice between psychometric
increments of the first and second kind. We know that this choice
is immaterial in Fechnerian Scaling\index{Fechnerian Scaling}, but
in MDS it is immaterial only if the matrix is symmetrical, 
\[
d_{ij}=d_{ji}.
\]
If this is not the case, one usually uses in MDS some symmetrization
procedure: e.g., one can replace each $d_{ij}$ with 
\[
\delta_{ij}=d_{ij}+d_{ji}-d_{ii}-d_{jj}=\left\{ \begin{array}{l}
\delta_{ij}^{\left(1\right)}+\delta_{ji}^{\left(1\right)}\\
\delta_{ij}^{\left(2\right)}+\delta_{ji}^{\left(2\right)}
\end{array}\right.,
\]
proposed by Roger Shepard in the 1950s for so-called \emph{confusion
matrices}\index{confusion matrix} (we will refer to it as \emph{Shepard
symmetrization}, SS)\index{Shepard symmetrization (SS)}. Following
these or similar modifications, the matrix $\delta_{ij}$ can be viewed
as a symmetric dissimilarity function\index{dissimilarity function}.

If in addition the entries of the matrix satisfy the triangle inequality\index{inequality!triangle},
the matrix represents a true metric\index{metric} on the set $\mathfrak{S}=\left\{ 1,2,\ldots,n\right\} $.
In such a case one can apply a procedure of \emph{metric }MDS (mMDS),\index{multidimensional scaling (MDS)!metric}
that consists in embedding the $n$ elements of $\mathfrak{S}$ in
an $\mathbb{R}^{k}$ so that the distances $\Delta_{ij}$ between
the points are as close as possible to the corresponding $\delta_{ij}$.
The quality of approximation is usually estimated by a measure called
\emph{stress}, one variant of which is 
\[
\left(\frac{\sum_{i,j}\left(\Delta_{ij}-\delta_{ij}\right)^{2}}{\sum_{i,j}\delta_{ij}^{2}}\right)^{\nicefrac{1}{2}}.
\]
Since one of the goals of MDS is to help one to visualize the data,
the distance in $\mathbb{R}^{k}$ is usually chosen to be Euclidean,
and $k$ chosen as small as possible (preferably 2 or 3).

However, in most applications $\delta_{ij}$ does not satisfy the
triangle inequality\index{inequality!triangle}, because of which
MDS is used in its \emph{nonmetric} version (nmMDS)\index{multidimensional scaling (MDS)!non-metric}:
here one seeks an embedding into a low-dimensional $\mathbb{R}^{k}$
in which the Euclidean distances match as close as possible not $\delta_{ij}$
but some monotonically increasing transformation of $\delta_{ij}$.
The stress measure then has the form\index{stress measure} 
\[
\left(\frac{\sum_{i,j}\left(\Delta_{ij}-g\left(\delta_{ij}\right)\right)^{2}}{\sum_{i,j}g\left(\delta_{ij}^{2}\right)}\right)^{\nicefrac{1}{2}},
\]
minimized across all possible monotone functions $g$.

\begin{figure}
\begin{raggedright}
\includegraphics[scale=0.3]{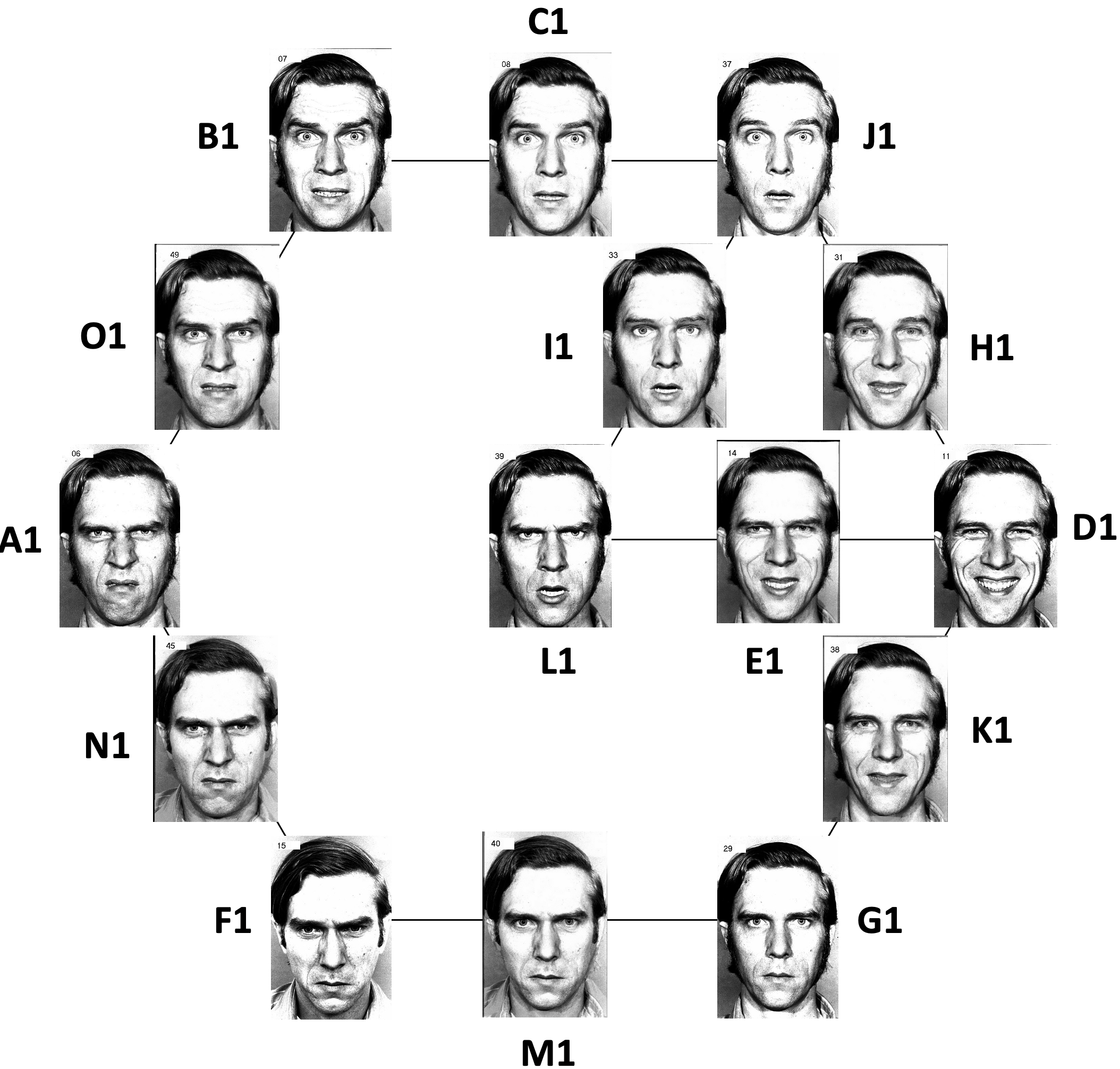} 
\par\end{raggedright}
\caption{\label{fig:faces} A sample of faces presented two at a time with
the question whether they represent the same emotion or different
emotions.}
\end{figure}

\begin{figure}
\index{scree plot} 
\begin{centering}
\includegraphics[scale=0.2]{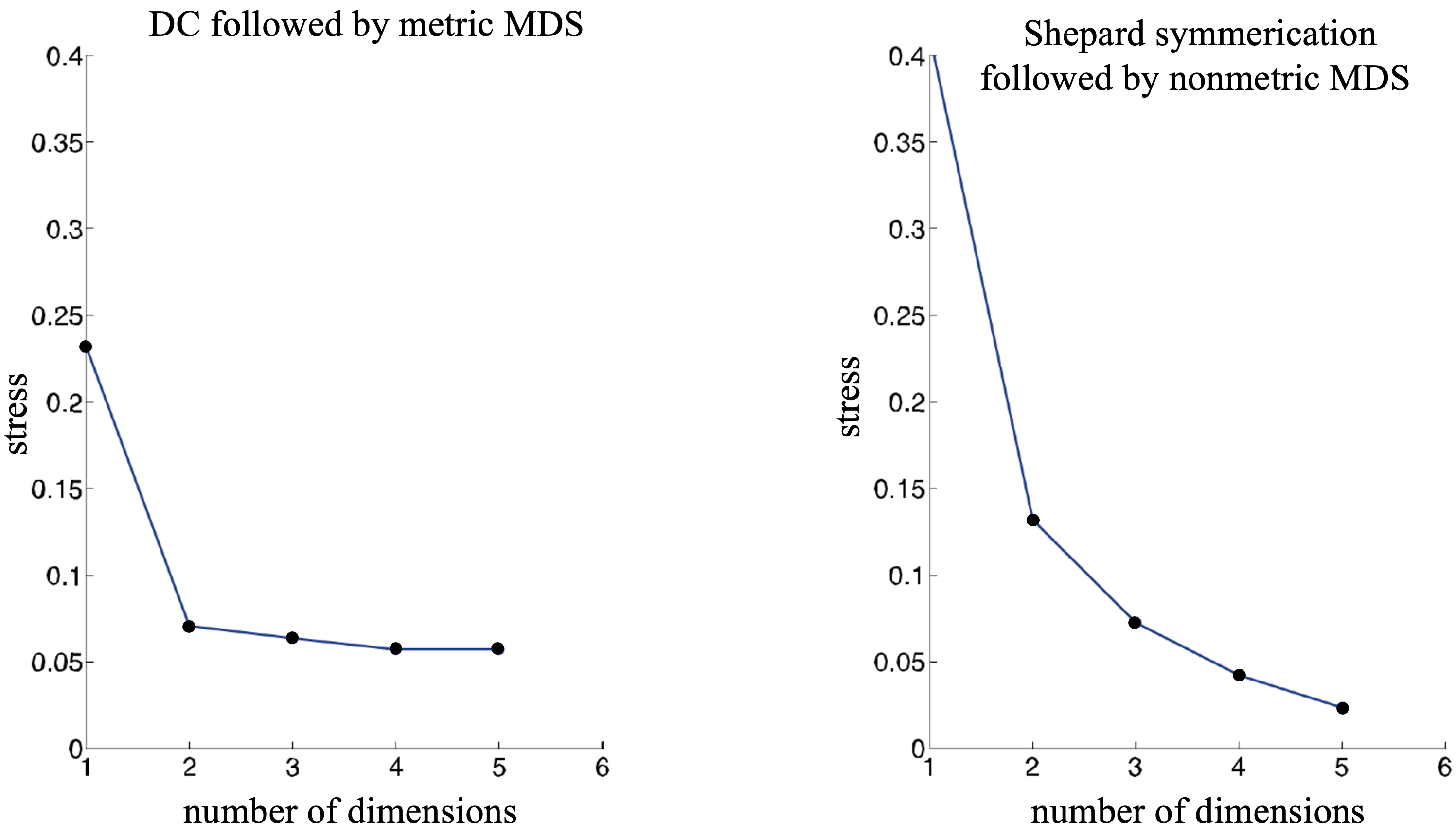} 
\par\end{centering}
\caption{\label{fig:scree} Scree plots of mMDS following Fechnerian Scaling\index{Fechnerian Scaling}
(left) and nmMDS following Shepard's symmetrization. The optimal number
of dimensions is usually chosen as one at which the scree plot visibly
decelerates (exhibits a ``knee''). }
\end{figure}

Dissimilarity cumulation offers a different approach to the same problem,
one that does not require any transformations. Once the original matrix
$d_{ij}$ is brought to a canonical form and replaced with $\delta_{ij}^{\left(1\right)}$
or $\delta_{ij}^{\left(2\right)}$, one computes from either of them
the Fechnerian distances $\overleftrightarrow{G}_{ij}$. Since these
are true distances, one can apply to them the metric version of MDS
to seek a low-dimensional Euclidean embedding. For illustration, consider
an experiment reported in Dzhafarov and Paramei (2010). Images of
faces shown Figure \ref{fig:faces} were presented two at a time,
and the observer was asked to determine whether they exhibited the
same emotion or different emotions. The data $d_{ij}$ were estimates
of the probabilities of the response ``different emotions.'' Figure
\ref{fig:scree} shows the value of stress as a function of $k$ in
the embedding space $\mathbb{R}^{k}$ (so-called \emph{scree plots}).
The comparison of the two procedures, 
\begin{description}
\item [{(DC-mMDS)}] metric MDS applied to the results of dissimilarity
cumulation, and 
\item [{(SS-nmMDS)}] non-metric MDS applied to Shepard-symmetrized data,
\index{multidimensional scaling (MDS)!non-metric} 
\end{description}
shows that the former seems to better identify the minimal dimensionality
of the embedding space. In DC-mMDS, acceptably small value of stress
is achieved at $k=2$ or $3$, and stress drops very slowly afterwards,
whereas in SS-nmMDS, the deceleration of the scree plot is less pronounced.
Having chosen, say, $k=3$, the results of both procedures can be
further subjected to cluster analysis, which groups the points in
$\mathbb{R}^{3}$ into a designated number of \emph{clusters} (the
K-means procedure) or constructs their \emph{dendrogram} (hierarchical
cluster analysis). We do not discuss these procedure, as our goal
is to merely point out that Fechnerian Scaling\index{Fechnerian Scaling}
allows one to base all of them on true distances, without resorting
to an unconstrained search of a monotone transformation. Moreover,
the example in the next section describes an alternative to the dissimilarity
cumulation approach that results in a cluster analysis representation.

There are two public-domain programs that perform MDS and clustering
of the results of dissimilarity cumulation. One of them is the Matlab-based
software package $\mathsf{FSCAMDS}$ (stands for \emph{F}echnerian
\emph{S}caling -- \emph{C}lustering -- \emph{a}nd -- \emph{M}ulti\emph{d}imensional
\emph{S}caling), the other is the R-language package $\mathsf{fechner}$
(see the next section for references). These data-analytic programs
have a variety of options of which we will mention the following.

It is sometimes the case, especially if the data are probabilities,
or if they are sampled from a path-connected space, that large values
of dissimilarity are unreliable, and the cumulation is to be restricted
only to smaller values. The software packages allow one to set a value
above which a dissimilarity $D\mathbf{ab}$ is replaced with infinity,
removing thereby the link $\mathbf{ab}$ from the cumulation process
(because it seeks the smallest cumulated value).

It is sometimes the case that Regular Minimality in the original data
set is violated. The software packages allow one to choose between
the following options: 
\begin{enumerate}
\item to ``doctor'' the data by designating the pairs of PSE and, following
the canonical transformation, to replace negative values of $d_{ij}-d_{ii}$
with zero; 
\item to perform Fechnerian Scaling separately for the two observation areas,
obtaining thereby $\overleftrightarrow{G_{1}}$ and $\overleftrightarrow{G_{2}}$
distances, not equal to each other. 
\end{enumerate}
The justifiability of the second option depends on one's position
with respect to the empirical status of the Regular Minimality law.
As mentioned in Section \ref{sec:Same-different-judgments}, Regular
Minimality in this chapter is not taken as an empirical claim. Rather
it has been part of the definition of the functions we have dealt
with in our mathematical theory.

\subsection{\label{subsec:ultra}Example 5: Ultrametric Fechnerian Scaling}

\index{Fechnerian Scaling!Ultrametric} There is a more direct way
to obtain a representation of dissimilarities by hierarchical clusters
(dendogram or \emph{rooted tree}). The basic idea consists in replacing
``dissimilarity cumulation'' by a ``dissimilarity maximization''
procedure.

Given a chain $\mathbf{X}=\mathbf{x}_{1}\ldots\mathbf{x}_{n}$ and
a binary (real-valued) function $F$, the notation $\Delta_{F}\mathbf{X}$
stands for 
\[
\max_{i=1,\ldots,n-1}F\mathbf{x_{i}x_{i+1}},
\]
again with the obvious convention that the quantity is zero if $n$
is 1 or 0. A dissimilarity function\index{dissimilarity function}
$M$ on a finite set $\mathfrak{S}$ is called a \textit{quasi-ultrametric}\index{quasi-ultrametric}
if it satisfies the \emph{ultrametric inequality},\index{inequality!ultrametric}

\begin{equation}
\max\{M\mathbf{ab},M\mathbf{bc}\}\geq M\mathbf{ac}\label{ultra}
\end{equation}
for all $\mathbf{a},\mathbf{b},\mathbf{c}\in\mathfrak{S}$.

The ultrametric inequality\index{inequality!ultrametric} is rather
restrictive: it is equivalent to postulating that, for any triple
of elements, two dissimilarities have to be equal and not smaller
than the third.
\begin{defn}
Given a dissimilarity $D$ on a finite set $\mathfrak{S}$, the \textit{quasi-ultrametric
$G^{\infty}$ induced by}\index{quasi-ultrametric} $D$ is defined
as

\begin{equation}
G^{\infty}\mathbf{ab}=\min_{\mathbf{X}\in\mathcal{C}}\Delta_{D}\mathbf{aXb},\label{Ginduced_max}
\end{equation}
for all $\mathbf{a},\mathbf{b}\in\mathfrak{S}$. 
\end{defn}

Thus, the value of $G^{\infty}\mathbf{ab}$ is obtained by taking
the minimum, across all chains $\mathbf{X}$ from $\mathbf{a}$ to
$\mathbf{b}$, of the maximum dissimilarity value of the chain. That
$G^{\infty}$ is a quasi-ultrametric\index{quasi-ultrametric} is
easy to prove. A reasonable symmetrization procedure, yielding a metric\index{metric}
is

\begin{equation}
G^{\infty^{*}}\mathbf{ab}=\max\{G^{\infty}\mathbf{ab},G^{\infty}\mathbf{ba}\}\label{overallultra}
\end{equation}
called the \textit{overall Fechnerian ultrametric} on $\mathfrak{S}$.

The ultrametric inequality\index{inequality!ultrametric} is often
violated in empirical data. However, in analogy to recursive corrections
for violations of the triangle inequality\index{inequality!triangle},
it can be shown that a corresponding series of recursive corrections
on the dissimilarity values for violations of the ultrametric inequality
would yield the induced quasi-ultrametric distances. This is in contrast
to applying the different standard hierarchical cluster algorithms
(like single-link, combined-link, etc.) to one and the same data set:
when violations exist, these algorithms will typically result in rather
different ultrametrics.

One can consider procedures intermediate between cumulation and maximization
of dissimilarities by defining, for any dissimilarity function\index{dissimilarity function}
$D$, the length of a chain $\mathbf{X}=\mathbf{x}_{1},\ldots\mathbf{x}_{n}$
by

\begin{equation}
D\mathbf{X}=((D\mathbf{x_{1}x_{2}})^{k}+\ldots+(D\mathbf{x}_{n-1}\mathbf{x}_{n})^{k})^{1/k}.\label{minkowski}
\end{equation}

For $k\rightarrow\infty$ this would result in the ultrametric approach
outlined above. For finite $k$, the procedure is generalizable to
arbitrary dissimilarity spaces. This follows from the fact the use
of (\ref{minkowski}) is equivalent to the use of the original dissimilarity
cumulation procedure in which one, first, redefines $D$ into $D^{k}$
(which yields another dissimilarity function)\index{dissimilarity function},
and then redefines the quasimetric $G$ induced by $D^{k}$ into $G^{1/k}$
(which yields another quasimetric).

\section{\label{sec:Related-Literature}Related Literature}

Fechner's original theory is presented in the \emph{Elemente der Psychophysik}
(Fechner, 1860), but important additions and clarifications can be
found in a later book (Fechner, 1877), and in a paper written shortly
before Fechner's death (Fechner, 1887). A detailed modern account
of Fechner's original theory, especially the ways he derived his logarithmic
psychophysical law, can be found in Dzhafarov and Colonius (2011).
For related interpretations of Fechner's theory, see Pfanzagl (1962),
Creelman (1967), Krantz (1971), and Falmagne (1971). A different interpretation
of Fechner's theory, one that finds it lacking in mathematical coherence
and with which we disagree, is presented in Luce \& Edwards (1958)
and Luce and Galanter (1963).

The theory of dissimilarity cumulation is presented in Dzhafarov and
Colonius (2007) and elaborated in Dzhafarov (2008a). The geometric
aspects of this theory are close to those of the distance and geodesics
theory developed in Blumenthal (1953), Blumenthal and Menger (1970),
and Busemann (2005). To better understand the topology and uniformity
aspects of dissimilarity cumulation, one can consult, e.g., Kelly
(1955) and Hocking and Young (1961). A proof of Theorem \ref{thm:The-only-function}
can be found in Dzhafarov and Colonius (2007). A proof of Theorem
\ref{thm:Intermediate points} is presented in Dzhafarov (2008a).

For stimuli spaces defined on regions of $\mathbb{R}^{n}$, the mathematical
theory essentially becomes a generalized form of Finsler geometry,\index{Finsler geometry}
as presented in Dzhafarov (2008b). A more detailed presentation, however,
and one closer to this chapter, is found in earlier work (Dzhafarov
\& Colonius, 1999, 2001). This part of the theory has its precursors
in Helmholtz (1891) and Schrödinger (1920/1970, 1926/1970), both of
whom, in different ways, used Fechner's cumulation of infinitesimal
differences to construct a Riemannian geometry (a special case of
Finsler geometry) of color space.

In this chapter we have entirely omitted the important topic of invariance
of length and distance under homeomorphic (for general path-connected
spaces) and diffeomorphic (for $\mathbb{R}^{n}$-based spaces) transformations
of space and reparametrizations of paths. These topics are discussed
in Dzhafarov (2008b, c) and Dzhafarov \& Colonius (2001). We have
also ignored the difference between paths and arcs, discussed in detail
in Dzhafarov (2008b).

Dissimilarity cumulation in discrete stimulus spaces is described
in Dzhafarov and Colonius (2006a, c) and Dzhafarov (2010a). The generalization
of the Floyd-Warshall algorithm to arbitrary spaces (Section \ref{subsec:DzhDzh})
is described in Dzhafarov and Dzhafarov (2011).

The notion of separate observation area\index{observation area} in
stimulus comparisons, as well as the Regular Minimality law have been
initially formulated in Dzhafarov (2002) and elaborated in Dzhafarov
(2006b), Dzhafarov and Colonius (2006b), and Kujala and Dzhafarov
(2008, 2009a). The application of the regularity and well-matchedness
principles to the comparative sorites ``paradox'' is presented in
Dzhafarov and Dzhafarov (2010, 2012), with a proof of Theorem \ref{thm:DzhDzh},
and in Dzhafarov and Perry (2014).

The application of these principles together with nonconstant self-dissimilarity
to Thurstonian-type modeling is presented in Dzhafarov (2003a, b),
where one can find proofs of the theorems in Section \ref{subsec:Thurstonian-type}.
This part of the theory has been generalized and greatly extended
in Kujala and Dzhafarov (2008, 2009a, b).

For Multidimensional Scaling see, e.g., Borg and Groenen (1997). Clustering
procedures, hierarchical and K-means, are described in standard textbooks
of multivariate statistics, e.g. Everitt et. al. (2011). The ultrametric
Fechnerian Scaling\index{Fechnerian Scaling} approach is presented
in Colonius \& Dzhafarov (2012).

The link and instructions to the R language software package $\mathsf{fechner}$
mentioned in Section \ref{subsec:Data} is available in Ünlü, Kiefer,
and Dzhafarov (2009). The link and instructions to the software package
$\mathsf{FSCAMDS}$ are available in Dzhafarov (2010b).

\renewcommand{\thesection}{A}

\setcounter{equation}{0}

\setcounter{thm}{0}

\section*{Appendix: Select proofs}

\textbf{Theorem \ref{thm:Submetric function}}\emph{. $F\left(\mathbf{x,u}\right)$
is well-defined for any $\left(\mathbf{x,u}\right)\in\mathbb{T}\cup\left\{ \left(\mathbf{x},\mathbf{0}\right):\mathbf{x}\in\mathfrak{S}\right\} $.
It is positive for $\mathbf{u\neq0,}$ continuous in $\left(\mathbf{x,u}\right)$,
and Euler homogeneous in $\mathbf{u}$. \index{Euler homogeneity} }
\begin{proof}
We first show that $F\left(\mathbf{x,}\overline{\mathbf{u}}\right)$
is continuous in $\left(\mathbf{x,}\overline{\mathbf{u}}\right).$
By Assumptions $\mathcal{E}2$, for any $\varepsilon>0$ there is
a $\delta=\delta\left(\mathbf{x,}\overline{\mathbf{u}},\varepsilon\right)>0$
such that 
\begin{multline*}
\max\left\{ \left\vert \mathbf{a-x}\right\vert ,\left\vert \mathbf{b-x}\right\vert ,\left\vert \overline{\mathbf{b}-\mathbf{a}}-\overline{\mathbf{u}}\right\vert \right\} \\
<\delta\left(\mathbf{x,}\overline{\mathbf{u}},\varepsilon\right)\Longrightarrow\left\vert \frac{D\mathbf{ab}}{\left\vert \mathbf{b}-\mathbf{a}\right\vert }-F\left(\mathbf{x,}\overline{\mathbf{u}}\right)\right\vert <\varepsilon.
\end{multline*}
Consider a sequence $\left(\mathbf{x}_{n}\mathbf{,}\overline{\mathbf{u}}_{n}\right)\rightarrow\left(\mathbf{x,}\overline{\mathbf{u}}\right),$
and let $\left(\mathbf{a}_{n},\mathbf{b}_{n}\right),$ $\mathbf{a}_{n}\neq\mathbf{b}_{n}$,
be any sequence satisfying 
\begin{multline*}
\max\left\{ \left\vert \mathbf{a}_{n}\mathbf{-x}_{n}\right\vert ,\left\vert \mathbf{b}_{n}\mathbf{-x}_{n}\right\vert ,\left\vert \overline{\mathbf{b}_{n}-\mathbf{a}_{n}}-\overline{\mathbf{u}}_{n}\right\vert \right\} \\
<\min\left\{ \delta\left(\mathbf{x}_{n}\mathbf{,}\overline{\mathbf{u}}_{n},\frac{1}{n}\right),\frac{1}{2}\delta\left(\mathbf{x,}\overline{\mathbf{u}},\varepsilon\right)\right\} .
\end{multline*}
Clearly, 
\[
\frac{D\mathbf{a}_{n}\mathbf{b}_{n}}{\left\vert \mathbf{b}_{n}-\mathbf{a}_{n}\right\vert }-F\left(\mathbf{x}_{n}\mathbf{,}\overline{\mathbf{u}}_{n}\right)\rightarrow0.
\]
At the same time, for all sufficiently large $n,$ 
\[
\max\left\{ \left\vert \mathbf{x}_{n}\mathbf{-x}\right\vert ,\left\vert \overline{\mathbf{u}}_{n}-\overline{\mathbf{u}}\right\vert \right\} <\frac{1}{2}\delta\left(\mathbf{x,}\overline{\mathbf{u}},\varepsilon\right),
\]
implying 
\[
\max\left\{ \left\vert \mathbf{a}_{n}\mathbf{-x}\right\vert ,\left\vert \mathbf{b}_{n}\mathbf{-x}\right\vert ,\left\vert \overline{\mathbf{b}_{n}-\mathbf{a}_{n}}-\overline{\mathbf{u}}\right\vert \right\} <\delta\left(\mathbf{x,}\overline{\mathbf{u}},\varepsilon\right).
\]
But then 
\[
\left\vert \frac{D\mathbf{a}_{n}\mathbf{b}_{n}}{\left\vert \mathbf{b}_{n}-\mathbf{a}_{n}\right\vert }-F\left(\mathbf{x,}\overline{\mathbf{u}}\right)\right\vert <\varepsilon,
\]
and, as $\varepsilon$ can be chosen arbitrarily small, we have 
\[
\frac{D\mathbf{a}_{n}\mathbf{b}_{n}}{\left\vert \mathbf{b}_{n}-\mathbf{a}_{n}\right\vert }-F\left(\mathbf{x,}\overline{\mathbf{u}}\right)\rightarrow0.
\]
The convergence 
\[
F\left(\mathbf{x}_{n}\mathbf{,}\overline{\mathbf{u}}_{n}\right)\rightarrow F\left(\mathbf{x,}\overline{\mathbf{u}}\right)
\]
follows, establishing the continuity of $F\left(\mathbf{x,}\overline{\mathbf{u}}\right)$.
Now, for $\mathbf{u\neq0}$, denoting $\mathbf{u}=\left\vert \mathbf{u}\right\vert \overline{\mathbf{u}}$,
\[
F\left(\mathbf{x,}\mathbf{u}\right)=\lim_{s\rightarrow0+}\frac{D\mathbf{x}\left[\mathbf{x+}\mathbf{u}s\right]}{s}=\left\vert \mathbf{u}\right\vert \lim_{\left\vert \mathbf{u}\right\vert s\rightarrow0+}\frac{D\mathbf{x}\left[\mathbf{x+}\overline{\mathbf{u}}\left\vert \mathbf{u}\right\vert s\right]}{\left\vert \mathbf{u}\right\vert s}=\left\vert \mathbf{u}\right\vert F\left(\mathbf{x,}\overline{\mathbf{u}}\right).
\]
It immediately follows that $F\left(\mathbf{x,u}\right)$ exists,
that it is positive and continuous, and that 
\[
F\left(\mathbf{x,u}\right)=\left\vert \mathbf{u}\right\vert F\left(\mathbf{x,}\overline{\mathbf{u}}\right).
\]
So, for $k>0$, 
\[
F\left(\mathbf{x,}k\mathbf{u}\right)=k\left\vert \mathbf{u}\right\vert F\left(\mathbf{x,}\overline{\mathbf{u}}\right)=kF\left(\mathbf{x,u}\right).
\]
Finally, since any convergence of $\left(\mathbf{x}_{n}\mathbf{,u}_{n}\right)\rightarrow\left(\mathbf{x,0}\right)$
with $\mathbf{u}_{n}\neq\mathbf{0}$ can be presented as $\left(\mathbf{x}_{n}\mathbf{,}\left\vert \mathbf{u}_{n}\right\vert \overline{\mathbf{u}}_{n}\right)\rightarrow\left(\mathbf{x,0}\right)$
with $\left\vert \mathbf{u}_{n}\right\vert \rightarrow0,$ we have
\[
F\left(\mathbf{x}_{n}\mathbf{,u}_{n}\right)=\left\vert \mathbf{u}_{n}\right\vert F\left(\mathbf{x}_{n}\mathbf{,\overline{u}}_{n}\right)\rightarrow0,
\]
because within a small ball around $\mathbf{x}$ and on a compact
set of unit vectors the function $F\left(\mathbf{x}_{n}\mathbf{,\overline{u}}_{n}\right)$
does not exceed some finite value. Thus $F\left(\mathbf{x}_{n}\mathbf{,u}_{n}\right)$
extends to $F\left(\mathbf{x,0}\right)=0$ by continuity.
\end{proof}
\textbf{Lemma \ref{lem:max prod n}}\emph{. For any $\mathbf{\left(\mathbf{a},\mathbf{u}\right)}\in\mathbb{T}$,
the maximal production of $\mathbf{u}$ in $\mathbb{I}_{\mathbf{a}}$
can be presented as a convex combination of $n$ (not necessarily
distinct) radius-vectors $\mathbf{v}_{1},\ldots,\mathbf{v}_{n}\in\delta\mathbb{I}_{\mathbf{a}}$. }
\begin{proof}
With no loss of generality, let $\mathbf{u\in\delta\mathbb{I}_{\mathbf{a}}},$
and let $\kappa$ stand for $\kappa\left(\mathbf{a},\mathbf{u}\right)$.
By Corollary \ref{cor:Carath}, for some $\mathbf{v}_{1},\ldots,\mathbf{v}_{n+1}\in\mathbb{I}_{\mathbf{a}}$,
the system of $n+1$ linear equations 
\[
\left\{ \begin{array}{c}
\kappa\mathbf{u}=\lambda_{1}\mathbf{v}_{1}+\ldots+\lambda_{n+1}\mathbf{v}_{n+1}\\
\lambda_{1}+\ldots+\lambda_{n+1}=1
\end{array}\right.
\]
has a solution $\lambda_{1},\ldots,\lambda_{n+1}\in\left[0,1\right]$.
Assume that $\lambda_{1},\ldots,\lambda_{n+1}$ are all positive (if
some of them are zero, the theorem's statement holds). If the determinant
of the matrix of coefficients for this system were nonzero, then,
for any $\varepsilon$, the modified system 
\[
\left\{ \begin{array}{c}
\left[\kappa+\varepsilon\right]\mathbf{u}=\lambda_{1}\mathbf{v}_{1}+\ldots+\lambda_{n+1}\mathbf{v}_{n+1}\\
\lambda_{1}+\ldots+\lambda_{n+1}=1
\end{array}\right.
\]
would also have a solution $\lambda'_{1},\ldots,\lambda'_{n+1}$,
and choosing $\varepsilon$ positive and sufficiently small, this
solution (by continuity) would also satisfy $\lambda'_{1}>0,\ldots,\lambda'_{n+1}>0$.
But this would mean that $\left[\kappa+\varepsilon\right]\mathbf{u}$
belongs to the convex hull of $\mathbb{I}_{\mathbf{a}}$, which is
impossible since $\kappa\mathbf{u}$ is the maximal production of
$\mathbf{u}$. Hence 
\[
\det\left[\begin{array}{ccc}
\mathbf{v}_{1} & \cdots & \mathbf{v}_{n+1}\\
1 & \cdots & 1
\end{array}\right]=0,
\]
where we treat $\mathbf{v}_{1},\ldots,\mathbf{v}_{n+1}$ as $n$-element
columns. But this means that, for some $\gamma_{1},\ldots,\gamma_{n+1}$,
not all zero, 
\[
\gamma_{1}\left[\begin{array}{c}
\mathbf{v}_{1}\\
1
\end{array}\right]+\ldots+\gamma_{n+1}\left[\begin{array}{c}
\mathbf{v}_{n+1}\\
1
\end{array}\right]=\mathbf{0},
\]
which indicates the affine dependence of $\mathbf{v}_{1},\ldots,\mathbf{v}_{n+1}$.
It follows from Lemma \ref{lem:almostCarath} that $\mathbf{u}$ can
be presented as a convex combination of some $m<n+1$ (not necessarily
distinct) nonzero vectors in $\mathbf{v}_{1},\ldots,\mathbf{v}_{n+1}\in\mathbb{I}_{\mathbf{a}}$.
Let them be the first $m$ vectors in the list, $\mathbf{v}_{1},\ldots,\mathbf{v}_{m}$.
We have now the system 
\[
\left\{ \begin{array}{c}
\kappa\mathbf{u}=\lambda_{1}\mathbf{v}_{1}+\ldots+\lambda_{m}\mathbf{v}_{m}\\
\lambda_{1}+\ldots+\lambda_{m}=1
\end{array}\right.
\]
with a solution $\lambda_{1}>0,\ldots,\lambda_{m}>0$ (zero values
here would simply decrease $m$). Rewriting it as 
\[
\left\{ \begin{array}{c}
\kappa\mathbf{u}=\lambda_{1}c_{1}\mathbf{\widetilde{v}}_{1}+\ldots+\lambda_{m}c_{m}\mathbf{\widetilde{v}}_{m}\\
\lambda_{1}+\ldots+\lambda_{m}=1
\end{array}\right.,
\]
where $\mathbf{\widetilde{v}}_{i}\in\delta\mathbb{I}_{\mathbf{a}}$
is codirectional with $\mathbf{v}_{i}$ ($i=1,\ldots,m$), it is clear
by Lemma \ref{lem:damned} that for $\kappa$ to have a maximal possible
value, all $c_{i}$ should have maximal possible values. In $\mathbb{I}_{\mathbf{a}}$
these values are $c_{1}=\ldots=c_{m}=1$, that is, all vectors $\mathbf{v}_{1},\ldots,\mathbf{v}_{m}$
are radius-vectors. This completes the proof.
\end{proof}
\textbf{Theorem \ref{thm:Fhat is submetric}.} \emph{The minimal submetric
function\index{submetric function!minimal} $\widehat{F}\left(\mathbf{a,u}\right)$
has all the properties of a submetric function\index{submetric function}:
it is positive for $\mathbf{u}\neq\mathbf{0}$, Euler homogeneous,
and continuous. }
\begin{proof}
We only prove the continuity, as the other properties follow trivially
from the definition of $\widehat{F}$ and the analogous properties
of $F$. Consider a sequence of line elements 
\[
\left(\mathbf{a}_{k},\mathbf{u}_{k}\right)\rightarrow\left(\mathbf{a},\mathbf{u}\right).
\]
Let $\left(\mathbf{v}_{1},...,\mathbf{v}_{n}\right)$ be a minimizing
chain for $\mathbf{\left(\mathbf{a},\mathbf{u}\right)}$ (or a sequence
of $n$ zero vectors if $\mathbf{u=}\mathbf{0}$). For every $k$,
consider the sequence $\mathbf{v}_{1}+\left(\mathbf{u}_{k}-\mathbf{u}\right),\mathbf{v}_{2},...,\mathbf{v}_{n}$,
which differs from the minimizing chain in the first element only.
Its elements sum to $\mathbf{u}_{k}$, because of which 
\[
F\left(\mathbf{a}_{k},\mathbf{v}_{1}+\left(\mathbf{u}_{k}-\mathbf{u}\right)\right)+F\left(\mathbf{a}_{k},\mathbf{v}_{2}\right)+...+F\left(\mathbf{a}_{k},\mathbf{v}_{n}\right)\geq\widehat{F}\left(\mathbf{a}_{k},\mathbf{u}_{k}\right).
\]
At the same time, by continuity of $F$, 
\[
\begin{array}{l}
F\left(\mathbf{a}_{k},\mathbf{v}_{1}+\left(\mathbf{u}_{k}-\mathbf{u}\right)\right)+F\left(\mathbf{a}_{k},\mathbf{v}_{2}\right)+...+F\left(\mathbf{a}_{k},\mathbf{v}_{n}\right)\\
\rightarrow F\left(\mathbf{a}_{k},\mathbf{v}_{1}\right)+F\left(\mathbf{a}_{k},\mathbf{v}_{2}\right)+...+F\left(\mathbf{a}_{k},\mathbf{v}_{n}\right)=\widehat{F}\left(\mathbf{a,u}\right),
\end{array}
\]
whence it follows that 
\[
\limsup_{k\rightarrow\infty}\widehat{F}\left(\mathbf{a}_{k},\mathbf{u}_{k}\right)\leq\widehat{F}\left(\mathbf{a,u}\right).
\]
To prove that at the same time 
\[
\liminf_{k\rightarrow\infty}\widehat{F}\left(\mathbf{a}_{k},\mathbf{u}_{k}\right)\geq\widehat{F}\left(\mathbf{a,u}\right),
\]
let $\left(\mathbf{v}_{1k},...,\mathbf{v}_{nk}\right)$ be a minimizing
chain for $\left(\mathbf{a}_{k},\mathbf{u}_{k}\right)$, for every
$k$, and consider the sequence $\mathbf{v}_{1k}+\left(\mathbf{u}-\mathbf{u}_{k}\right),\mathbf{v}_{2k},...,\mathbf{v}_{nk}$,
which differs from the minimizing chain in the first element only.
Its elements sum to $\mathbf{u}$, because of which 
\[
F\left(\mathbf{a},\mathbf{v}_{1k}+\left(\mathbf{u}-\mathbf{u}_{k}\right)\right)+F\left(\mathbf{a},\mathbf{v}_{2k}\right)+...+F\left(\mathbf{a},\mathbf{v}_{nk}\right)\geq\widehat{F}\left(\mathbf{a},\mathbf{u}\right).
\]
We will arrive at the desired inequality for $\liminf$ if we show
that 
\[
\left[F\left(\mathbf{a},\mathbf{v}_{1k}+\left(\mathbf{u}-\mathbf{u}_{k}\right)\right)+F\left(\mathbf{a},\mathbf{v}_{2k}\right)+...+F\left(\mathbf{a},\mathbf{v}_{nk}\right)\right]-\widehat{F}\left(\mathbf{a}_{k},\mathbf{u}_{k}\right)\rightarrow0.
\]
The left-hand side difference here is 
\begin{multline*}
\left[F\left(\mathbf{a},\mathbf{v}_{1k}+\left(\mathbf{u}-\mathbf{u}_{k}\right)\right)+F\left(\mathbf{a},\mathbf{v}_{2k}\right)+...+F\left(\mathbf{a},\mathbf{v}_{nk}\right)\right]\\
-\left[F\left(\mathbf{a}_{k},\mathbf{v}_{1k}\right)+F\left(\mathbf{a}_{k},\mathbf{v}_{2k}\right)+...+F\left(\mathbf{a}_{k},\mathbf{v}_{nk}\right)\right]\\
=\left[F\left(\mathbf{a},\mathbf{v}_{1k}+\left(\mathbf{u}-\mathbf{u}_{k}\right)\right)-F\left(\mathbf{a}_{k},\mathbf{v}_{1k}\right)\right]+\left[F\left(\mathbf{a},\mathbf{v}_{2k}\right)-F\left(\mathbf{a}_{k},\mathbf{v}_{2k}\right)\right]\\
+\ldots+\left[F\left(\mathbf{a},\mathbf{v}_{nk}\right)-F\left(\mathbf{a}_{k},\mathbf{v}_{nk}\right)\right],
\end{multline*}
where 
\begin{multline*}
F\left(\mathbf{a},\mathbf{v}_{1k}+\left(\mathbf{u}-\mathbf{u}_{k}\right)\right)-F\left(\mathbf{a}_{k},\mathbf{v}_{1k}\right)\\
=\left(\left|\mathbf{v}_{1k}+\left(\mathbf{u}-\mathbf{u}_{k}\right)\right|-\left|\mathbf{v}_{1k}\right|\right)F\left(\mathbf{a},\overline{\mathbf{v}_{1k}+\left(\mathbf{u}-\mathbf{u}_{k}\right)}\right)\\
+\left|\mathbf{v}_{1k}\right|\left[F\left(\mathbf{a},\overline{\mathbf{v}_{1k}+\left(\mathbf{u}-\mathbf{u}_{k}\right)}\right)-F\left(\mathbf{a}_{k},\mathbf{\overline{v}}_{1k}\right)\right],
\end{multline*}
and 
\[
F\left(\mathbf{a},\mathbf{v}_{ik}\right)-F\left(\mathbf{a}_{k},\mathbf{v}_{ik}\right)=\left|\mathbf{v}_{ik}\right|\left[F\left(\mathbf{a},\mathbf{\overline{v}}_{ik}\right)-F\left(\mathbf{a}_{k},\mathbf{\overline{v}}_{ik}\right)\right],i=2,\ldots,n.
\]
Since $\mathbf{u}_{k}\rightarrow\mathbf{u}$, $\mathbf{a}_{k}\rightarrow\mathbf{a}$,
and $F$ is uniformly continuous and bounded on the compact set of
unit vectors, we have 
\[
\begin{array}{c}
\left|\mathbf{v}_{1k}+\left(\mathbf{u}-\mathbf{u}_{k}\right)\right|-\left|\mathbf{v}_{1k}\right|\rightarrow0,\\
F\left(\mathbf{a},\overline{\mathbf{v}_{1k}+\left(\mathbf{u}-\mathbf{u}_{k}\right)}\right)-F\left(\mathbf{a}_{k},\mathbf{\overline{v}}_{1k}\right)\rightarrow0,\\
\left(\left|\mathbf{v}_{1k}+\left(\mathbf{u}-\mathbf{u}_{k}\right)\right|-\left|\mathbf{v}_{1k}\right|\right)F\left(\mathbf{a},\overline{\mathbf{v}_{1k}+\left(\mathbf{u}-\mathbf{u}_{k}\right)}\right)\rightarrow0,\\
F\left(\mathbf{a},\mathbf{\overline{v}}_{ik}\right)-F\left(\mathbf{a}_{k},\mathbf{\overline{v}}_{ik}\right)\rightarrow0.
\end{array}
\]
To see that 
\[
\begin{array}{c}
F\left(\mathbf{a},\mathbf{v}_{1k}+\left(\mathbf{u}-\mathbf{u}_{k}\right)\right)-F\left(\mathbf{a}_{k},\mathbf{v}_{1k}\right)\rightarrow0,\\
F\left(\mathbf{a},\mathbf{v}_{ik}\right)-F\left(\mathbf{a}_{k},\mathbf{v}_{ik}\right)\rightarrow0,\qquad i=2,\ldots,n,
\end{array}
\]
it remains to show that $\left|\mathbf{v}_{ik}\right|$ is bounded
for $i=2,\ldots,n$. But this follows from the fact that 
\[
F\left(\mathbf{a}_{k},\mathbf{v}_{1k}\right)+\ldots+F\left(\mathbf{a}_{k},\mathbf{v}_{nk}\right)\leq F\left(\mathbf{a}_{k},\mathbf{u}_{k}\right)\rightarrow F\left(\mathbf{a},\mathbf{u}\right),
\]
because of which 
\[
F\left(\mathbf{a}_{k},\mathbf{v}_{ik}\right)=\left|\mathbf{v}_{ik}\right|F\left(\mathbf{a}_{k},\mathbf{\overline{v}}_{ik}\right)\leq F\left(\mathbf{a},\mathbf{u}\right)+C,
\]
where $C$ is some positive constant. 
\end{proof}
\textbf{Theorem} \textbf{\ref{thm:G(x,x+us)/s}}. \emph{The distance
$G\left(\mathbf{x},\mathbf{x}+\mathbf{u}s\right)$ is differentiable
at $s=0+$ for any $\left(\mathbf{x},\mathbf{u}\right)\in\mathbb{T}$,
and 
\[
\left.\frac{\mathrm{d}G\left(\mathbf{x},\mathbf{x}+\mathbf{u}s\right)}{\mathrm{d}s+}\right|_{s=0}=\lim_{s\rightarrow0+}\frac{G\left(\mathbf{x},\mathbf{x}+\mathbf{u}s\right)}{s}=\widehat{F}\left(\mathbf{x},\mathbf{u}\right).
\]
}
\begin{proof}
We prove first that 
\[
\limsup_{s\rightarrow0+}\frac{G\left(\mathbf{x},\mathbf{x}+\mathbf{u}s\right)}{s\widehat{F}\left(\mathbf{x},\mathbf{u}\right)}\leq1.
\]
Let $\left(\mathbf{u}_{1},\ldots,\mathbf{u}_{n}\right)$ be a minimizing
vector chain for $\left(\mathbf{x},\mathbf{u}\right)$, so that 
\[
\widehat{F}\left(\mathbf{x},\mathbf{u}\right)=F\left(\mathbf{x},\mathbf{u}_{1}\right)+\ldots+F\left(\mathbf{x},\mathbf{u}_{n}\right).
\]
Consider the chain of points 
\[
\mathbf{x}\:\left[\mathbf{x}+\mathbf{u}_{1}s\right]\:\left[\mathbf{x}+\left(\mathbf{u}_{1}+\mathbf{u}_{2}\right)s\right]\:\ldots\:\left[\mathbf{x}+\left(\mathbf{u}_{1}+\ldots+\mathbf{u}_{n}\right)s\right],
\]
in which the last point coincides with $\mathbf{x}+\mathbf{u}s$.
We will generically refer to a point in this chain as 
\[
\mathbf{x}+\left(\mathbf{u}_{1}+\ldots+\mathbf{u}_{i}\right)s,\qquad i=0,1,\ldots,n,
\]
with the obvious convention for $i=0$. For all sufficiently small
$s$, all these points belong to a compact ball in $\mathfrak{S}$
centered at $\mathbf{x}$. Then, by Theorem \ref{thm:DtoF and GtoF ratios}
and the continuity of $F$, we have, as $s\rightarrow0+$, 
\begin{multline*}
\frac{D\left[\mathbf{x}+\left(\mathbf{u}_{1}+\ldots\mathbf{u}_{i}\right)s\right]\left[\mathbf{x}+\left(\mathbf{u}_{1}+\ldots+\mathbf{u}_{i+1}\right)s\right]}{sF\left(\mathbf{x},\mathbf{u}_{i+1}\right)}\\
=\frac{D\left[\mathbf{x}+\left(\mathbf{u}_{1}+\ldots\mathbf{u}_{i}\right)s\right]\left[\mathbf{x}+\left(\mathbf{u}_{1}+\ldots+\mathbf{u}_{i+1}\right)s\right]}{F\left(\mathbf{x}+\left(\mathbf{u}_{1}+\ldots\mathbf{u}_{i}\right)s,\mathbf{u}_{i+1}s\right)}\\
\times\frac{sF\left(\mathbf{x}+\left(\mathbf{u}_{1}+\ldots\mathbf{u}_{i}\right)s,\mathbf{u}_{i+1}\right)}{sF\left(\mathbf{x},\mathbf{u}_{i+1}\right)}\rightarrow1,
\end{multline*}
whence 
\begin{multline*}
\frac{D\mathbf{x}\left[\mathbf{x}+\mathbf{u}_{1}s\right]\ldots\left[\mathbf{x}+\mathbf{u}s\right]}{s\widehat{F}\left(\mathbf{x},\mathbf{u}\right)}\\
=\frac{\sum_{i=0}^{n-1}D\left[\mathbf{x}+\left(\mathbf{u}_{1}+\ldots\mathbf{u}_{i}\right)s\right]\left[\mathbf{x}+\left(\mathbf{u}_{1}+\ldots+\mathbf{u}_{i+1}\right)s\right]}{s\sum_{i=1}^{n}F\left(\mathbf{x},\mathbf{u}_{i}\right)}\rightarrow1.
\end{multline*}
But then 
\[
\limsup_{s\rightarrow0+}\frac{G\left(\mathbf{x},\mathbf{x}+\mathbf{u}s\right)}{s\widehat{F}\left(\mathbf{x},\mathbf{u}\right)}=\limsup_{s\rightarrow0+}\frac{G\left(\mathbf{x},\mathbf{x}+\mathbf{u}s\right)}{D\mathbf{x}\left[\mathbf{x}+\mathbf{u}_{1}s\right]\ldots\left[\mathbf{x}+\mathbf{u}s\right]}\leq1,
\]
by the definition of $G$. We prove next that 
\[
\liminf_{s\rightarrow0+}\frac{G\left(\mathbf{x},\mathbf{x}+\mathbf{u}s\right)}{s\widehat{F}\left(\mathbf{x},\mathbf{u}\right)}\geq1.
\]
Consider a sequence of chains 
\[
\mathbf{x}\:\left[\mathbf{x}+\mathbf{v}_{1k}s_{k}\right]\:\left[\mathbf{x}+\left(\mathbf{v}_{1k}+\mathbf{v}_{2k}\right)s_{k}\right]\:\ldots\:\left[\mathbf{x}+\left(\mathbf{v}_{1k}+\ldots+\mathbf{v}_{m_{k}k}\right)s_{k}\right],\qquad k=1,2,\ldots,
\]
such that 
\[
s_{k}\rightarrow0+,
\]
\[
\mathbf{v}_{1k}+\ldots+\mathbf{v}_{m_{k}k}=\mathbf{u},\qquad k=1,2,\ldots,
\]
and 
\[
\frac{D\mathbf{x}\left[\mathbf{x}+\mathbf{v}_{1k}s_{k}\right]\ldots\left[\mathbf{x}+\mathbf{u}s_{k}\right]}{G\left(\mathbf{x},\mathbf{x}+\mathbf{u}s\right)}\rightarrow1.
\]
Again, it is easy to see that for all all $k$ sufficiently large
(i.e., $s_{k}$ sufficiently small) all these chains fall within a
compact ball in $\mathfrak{S}$ centered at $\mathbf{x}$. Then, for
$i=0,1,\ldots,m_{k}-1$, by Theorem \ref{thm:DtoF and GtoF ratios}
and the continuity of $F$, as $k\rightarrow\infty$, 
\begin{multline*}
\frac{D\left[\mathbf{x}+\left(\mathbf{v}_{1k}+\ldots\mathbf{v}_{ik}\right)s_{k}\right]\left[\mathbf{x}+\left(\mathbf{v}_{1k}+\ldots+\mathbf{v}_{i+1,k}\right)s_{k}\right]}{s_{k}F\left(\mathbf{x},\mathbf{v}_{i+1,k}\right)}\\
=\frac{D\left[\mathbf{x}+\left(\mathbf{v}_{1k}+\ldots\mathbf{v}_{ik}\right)s_{k}\right]\left[\mathbf{x}+\left(\mathbf{v}_{1k}+\ldots+\mathbf{v}_{i+1,k}\right)s_{k}\right]}{F\left(\mathbf{x}+\left(\mathbf{v}_{1k}+\ldots\mathbf{v}_{ik}\right)s_{k},\mathbf{v}_{i+1,k}s_{k}\right)}\\
\times\frac{s_{k}F\left(\mathbf{x}+\left(\mathbf{v}_{1k}+\ldots\mathbf{v}_{ik}\right)s_{k},\mathbf{v}_{i+1,k}\right)}{s_{k}F\left(\mathbf{x},\mathbf{v}_{i+1,k}\right)}\rightarrow1
\end{multline*}
uniformly across all choices of $\left(\mathbf{v}_{1k}+\ldots\mathbf{v}_{m_{k}k}\right)$.
It follows that
\begin{multline*}
\frac{D\mathbf{x}\left[\mathbf{x}+\mathbf{v}_{1k}s_{k}\right]\ldots\left[\mathbf{x}+\mathbf{u}s_{k}\right]}{s_{k}\sum_{i=1}^{m_{k}}F\left(\mathbf{x},\mathbf{v}_{ik}\right)}\\
=\frac{\sum_{i=0}^{m_{k}-1}D\left[\mathbf{x}+\left(\mathbf{v}_{1k}+\ldots\mathbf{v}_{ik}\right)s_{k}\right]\left[\mathbf{x}+\left(\mathbf{v}_{1k}+\ldots+\mathbf{v}_{i+1,k}\right)s_{k}\right]}{s_{k}\sum_{i=1}^{m_{k}}F\left(\mathbf{x},\mathbf{v}_{ik}\right)}\rightarrow1.
\end{multline*}
But then 
\begin{multline*}
\liminf_{s\rightarrow0+}\frac{G\left(\mathbf{x},\mathbf{x}+\mathbf{u}s\right)}{s\widehat{F}\left(\mathbf{x},\mathbf{u}\right)}=\liminf_{k\rightarrow\infty}\frac{D\mathbf{x}\left[\mathbf{x}+\mathbf{v}_{1k}s_{k}\right]\ldots\left[\mathbf{x}+\mathbf{u}s_{k}\right]}{s_{k}\widehat{F}\left(\mathbf{x},\mathbf{u}\right)}\\
=\liminf_{k\rightarrow\infty}\frac{\sum_{i=1}^{m_{k}}F\left(\mathbf{x},\mathbf{v}_{ik}\right)}{\widehat{F}\left(\mathbf{x},\mathbf{u}\right)}\geq1,
\end{multline*}
by the definition of $\widehat{F}$ in terms of minimizing chains.
This establishes 
\[
\lim_{s\rightarrow0+}\frac{G\left(\mathbf{x},\mathbf{x}+\mathbf{u}s\right)}{s\widehat{F}\left(\mathbf{x},\mathbf{u}\right)}=1,
\]
and the theorem is proved.
\end{proof}
\textbf{Theorem} \textbf{\ref{thm:MDFS}}. \emph{For every path $\mathbf{h}|\left[a,b\right]$
connecting $\mathbf{a}$ to $\mathbf{b}$ one can find a piecewise
linear path from $\mathbf{a}$ to $\mathbf{b}$ which is arbitrarily
close to $\mathbf{h}|\left[a,b\right]$ pointwise and in its length. }
\begin{proof}
Let 
\[
\mu_{n}=\left\{ a=t_{n0},...,t_{ni},t_{n,i+1},...,t_{n,k_{n}+1}=b\right\} 
\]
be a sequence of nets with $\delta\mu_{n}\rightarrow0$. Since the
set $\mathbf{h}\left(\left[a,b\right]\right)$ is compact, $n$ can
be chosen sufficiently large so that any two successive $\mathbf{h}\left(\alpha=t_{ni}\right)$
and $\mathbf{h}\left(\beta=t_{n,i+1}\right)$ can be connected by
a straight line segment 
\[
\mathbf{s}_{ni}\left(t\right)=\mathbf{h}\left(\alpha\right)+\frac{\mathbf{h}\left(\beta\right)-\mathbf{h}\left(\alpha\right)}{\beta-\alpha}\left(t-\alpha\right).
\]
Then $n$ can further be increased to ensure
\[
1-\varepsilon<\frac{G\mathbf{\mathbf{h}}\left(\alpha\right)\mathbf{h}\left(\beta\right)}{\widehat{F}\left(\mathbf{\mathbf{h}}\left(\alpha\right),\mathbf{h}\left(\beta\right)\mathbf{-h}\left(\alpha\right)\right)}<1+\varepsilon
\]
and 
\[
1-\varepsilon<\frac{D\mathbf{s}_{ni}|\left[\alpha,\beta\right]}{\widehat{F}\left(\mathbf{\mathbf{h}}\left(\alpha\right),\mathbf{h}\left(\beta\right)\mathbf{-h}\left(\alpha\right)\right)}<1+\varepsilon.
\]
The latter follows from 
\[
D\mathbf{s}_{ni}|\left[\alpha,\beta\right]=\int_{\alpha}^{\beta}\widehat{F}\left(\mathbf{\mathbf{h}}\left(x\right),\mathbf{\dot{h}}\left(x\right)\right)\mathrm{d}x=\widehat{F}\left(\mathbf{\mathbf{h}}\left(\xi\right),\frac{\mathbf{h}\left(\beta\right)\mathbf{-h}\left(\alpha\right)}{\beta-\alpha}\right)\left(\beta-\alpha\right),
\]
for some $\alpha\leq\xi\leq\beta$. Combining the two double-inequalities,
for any $\delta>0$ and all sufficiently large $n$, 
\[
1-\delta<\frac{G\mathbf{\mathbf{h}}\left(t_{ni}\right)\mathbf{h}\left(t_{n,i+1}\right)}{D\mathbf{s}_{ni}|\left[t_{ni},t_{n,i+1}\right]}<1+\delta,
\]
whence 
\[
1-\delta<\frac{\sum_{i=0}^{k_{n}}G\mathbf{\mathbf{h}}\left(t_{ni}\right)\mathbf{h}\left(t_{n,i+1}\right)}{D\mathbf{s}_{n}|\left[a,b\right]}<1+\delta,
\]
where $\mathbf{s}_{n}|\left[a,b\right]$ is the \emph{piecewise linear}
path concatenating together $\mathbf{s}_{ni}|\left[t_{ni},t_{n,i+1}\right]$,
$i=0,\ldots,k_{n}$. By the definition of $D\mathbf{\mathbf{h}}|\left[a,b\right]$,
we have then 
\[
\lim_{n\rightarrow\infty}D\mathbf{s}_{n}|\left[a,b\right]=D\mathbf{\mathbf{h}}|\left[a,b\right].
\]
Since it is obvious that, as $n\rightarrow\infty$, $\mathbf{s}_{n}|\left[a,b\right]$
tends to $\mathbf{\mathbf{h}}|\left[a,b\right]$ pointwise, the theorem
is proved.
\end{proof}

\printindex

\end{document}